\documentclass[twocolumn, times, trackchanges]{aastex63_bren}

\usepackage[ruled,vlined]{algorithm2e}
\usepackage{graphicx}
\usepackage{bbm}
\usepackage{float}
\usepackage{xspace}
\usepackage[hypertexnames=false]{hyperref}
\usepackage{subfiles} 
\usepackage{amsmath}
\hypersetup{linkcolor=red,citecolor=green,filecolor=cyan,urlcolor=magenta}

\defcitealias{baron19}{B19}
\defcitealias{baron24}{B24}

\newcommand{\halpha}{H$\alpha$\xspace}
\newcommand{\hbeta}{H$\beta$\xspace}

\newcommand{\oiii}{$\text{[O\,{\sc iii}]}$\xspace}

\newcommand{\oi}{$\text{[O\,{\sc i}]}$\xspace}

\newcommand{\nii}{$\text{[N\,{\sc ii}]}$\xspace}

\newcommand{\sii}{$\text{[S\,{\sc ii}]}$\xspace}
\newcommand{\siii}{$\text{[S\,{\sc iii}]}$\xspace}
\newcommand{\hi}{$\text{H\,{\sc i}}$\xspace}
\newcommand{\hei}{$\text{He\,{\sc i}}$\xspace}

\newcommand{\oiiihbeta}{$\log (\text{[O\,{\sc iii}]}/\text{H}\beta)$\xspace}
\newcommand{\niihalpha}{$\log (\text{[N\,{\sc ii}]}/\text{H}\alpha)$\xspace}
\newcommand{\siihalpha}{$\log (\text{[S\,{\sc ii}]}/\text{H}\alpha)$\xspace}
\newcommand{\oihalpha}{$\log (\text{[O\,{\sc i}]}/\text{H}\alpha)$\xspace}

\newcommand{\mic}{$\mathrm{\mu m}$\xspace}

\shorttitle{DIGging into HII regions}

\begin{document}

\title{DIGging into HII regions: the role of ionizing photon leakage and spectral hardening in coupling HII regions and diffuse ionized gas}

\author[0000-0003-4974-3481]{Dalya Baron}
\email{dalyabaron@gmail.com}
\affiliation{Kavli Institute for Particle Astrophysics \& Cosmology, Stanford University, CA 94305, USA}
\affiliation{Center for Decoding the Universe, Stanford University, CA 94305, USA}
\affiliation{Racah Institute of Physics, The Hebrew University of Jerusalem, 9190401, Israel}

\author[0000-0001-6551-3091]{Kathryn Kreckel}
\affiliation{Universit\"at Heidelberg, Zentrum f\"ur Astronomie, Astronomisches Rechen-Institut, M\"onchhofstra{\ss}e 12-14, 69120 Heidelberg, Germany}

\author[0000-0003-4974-3481]{Gwen C. Rudie}
\affiliation{Observatories of the Carnegie Institution for Science, 813 Santa Barbara Street, Pasadena, CA 91101, USA}

\author[0000-0002-4378-8534]{Karin M. Sandstrom}
\affiliation{Department of Astronomy \& Astrophysics, University of California, San Diego. 9500 Gilman Drive, La Jolla, CA 92093, USA}

\author[0000-0002-5480-5686]{Alberto D. Bolatto}
\affiliation{Department of Astronomy and Joint Space-Science Institute, University of Maryland, College Park, MD 20742, USA}

\author[0000-0003-4218-3944]{Guillermo A. Blanc}
\affiliation{Observatories of the Carnegie Institution for Science, 813 Santa Barbara Street, Pasadena, CA 91101, USA}
\affiliation{Departamento de Astronom\'{i}a, Universidad de Chile, Camino del Observatorio 1515, Las Condes, Santiago, Chile}

\author[0000-0002-4755-118X]{Oleg V. Egorov}
\affiliation{Universit\"at Heidelberg, Zentrum f\"ur Astronomie, Astronomisches Rechen-Institut, M\"onchhofstra{\ss}e 12-14, 69120 Heidelberg, Germany}

\author[0000-0002-3933-7677]{Eva Schinnerer}
\affiliation{Max-Planck-Institut f\"{u}r Astronomie, K\"{o}nigstuhl 17, D-69117 Heidelberg, Germany}

\author[0000-0003-0410-4504]{A.~T.~Barnes}
\affiliation{European Southern Observatory (ESO), Karl-Schwarzschild-Stra{\ss}e 2, 85748 Garching, Germany}

\author[0000-0002-8549-4083]{Enrico Congiu}
\affiliation{European Southern Observatory (ESO), Alonso de C\'ordova 3107, Casilla 19, Santiago 19001, Chile}

\author[0000-0002-5782-9093]{Daniel~A.~Dale}
\affiliation{Department of Physics and Astronomy, University of Wyoming, Laramie, WY 82071, USA}

\author[0000-0001-6708-1317]{Simon C.~O.\ Glover}
\affiliation{Universit\"{a}t Heidelberg, Zentrum f\"{u}r Astronomie, Institut f\"{u}r Theoretische Astrophysik, Albert-Ueberle-Str.\ 2, 69120 Heidelberg, Germany}

\author[0000-0002-0432-6847]{Jaeyeon Kim}
\affiliation{School of Space Research, Kyung Hee University, 1732 Deogyeong-daero, Yongin-si, Gyeonggi-do 17104, Korea}
\affiliation{Department of Astronomy \& Space Science, Kyung Hee University, 1732 Deogyeong-daero, Yongin-si, Gyeonggi-do 17104, Korea}

\author[0000-0002-0560-3172]{Ralf S.\ Klessen}
\affiliation{Universit\"{a}t Heidelberg, Zentrum f\"{u}r Astronomie, Institut f\"{u}r Theoretische Astrophysik, Albert-Ueberle-Str.\ 2, 69120 Heidelberg, Germany}
\affiliation{Universit\"{a}t Heidelberg, Interdisziplin\"{a}res Zentrum f\"{u}r Wissenschaftliches Rechnen, Im Neuenheimer Feld 225, 69120 Heidelberg, Germany}

\author[0000-0002-6972-6411]{Jos\'e Eduardo M\'endez-Delgado}
\affiliation{Instituto de Astronom\'ia, Universidad Nacional Aut\'onoma de M\'exico, Circuito Exterior s/n, Ciudad Universitaria, Coyoac\'an, 04510 Ciudad de M\'exico, Mexico}

\author[0000-0002-3289-8914]{Justus Neumann}
\affiliation{Max-Planck-Institut f\"{u}r Astronomie, K\"{o}nigstuhl 17, D-69117 Heidelberg, Germany}

\author[0000-0002-0873-5744]{Ismael Pessa}
\affiliation{Leibniz-Institut for Astrophysik Potsdam (AIP), An der Sternwarte 16, 14482 Potsdam, Germany}

\author[0000-0002-0012-2142]{Thomas G. Williams}
\affiliation{UK ALMA Regional Centre Node, Jodrell Bank Centre for Astrophysics, Department of Physics and Astronomy, The University of Manchester, Oxford Road, Manchester M13 9PL, UK}

\suppressAffiliations



\begin{abstract}

The optical emission-line ratios of star-forming galaxies are broadly understood in the framework of photoionization, yet some systematic trends remain unexplained. Using 50--100 pc resolution PHANGS-MUSE observations of 19 nearby galaxies, we investigate how escaping ionizing radiation couples HII regions to the diffuse ionized gas (DIG), analyzing thousands of HII regions and DIG sightlines. HII regions and the DIG form a continuous sequence across standard line-ratio diagnostic diagrams (e.g., BPT diagrams), where progressively fainter regions show elevated high- and low-ionization line ratios, and the DIG extends the same trends into the composite--LINER regime. Line-ratio variations with global specific star formation rate (sSFR) suggest progressively harder ionizing radiation with increasing star formation activity, contrary to expectations for emission increasingly dominated by older stellar populations. Line-of-sight mixing explains part of the behavior but cannot reproduce all line-ratio trends. We therefore develop a model in which ionizing photons escape HII regions and propagate through the diffuse ISM, undergoing geometric dilution, attenuation, and spectral hardening. The model predicts that progressively fainter HII regions are increasingly ionized by leaked radiation from neighboring luminous regions, implying radiative coupling between HII regions. Joint modeling of HII regions and the DIG reproduces the HII region luminosity distribution and DIG surface brightness for escape fractions of $\sim$0.4$-$0.6 and ionizing-photon mean free paths of $\sim$0.5$-$1 kpc, and qualitatively reproduces the observed line-ratio sequences and their dependence on sSFR. Our results suggest that ionizing photon leakage may provide an important non-local feedback channel linking compact star-forming sites to large-scale galaxy disks.
\end{abstract}


\section{Introduction}\label{sec:intro}

Massive stars are key drivers of interstellar medium (ISM) evolution in star-forming galaxies. Through ionizing radiation, winds, and supernova explosions, they inject energy and momentum that shape ISM structure and regulate subsequent star formation \citep{hopkins14, klessen16, gatto17}. Among these feedback channels, ionizing radiation provides a particularly direct link between young stellar populations and the surrounding gas, acting immediately over the lifetimes of massive stars and influencing both dense star-forming regions and gas on larger galactic scales.

The immediate ionized environments of young massive stars are observed as HII regions, whose optical line emission traces nebular gas conditions and recent star formation \citep{kennicutt89, maiolino19, kewley19}. Studies of HII regions in nearby galaxies have constrained local feedback and cloud-dispersal timescales \citep{lopez11, lopez14, mcleod20, barnes21, DellaBruna21, chevance20, chevance22, kimJ25, pathak25}, calibrated star formation timescales and relations \citep{kennicutt89, schinnerer19, chevance20}, and mapped metallicities and abundance gradients \citep{pilyugin16, berg20, kreckel19, groves23, brazzini24}. HII regions therefore have a central role in connecting star formation, massive-star feedback, and the chemical evolution of galaxies.

Some fraction of ionizing photons, however, escapes beyond these dense environments. Observations suggest that 10\% to 80\% may leak into the surrounding medium \citep{zurita02, oey07, pellegrini12, mcleod19, mcleod20, DellaBruna21, belfiore22, teh23, scheuermann26}, plausibly powering the DIG (\citealt{haffner09}): a warm ($T\sim10^4$~K), tenuous ($n_e\sim10^{-2}$--$10^{-1}\,\mathrm{cm^{-3}}$) plasma that permeates galactic disks, reaches scale heights of order kpc \citep{reynolds91, ferriere01, miller03, gaensler08, levy19}, and that can account for a large fraction of the total H$\alpha$ luminosity of some galaxies \citep{oey07, kreckel16, belfiore22}. The DIG is spatially associated with concentrations of HII regions \citep{miller03, madsen06, seon09, howard18, pellegrini20a, belfiore22}, supporting the leaked-radiation picture. 

A central uncertainty is whether leaked radiation alone can explain the DIG ionization state or whether additional sources of ionization and heating are required. Proposed additional contributors include stellar populations with harder ionizing spectra, such as hot low-mass evolved stars (HOLMES; \citealt{stasinska08, cidfernandes10, flores-fajardo11, belfiore16, zhang17, byler19, belfiore22}) and 5--25 Myr stellar populations \citep{mcclymont24}, as well as non-stellar heating mechanisms including shocks \citep{dopita95, allen08}, cosmic rays \citep{wiener13, vandenbroucke18}, and turbulent mixing layers \citep{rand98, binette09}.

Determining what powers the DIG is important for constraining the local escape of ionizing photons from HII regions and, thus, the feedback energy budget available for dispersing natal clouds and regulating the star formation cycle (e.g., \citealt{chevance20, mcleod20, chevance22, kimJ23, kimJ25}). It may also provide insight into how escaping radiation is spectrally reshaped as it propagates through the ISM \citep{weber19, pellegrini20a, tacchella22}, a process that influences the interaction of ionizing photons with multiphase gas on galactic scales (e.g., \citealt{hopkins14, klessen16, naab17}).

Leaking radiation implies a physical coupling between compact HII regions and the large-scale diffuse ionized medium. Characterizing this connection observationally requires (i) identifying individual HII regions, (ii) mapping the faint DIG across entire galactic disks, and (iii) obtaining spectroscopic diagnostics for both components. Previous surveys approached these requirements with complementary strengths. Targeted slit spectroscopy of nearby galaxies cataloged $\sim$100s of HII regions per system \citep{berg15, croxall15, croxall16, berg20}, enabling detailed measurements of HII region properties but with limited information on the surrounding environments. Narrow-band imaging surveys mapped HII regions and the DIG down to low surface brightness levels (e.g., \citealt{walterbos92, oey07}), but often lacked the spectral information needed to constrain ionization conditions. Wide-field integral-field spectroscopic surveys such as CALIFA \citep{sanchez12}, SAMI \citep{croom12}, and MaNGA \citep{bundy15} subsequently provided galaxy-wide coverage with spatially resolved spectroscopy, but their $\sim$kpc resolution blended emission from multiple HII regions and the surrounding DIG within individual spaxels, limiting direct tests of how compact regions and diffuse gas are physically connected.

With MUSE observations of nearby galaxies ($D \lesssim 20$ Mpc), surveys such as PHANGS (The Physics at High Angular resolution in Nearby GalaxieS; \citealt{emsellem22}) and MAUVE (Multiphase Astrophysics to Unveil the Virgo Environment; \citealt{catinella25, brown26}) resolve the ionized ISM on $\sim$50--100 pc scales while covering entire disks. This resolution enables a cleaner separation between compact HII regions and the DIG, reducing (though not eliminating) the component blending that is unavoidable in kpc-scale surveys. In particular, PHANGS-MUSE simultaneously catalogs hundreds to thousands of HII regions per galaxy \citep{santoro22, congiu23, groves23} and maps the surrounding DIG across the full field of view \citep{belfiore22}, providing both the statistical power to identify trends and the spatial context needed to connect HII regions to their diffuse environments.

At PHANGS-MUSE resolution, previous work has shown that HII regions and the DIG each form tight diagonal sequences in standard line-diagnostic diagrams (e.g., BPT diagrams). Integrated HII regions occupy such sequences within the star-forming locus (\citealt{groves23, congiu23}), while the DIG follows a diagonal sequence from the star-forming locus to the LINER regime \citep{belfiore22}. Here, we show that these are not separate relations, but a single continuous sequence spanning HII regions and the DIG, which is also reflected in their trends with \halpha\ surface brightness.

A natural starting point for interpreting these trends is the leaking-radiation framework. Using PHANGS-MUSE, \citet{belfiore22} constructed a radiation propagation model of the DIG in which ionizing photons leaking from the observed HII region population reproduce the \halpha\ surface-brightness distribution, for an escape fraction of $\sim$40\% and an ionizing-photon mean free path of $\sim$2 kpc. However, their framework, which accounted for attenuation and geometric dilution of the escaping radiation, could not reproduce the observed line-ratio sequence in the DIG, requiring an additional hard ionizing component that they attributed to HOLMES.

Here we build on and extend the framework of \citet{belfiore22} by analyzing HII regions and the DIG jointly and investigating the physical origin of their continuous sequence. We examine two mechanisms that may contribute to the observed continuity. First, line-of-sight mixing can produce intermediate line ratios, as the vertically extended DIG overlaps HII regions in projection and biases their integrated spectra toward DIG-like emission. Second, radiative coupling through leaked ionizing photons can generate a physically connected sequence, as photons escaping HII regions propagate through the diffuse ISM, undergoing attenuation and spectral hardening, and create a non-local radiation field that influences both the DIG and neighboring HII regions. Relative to \citet{belfiore22}, we extend the radiation propagation model by allowing leaked radiation from HII regions to influence neighboring HII regions and by incorporating spectral hardening as the escaping radiation propagates through the ISM.

The paper is organized as follows. Section~\ref{sec:data} describes the PHANGS-MUSE data products used here. In Section~\ref{sec:line_variations_HII_DIG} we characterize the connected emission-line sequences and their dependence on H$\alpha$ surface brightness, local environment, and global galaxy properties. In Section~\ref{sec:mixing} we test the line-of-sight mixing scenario with composite spectral models. In Section~\ref{sec:leaking} we examine a radiation propagation model including attenuation and spectral hardening and assess whether it reproduces the observed trends. We discuss implications in Section~\ref{sec:discussion} and summarize in Section~\ref{sec:summary}.

\floattable
\begin{deluxetable}{ccCClC CCCCC Crr}
\tablecaption{Galaxy sample\label{tab:galaxy_properties}}
\tablecolumns{13}
\tablenum{1}
\tablewidth{0pt}
\tablehead{
\colhead{(1)} & \colhead{(2)} & \colhead{(3)} & \colhead{(4)} & \colhead{(5)} & \colhead{(6)} & \colhead{(7)} & \colhead{(8)} & \colhead{(9)} & \colhead{(10)} & \colhead{(11)} & \colhead{(12)} & \colhead{(13)} \\
\colhead{Galaxy} & \colhead{D} & \colhead{$\log M_{*}$} & \colhead{SFR} & \colhead{AGN?} & \colhead{$i$} & \colhead{PA} & \colhead{$C_{\rm opt}$} & \colhead{$C_{\rm opt}$} & \colhead{12+log(O/H)} & \colhead{12+log(O/H)} & \colhead{$N_{\mathrm{HII}}$} & \colhead{$N_{\mathrm{DIG}}$} \\
\colhead{} & \colhead{[Mpc]} & \colhead{[$\log \mathrm{M_{\odot}}$]} & \colhead{[$\mathrm{M_{\odot}\,yr^{-1}}$]} & \colhead{} & \colhead{[deg]} & \colhead{[deg]} & \colhead{[arcsec]} & \colhead{[pc]} & \colhead{median} & \colhead{std} & \colhead{} & \colhead{}
}
\startdata
NGC~5068 & 5.2 & 9.4 & 0.3 &  & 36 & 342 & 1.04 & 26 & 8.308 & 0.081 & 1484 & 16599 \\
IC~5332 & 9.0 & 9.7 & 0.4 &  & 27 & 74 & 0.87 & 38 & 8.381 & 0.085 & 470 & 4611 \\
NGC~1087 & 15.9 & 9.9 & 1.3 &  & 43 & 359 & 0.92 & 71 & 8.408 & 0.048 & 868 & 6967 \\
NGC~2835 & 12.2 & 10.0 & 1.3 &  & 41 & 1 & 1.15 & 68 & 8.380 & 0.104 & 836 & 12259 \\
NGC~1385 & 17.2 & 10.0 & 2.1 &  & 44 & 181 & 0.77 & 64 & 8.422 & 0.088 & 914 & 5632 \\
NGC~7496 & 18.7 & 10.0 & 2.2 & yes & 36 & 194 & 0.89 & 81 & 8.499 & 0.059 & 479 & 4356 \\
NGC~628 & 9.8 & 10.3 & 1.7 &  & 9 & 21 & 0.92 & 44 & 8.483 & 0.052 & 2182 & 18068 \\
NGC~3351 & 10.0 & 10.4 & 1.3 &  & 45 & 193 & 1.05 & 51 & 8.587 & 0.044 & 537 & 11033 \\
NGC~4254 & 13.1 & 10.4 & 3.1 &  & 34 & 68 & 0.89 & 57 & 8.532 & 0.041 & 2531 & 20552 \\
NGC~4303 & 17.0 & 10.5 & 5.4 & yes & 24 & 312 & 0.78 & 64 & 8.560 & 0.038 & 2161 & 15860 \\
NGC~4535 & 15.8 & 10.5 & 2.2 &  & 45 & 180 & 0.56 & 43 & 8.542 & 0.041 & 1033 & 9519 \\
NGC~1300 & 19.0 & 10.6 & 1.2 &  & 32 & 278 & 0.89 & 82 & 8.517 & 0.050 & 650 & 8337 \\
NGC~1512 & 18.8 & 10.7 & 1.3 &  & 43 & 262 & 1.25 & 114 & 8.554 & 0.042 & 434 & 6388 \\
NGC~1672 & 19.4 & 10.7 & 7.6 & yes & 43 & 134 & 0.96 & 90 & 8.536 & 0.038 & 1069 & 13417 \\
NGC~1566 & 17.7 & 10.8 & 4.6 & yes & 30 & 215 & 0.8 & 69 & 8.550 & 0.047 & 1590 & 13691 \\
NGC~3627 & 11.3 & 10.8 & 3.9 &  & 57 & 173 & 1.05 & 58 & 8.545 & 0.033 & 1015 & 13355 \\
NGC~4321 & 15.2 & 10.8 & 3.5 &  & 39 & 156 & 1.16 & 85 & 8.558 & 0.038 & 1141 & 17705 \\
NGC~1433 & 18.6 & 10.9 & 1.1 &  & 29 & 200 & 0.91 & 82 & 8.539 & 0.074 & 663 & 11406 \\
NGC~1365 & 19.6 & 11.0 & 17 & yes & 55 & 201 & 1.15 & 109 & 8.534 & 0.053 & 845 & 18153 \\
\enddata
\tablecomments{(1) Galaxy name. (2) Distance from \citet{lee23}, based on \citet{anand21a, anand21b, kourkchi17, shaya17}. (3) Stellar mass \citep{leroy21a}. (4) Star formation rate \citep{leroy21a}. (5) AGN classification from \citet{veron_cetty10} (only Seyfert nuclei are considered AGN).  (6)–(7) Inclination and position angle \citep{lang20, leroy21a}. (8)–(9) FWHM of the homogenized MUSE PSF, in arcsec and pc, respectively.  (10)–(11) Median and standard deviation of HII region metallicities from the \citealt{groves23} catalog, computed using the $S$-cal strong-line calibration \citep{pilyugin16} (12) Number of HII regions with \halpha, \hbeta, \oiii, \oi, \nii, and \sii\ detected at S/N $> 5$. (13) Number of DIG Voronoi bins considered in this work, based on \citealt{belfiore22}.}
\end{deluxetable}

\begin{figure*}
	\centering
\includegraphics[width=1\textwidth]{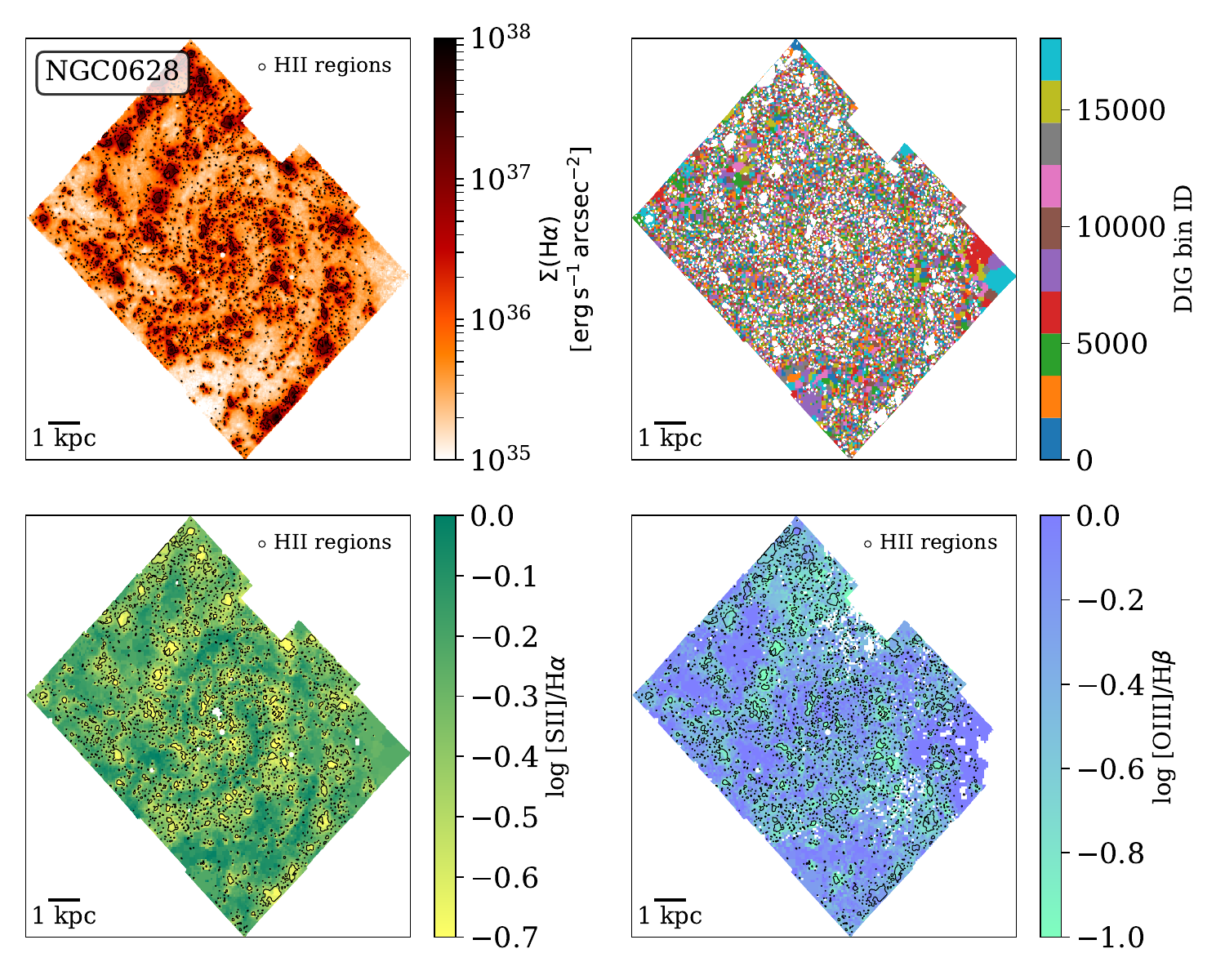}
\caption{\textbf{HII-region and DIG data for NGC~628 from PHANGS-MUSE.} The top-left panel shows the \halpha\ surface-brightness ($\Sigma_{\mathrm{H}\alpha}$) map across the field of view, after summing $4\times4$ pixels so that the displayed pixels approximately match the angular resolution ($C_{\rm opt}=0.92''$; original pixel size $0.2''$). HII regions identified in the PHANGS nebular catalog are outlined in black. Throughout this work, HII-region measurements are obtained by integrating the spectra within these masks, while the DIG is defined as all emission outside the HII-region masks. The top-right panel shows the Voronoi tessellation used to bin the DIG spectra after masking the HII regions, illustrating the spatial scale of the DIG bins used to recover faint metal-line emission (\oiii, \oi, \nii, and \sii). The bottom panels show the \sii/\halpha\ (left) and \oiii/\hbeta\ (right) line-ratio maps, measured in individual $0.2''$ spaxels within HII regions and in Voronoi bins within the DIG. The x- and y-axes correspond to right ascension and declination, respectively. }
\label{f:NGC0628_IFU_data}
\end{figure*}

\section{Data}\label{sec:data}

The PHANGS-MUSE survey obtained spatially resolved optical spectroscopy of 19 nearby ($D < 20$ Mpc), moderately inclined galaxies with stellar masses in the range $\log (M/M_{\odot}) = 9.4$--11, lying close to the star-forming main sequence \citep{emsellem22}. The observations were carried out with the MUSE integral-field unit (IFU; \citealt{bacon10}) on the Very Large Telescope (VLT), covering 4750--9350 $\mathrm{\AA}$ at a spectral resolution of $\sim$2.5 $\mathrm{\AA}$. The angular resolution of the cubes ranges from 0.56$''$ to 1.25$''$, corresponding to physical scales of 26--114 pc at the distances of the galaxies. Table \ref{tab:galaxy_properties} summarizes the main properties of the galaxies and the derived quantities used in this paper.

The data reduction and analysis pipelines are described in \citet{emsellem22}. In brief, the individual MUSE pointings were homogenized and combined into a single data cube per galaxy with a uniform point-spread function (PSF), characterized by a full width at half maximum (FWHM) of $C_{\mathrm{opt}}$; stellar continuum was fitted with stellar population synthesis models to derive stellar population and kinematic properties; and the optical emission lines were fitted using the stellar-subtracted spectra to estimate their fluxes and kinematics. We use two products derived from the MUSE cubes that are designed to optimally characterize HII regions and the DIG, as described below. These are illustrated in Figure~\ref{f:NGC0628_IFU_data} for NGC~628. Throughout the paper, we refer to the \halpha luminosity surface density, measured in units of $\mathrm{erg\,s^{-1}\,arcsec^{-2}}$ or $\mathrm{erg\,s^{-1}\,pc^{-2}}$, as $\Sigma$(\halpha). 

\subsection{HII region catalog}\label{sec:data:HII_region_catalogs}

We use the HII region catalog constructed by \citet{santoro22} and \citet{groves23}. This catalog is based on emission-line maps derived from the MUSE cubes, in which line-emitting nebulae are identified by applying {\sc hiiphot} \citep{thilker00} to \halpha surface brightness maps. Emission lines are integrated across the identified nebulae. Nebulae are then classified as HII regions, planetary nebulae, or supernova remnants using standard diagnostic diagrams \citep{baldwin81, veilleux87}. Specifically, HII regions are defined as those lying below the \citet{kauff03a} curve in the \niihalpha--\oiiihbeta diagram and below the \citet{kewley06} curve in the \siihalpha--\oiiihbeta and \oihalpha--\oiiihbeta diagrams. In total, 23,244 HII regions are detected across the 19 galaxies. Here we use the HII regions with line fluxes measured at S/N$>5$ in \halpha, \hbeta, \oiii, \oi, \nii, and \sii. We list the number of HII regions per galaxy in Table~\ref{tab:galaxy_properties}. In addition to line fluxes, we use the dust-corrected \halpha luminosities, centroid coordinates, the areas (in pixels and arcsec$^{2}$) of the nebulae, and the metallicities derived using the $S$-cal prescription by \citet{pilyugin16}.


\subsection{DIG surface brightness maps}\label{sec:data:dig_voronoi_bins}

The depth of the MUSE observations, corresponding to an \halpha surface brightness of $1.5 \times 10^{-17}\, \mathrm{erg\,s^{-1}\,cm^{-2}\,arcsec^{-2}}$ for a 3$\sigma$ detection per 0.2$''$ pixel, is sufficient to detect \halpha emission in 95\% of the MUSE spaxels \citep{emsellem22}. The metal lines (e.g., \oiii, \oi, \sii, and \nii), on the other hand, are not detected with comparable significance across the field of view, especially in low-surface-brightness regions. Because a proper characterization of the DIG requires measurements of these metal lines at low surface brightness, stacking of neighboring pixels is needed to increase the S/N. We use the DIG surface-brightness maps produced by \citet{belfiore22}, who performed spatial stacking in two steps. First, they created DIG-only maps by masking out nebulae from the \citet{groves23} nebular catalog. The DIG-only maps were then re-gridded to 1.4$''$ pixels and Voronoi binned to a target \halpha S/N of 60, which ensures $>5\sigma$ detections of the metal lines even in low-surface-brightness regions. We use the Voronoi-binned maps from \citet{belfiore22}, and list the number of DIG bins in Table~\ref{tab:galaxy_properties}.

\begin{figure*}
	\centering
\includegraphics[width=1\textwidth]{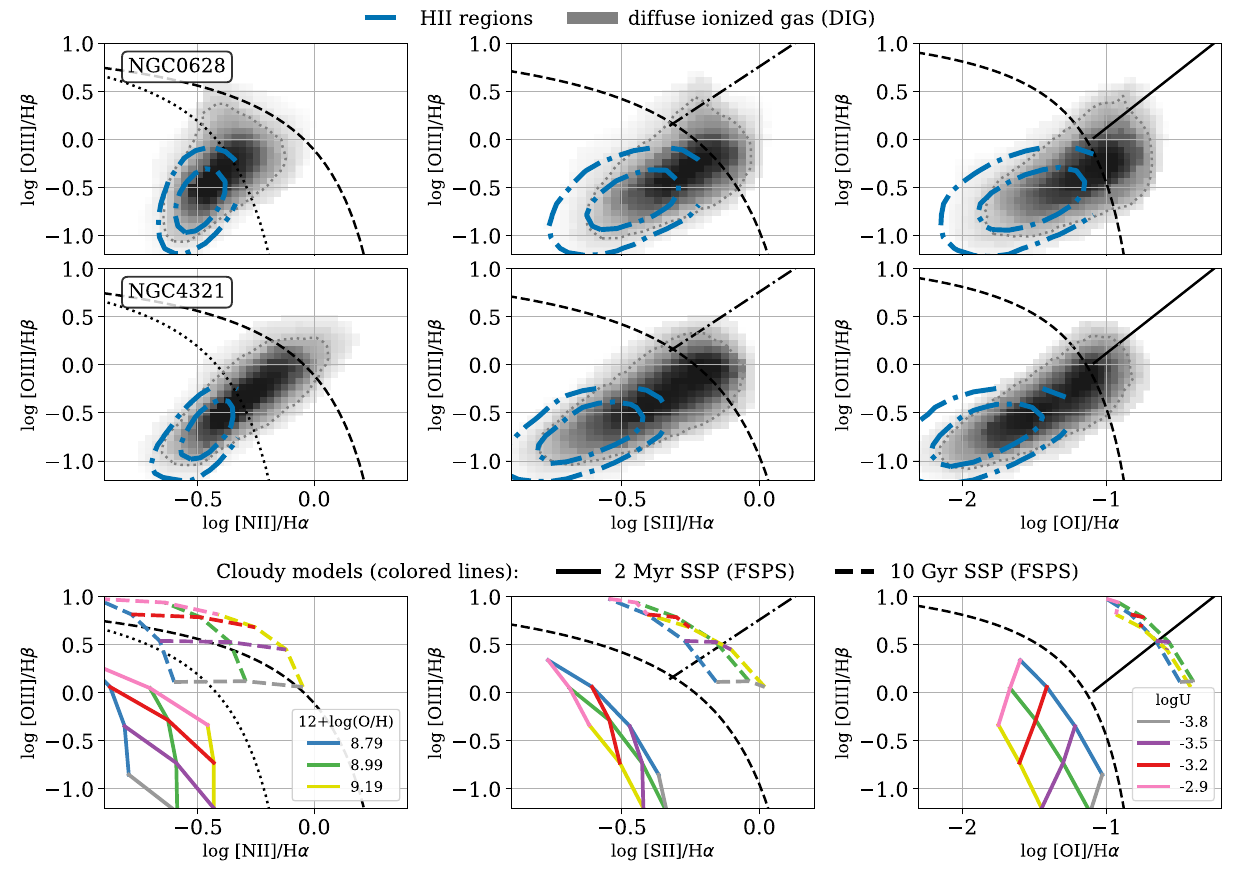}
\caption{\textbf{Optical line-ratio sequences from HII regions to the DIG compared to model expectations.} 
\textbf{Top two rows:} BPT diagrams of \nii/\halpha\ (left), \sii/\halpha\ (middle), and \oi/\halpha\ (right) versus \oiii/\hbeta\ for NGC~628 (top) and NGC~4321 (middle). Blue contours show the 50th and 95th percentiles of the HII region distributions, based on 2132 (NGC~628) and 1141 (NGC~4321) individual regions. The DIG is shown as a logarithmically scaled grayscale density map, with the gray dotted contour marking its 95th percentile. 
\textbf{Bottom row:} {\sc cloudy} model grids illustrating the dependence of the line ratios on gas-phase metallicity (and stellar metallicity), ionization parameter ($\log U$), and the shape of the ionizing radiation field. The models use FSPS spectra \citep{conroy09} for 2 Myr stellar populations (solid lines; softer ionizing spectra with lower $\mathrm{Q(O^{++})/Q(H^{+})}$ ratios) and 10 Gyr stellar populations (dashed lines; harder ionizing spectra with higher $\mathrm{Q(O^{++})/Q(H^{+})}$ ratios). Black demarcation curves from the literature are shown as dashed \citep{kewley01}, dotted \citep{kauff03a}, dash-dotted \citep{kewley06}, and solid \citep{cidfernandes10}.
\textbf{Summary:} HII regions and the DIG form a single continuous sequence across emission-line diagnostic diagrams, consistent with progressive hardening of the ionizing radiation field.
}
\label{f:BPT_tracks}
\end{figure*}

\section{Galaxy-wide emission-line sequences connecting HII regions to the DIG}\label{sec:line_variations_HII_DIG}

HII regions and the DIG represent two physically distinct regimes: compact, high-surface-brightness nebulae ($\sim$10–100 pc) versus an extended, faint ionized component permeating galactic disks. Motivated by the possibility that they are physically linked through ionizing radiation leakage, we analyze their emission-line properties jointly to assess this connection. 

In Section~\ref{sec:global_HII_DIG_variations_line_ratios} we characterize the sequences across line ratio diagnostic diagrams from HII regions to the DIG. In Section~\ref{sec:global_HII_DIG_variations_line_ratios_with_sigma} we study the dependence of line ratios on $\Sigma$(\halpha) and global galaxy properties. In Section~\ref{sec:HII_env_dependence} we examine how the line ratios trends with $\Sigma$(\halpha) for HII regions further depend on the local (radiative) environment.

\subsection{Line ratio sequences from HII regions to the DIG}\label{sec:global_HII_DIG_variations_line_ratios}

Figure~\ref{f:BPT_tracks} shows the distribution of HII regions and DIG emission, integrated within HII-region apertures and DIG bins, respectively, in the \nii/\halpha, \sii/\halpha, and \oi/\halpha\ versus \oiii/\hbeta\ line-ratio diagrams for NGC~628 and NGC~4321. The corresponding diagrams for the full PHANGS-MUSE sample are presented in Figures~\ref{f:BPT_tracks_all_gals_NII}--\ref{f:BPT_tracks_all_gals_OI} in \ref{app:bpt_trends_all_galaxies}. Many galaxies exhibit well-defined diagonal sequences in which the low-ionization line ratios \nii/\halpha, \sii/\halpha, and \oi/\halpha\ increase in tandem with the high-ionization ratio \oiii/\hbeta. Across the sample, HII regions (numbering from hundreds to thousands per galaxy) and the DIG broadly trace the same sequence: HII regions define the trend across the classical star-forming locus, while the DIG extends the same trend toward the composite and LINER regimes.

The bottom panel of Figure~\ref{f:BPT_tracks} presents a grid of {\sc cloudy} \citep{ferland17, chatzikos23} photoionization models illustrating the dependence of these line ratios on gas-phase (and stellar) metallicity, ionization parameter ($\log U$), and the spectral shape of the ionizing radiation field. The photoionization model largely follows standard assumptions, except for the adopted chemical abundances, which were optimized to reproduce the observed HII-region distribution across the PHANGS galaxies (see \ref{app:cloudy_grids} for details). Variations in metallicity and ionization parameter produce well-known trends across BPT space that have been extensively characterized in the literature (e.g., review by \citealt{kewley19}). 

The metallicities shown in the figure are $\sim$0.4 dex higher than those inferred for PHANGS HII regions using the $S$-cal calibration \citep{pilyugin16}. This offset reflects the well-known systematic differences between strong-line calibrations anchored to auroral-line measurements and those based on photoionization models (e.g., \citealt{kewley02, stasinska05, blanc15, dopita16}). Such discrepancies may arise from biases in auroral-line metallicities at high metallicity \citep{stasinska05}. Adopting instead the photoionization-based calibration of \citet{dopita16} brings the PHANGS metallicity scale into closer agreement with the range explored by the models, $12+\log(\mathrm{O/H}) \sim 9$ (see figure 5 in \citealt{groves23}).

To illustrate the role of spectral hardness, we adopt {\sc fsps} stellar population synthesis models \citep{conroy09, conroy10}, considering single stellar populations (SSPs) with ages of 2 Myr and 10 Gyr. The young SSP approximates the radiation field of massive stars powering HII regions and is characterized by a relatively soft ionizing spectrum, with $\mathrm{Q(O^{++})/Q(H^{+})} \sim 0.1$, where $\mathrm{Q(H^{+})}$ and $\mathrm{Q(O^{++})}$ denote the production rates of photons capable of ionizing H and O$^{+}$, respectively. In contrast, the old SSP represents evolved stellar populations, including contributions from HOLMES, and produces a harder ionizing spectrum with $\mathrm{Q(O^{++})/Q(H^{+})} \sim 0.4$. These populations also differ in their ionizing photon output, which is reflected in their predicted EW(\halpha). The young SSP yields EW(\halpha)$\sim$2000-2500~$\mathrm{\AA}$, while the old SSP produces much lower values of EW(\halpha)$\sim$0.5-2.5~$\mathrm{\AA}$.

Increasing the hardness of the radiation field produces a simultaneous rise in both low- and high-ionization line ratios, qualitatively reproducing the observed diagonal sequences. Since the rise in low- and high-ionization lines is also accompanied by decreasing EW(\halpha) (see e.g., \citealt{zhang17, belfiore22}), this behavior has motivated studies to attribute the increasingly LINER-like emission across the DIG to an increasing contribution from HOLMES (e.g., \citealt{stasinska08, cidFernandes11, flores-fajardo11, zhang17, byler19, belfiore22}). 

\begin{figure*}
	\centering
\includegraphics[width=1\textwidth]{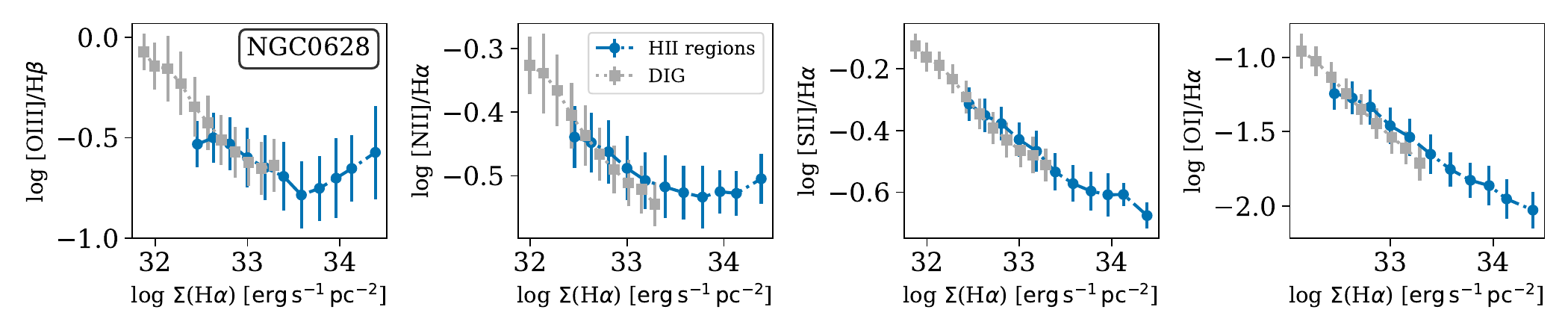}
\caption{\textbf{Line-ratio variations with the \halpha luminosity surface density from HII regions to the DIG in NGC~628.} Each panel shows the median and median absolute deviation of line ratios in bins of \halpha luminosity surface density, $\Sigma$(\halpha). The blue round markers and dash-dotted lines represent stacks of HII regions, while the gray square markers and dotted lines represent stacks across the DIG.
\textbf{Summary:} HII regions and the DIG form continuous sequences as a function of \halpha\ surface brightness, with line ratios increasing toward lower $\Sigma$(\halpha). This suggests that the trends seen in the BPT diagrams (Figure~\ref{f:BPT_tracks}) are luminosity-dependent.
 }
\label{f:line_ratios_vs_LHa}
\end{figure*}

However, interpreting the continuity of the HII-region and DIG sequences is less straightforward. While an increasing contribution from HOLMES may account for the progressively LINER-like line ratios of the DIG, the same explanation is unlikely to apply to the HII-region population, which follows the same sequence. HII regions are typically an order of magnitude brighter than the surrounding DIG and have EW(\halpha) values far above those expected from HOLMES. HOLMES-powered emission has EW(\halpha)$\sim0.5$--$2.5\,\mathrm{\AA}$ (e.g., \citealt{cidFernandes11}), whereas PHANGS HII regions have a median EW(\halpha) of $30\,\mathrm{\AA}$ (10th--90th percentiles: $10$--$100\,\mathrm{\AA}$), implying negligible HOLMES contributions for most of the population. The continuity of the HII-region and DIG sequences therefore points to a different mechanism capable of linking both components. In Sections~\ref{sec:mixing} and \ref{sec:leaking}, we investigate whether projection effects and radiative coupling through leaked ionizing photons can account for the observed sequence.

Several galaxies deviate from these otherwise well-defined trends for different reasons (Figures~\ref{f:BPT_tracks_all_gals_NII}--\ref{f:BPT_tracks_all_gals_OI} in \ref{app:bpt_trends_all_galaxies}). First, a subset of galaxies exhibits significantly broader gas-phase metallicity distributions (by a factor of $\sim$2–3) compared to the rest of the sample. These are generally lower-mass systems ($\log M/\mathrm{M_{\odot}} \sim 9.4$–10), including NGC~5068, IC~5332, NGC~2835, and NGC~1385. In these galaxies, the strong dependence of line ratios on metallicity dominates, which may wash out the diagonal sequences in BPT space. Second, galaxies hosting AGN display some deviation from the star-forming to LINER sequence, with spaxels extending into the Seyfert regime. This effect is most pronounced in NGC~1365 and NGC~7496, and more weakly in NGC~1672. Third, several galaxies contain extended ``star-formation deserts", where little or no ongoing star formation is present and the emission is dominated by DIG over large galactocentric radii. In these systems, HII regions and DIG do not trace the same spatial regions, potentially introducing systematic offsets due to radial metallicity gradients (e.g., \citealt{kreckel19}). Galaxies in this category, including NGC~3351, NGC~4535, NGC~1300, NGC~1512, and NGC~1433, still show diagonal trends, but these appear more bimodal than continuous. In summary, only galaxies with broad metallicity distributions fail to exhibit clear sequences in BPT space, while the remainder show such sequences, albeit with varying degrees of deviation.

\subsection{Dependence of line ratios on $\Sigma$(\halpha) and global galaxy properties}\label{sec:global_HII_DIG_variations_line_ratios_with_sigma}

How do these BPT sequences depend on \halpha\ luminosity? For the DIG, \citet{belfiore22} showed that the line ratios increase systematically with decreasing $\Sigma$(\halpha). In Figure~\ref{f:line_ratios_vs_LHa}, we examine the dependence of line ratios on $\Sigma$(\halpha) for both HII regions and the DIG in NGC~628. For HII regions, $\Sigma$(\halpha) is defined as the dust-corrected \halpha\ luminosity divided by the nebula area in $\mathrm{pc}^2$ units, while for the DIG it is divided by the DIG Voronoi-bin area. We show the median and median absolute deviation of \oiii/\hbeta, \nii/\halpha, \sii/\halpha, and \oi/\halpha\ in bins of $\Sigma$(\halpha).

Figure~\ref{f:line_ratios_vs_LHa} shows that line ratios generally increase toward lower $\Sigma$(\halpha): the brightest HII regions exhibit the lowest line ratios, progressively fainter HII regions show increasingly elevated ratios, and the DIG extends these trends to still lower $\Sigma$(\halpha). The strength and detailed shape of these trends depend on the diagnostic, with the smoothest trends observed for low-ionization transitions (\sii/\halpha and \oi/\halpha), while higher-ionization ratios (\oiii/\hbeta and \nii/\halpha) show a flattening toward the brightest HII regions. Overall, the relation between line ratios and $\Sigma$(\halpha) connects HII regions and the DIG continuously, with no clear separation between the two populations in this space. \ref{app:sigma_global_appendix} presents the corresponding galaxy-by-galaxy relations for the full sample and discusses the diagnostic-dependent behavior in more detail.

\begin{figure}
	\centering
\includegraphics[width=1\columnwidth]{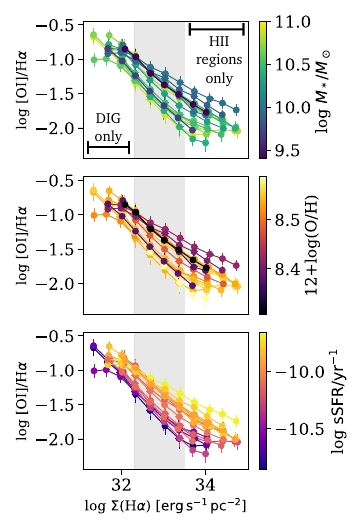}
\caption{\textbf{\oi/\halpha\ variation with $\Sigma$(\halpha) and its dependence on global galaxy properties across the PHANGS-MUSE sample.} The panels show the median \oi/\halpha\ line ratio and median absolute deviation in bins of $\Sigma$(\halpha), with each curve representing a different galaxy. The curves are color-coded by global galaxy properties: stellar mass (top), median gas-phase metallicity (middle), and sSFR (bottom). The gray band marks the range in $\Sigma$(\halpha) where both bright DIG and faint HII regions are present. At lower $\Sigma$(\halpha) (left of the band), the emission is exclusively from the DIG, while at higher $\Sigma$(\halpha) (right of the band), it is exclusively from HII regions.
\textbf{Summary:} At fixed $\Sigma$(\halpha), HII region and DIG line ratios show systematic dependence on global galaxy properties.
}
\label{f:OI_Halpha_vs_Sigma_Ha_color_coded_by_global_wlabels}
\end{figure}

We next ask whether galaxies exhibit systematic offsets in these relations, and whether these offsets correlate with global galaxy properties. Figure~\ref{f:OI_Halpha_vs_Sigma_Ha_color_coded_by_global_wlabels} shows the stacked \oi/\halpha\ line ratio as a function of $\Sigma$(\halpha) for individual galaxies, combining HII-region and DIG emission. The curves are color-coded by stellar mass, median gas-phase metallicity, and global sSFR. \ref{app:sigma_global_appendix} presents the corresponding relations for all four line ratios (Figure~\ref{f:line_ratios_vs_Lha_vs_galaxy_properties}) and discusses their dependence on global galaxy properties.

\begin{figure*}
	\centering
\includegraphics[width=1\textwidth]{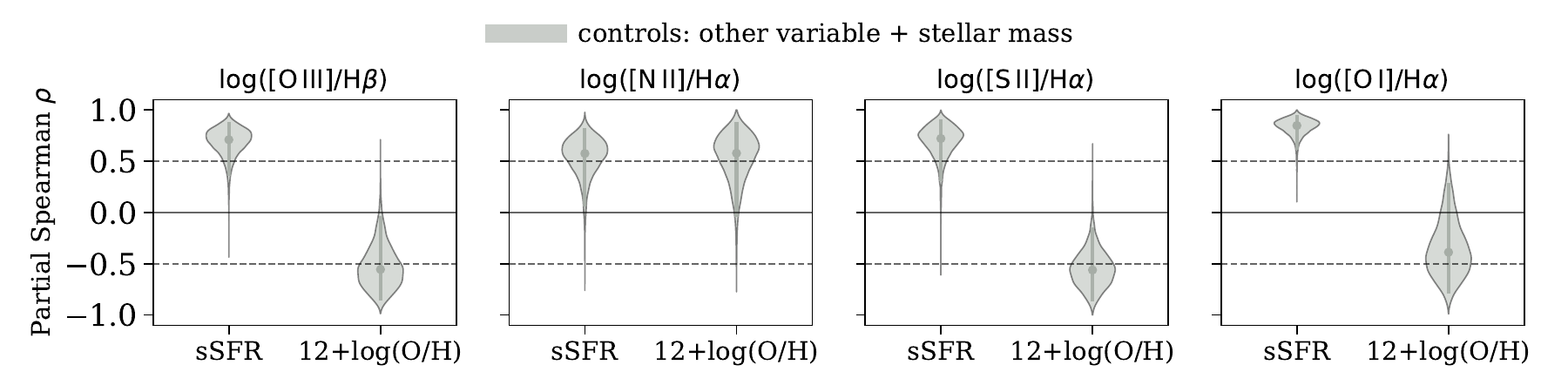}
\caption{\textbf{Partial correlations of line-ratio offsets with global galaxy properties.} Each panel shows one line ratio (from left to right: \oiii/\hbeta, \nii/\halpha, \sii/\halpha, and \oi/\halpha). We quantify how the stacked line ratios vary across the galaxy sample with global sSFR or gas-phase metallicity at fixed $\Sigma$(\halpha). For each $\Sigma$(\halpha) bin, we compute the partial Spearman correlation coefficient across galaxies and summarize the results by taking the median of the per-bin coefficients. The violins show the bootstrap distribution (galaxy bootstrap) of this median partial correlation. When evaluating the dependence on sSFR, we control for metallicity and stellar mass; when evaluating the dependence on metallicity, we control for sSFR and stellar mass.
\textbf{Summary:} Optical line ratios show the expected dependence on gas-phase metallicity, while all exhibit positive correlations with sSFR, indicating progressively harder ionizing radiation with increasing star formation activity.
}
\label{f:BPT_tracks_partial_correlation_all_gals_control_all}
\end{figure*}

At fixed $\Sigma$(\halpha), all line ratios show systematic variations with global galaxy properties. Because stellar mass, gas-phase metallicity, and sSFR are themselves correlated across the PHANGS-MUSE sample, we perform a partial-correlation analysis to isolate the independent effects of sSFR and metallicity. Using the stacked line ratios in fixed $\Sigma$(\halpha) bins (Figures~\ref{f:OI_Halpha_vs_Sigma_Ha_color_coded_by_global_wlabels} and \ref{f:line_ratios_vs_Lha_vs_galaxy_properties}), we treat galaxies as the statistical units and compute partial Spearman rank correlation coefficients \citep{feigelson12} across the galaxy sample. When evaluating the dependence on sSFR, we control for metallicity and stellar mass; when evaluating the dependence on metallicity, we control for sSFR and stellar mass. We summarize the results by taking the median of the per-$\Sigma$(\halpha)-bin partial-correlation coefficients. Uncertainties are estimated using a galaxy bootstrap, resampling galaxies with replacement and repeating the full analysis 10,000 times. The resulting bootstrap distributions are shown as violin plots in Figure~\ref{f:BPT_tracks_partial_correlation_all_gals_control_all}.

Figure~\ref{f:BPT_tracks_partial_correlation_all_gals_control_all} shows the expected correlations between line ratios and metallicity from photoionization models. \oiii/\hbeta, \sii/\halpha, and \oi/\halpha\ exhibit negative correlations, as lower metallicity reduces metal-line cooling, raising the gas temperature and strengthening collisionally excited emission (e.g., \citealt{byler17}). In contrast, \nii/\halpha\ shows a positive correlation, reflecting the secondary nucleosynthetic origin of nitrogen, which causes N/H to increase super-linearly with metallicity (e.g., \citealt{kewley08, dopita13}).

Interestingly, all four line ratios show positive correlations with sSFR. A simultaneous increase of all four line ratios cannot be explained by variations in the ionization parameter: while higher $\log U$ enhances \oiii/\hbeta, it leads to lower \sii/\halpha\ and \oi/\halpha\ ratios (see, e.g., Figure~\ref{f:BPT_tracks}). Instead, the simultaneous increase of both high- and low-ionization line ratios points to a progressively harder ionizing radiation field. A harder spectrum both increases the production of high-energy photons and extends partially ionized regions, naturally producing a coherent rise in all four optical line ratios (e.g., \citealt{ho08, netzer13, woods14}). The observed positive correlation with sSFR therefore suggests that the ionizing radiation field hardens with increasing global star formation activity.

Among proposed sources of harder ionizing radiation powering the DIG are older stellar populations. This includes the proposal by \citet{mcclymont24} that 5--25 Myr stellar populations, showing harder ionizing spectra than the 1--5 Myr populations that power classical HII regions, may dominate DIG ionization and drive line-ratio variations, as well as 1--10 Gyr populations including HOLMES \citep{zhang17, belfiore22}. However, this interpretation is in tension with the observed positive correlation with sSFR, which instead suggests harder ionizing radiation in systems with a larger contribution from the youngest (1--5 Myr) stellar populations.\footnote{The SFRs reported in Table~\ref{tab:galaxy_properties} are based on GALEX FUV and WISE W4 observations \citep{lee23}, which in principle trace star formation averaged over $\sim 100$ Myr. However, we confirm that $L_{\mathrm{H\alpha}}$-based SFRs (from either the summed luminosity of all HII regions or integrated over the full MUSE cube) are tightly correlated with these values, suggesting that systems with higher reported SFRs also have higher instantaneous SFRs traced by the youngest stellar populations.}

A plausible mechanism linking changes in the spectral shape of the radiation field to sSFR is the leakage of ionizing photons from HII regions. As ionizing radiation propagates through the ISM, it is progressively filtered because the hydrogen photoionization cross section decreases strongly with photon energy above 13.6 eV; consequently, photons closer to this energy are preferentially absorbed during propagation (e.g., \citealt{osterbrock06}). The transmitted spectrum is therefore hardened, which can in principle enhance both low- and high-ionization line ratios simultaneously.

\begin{figure*}
	\centering
\includegraphics[width=1\textwidth]{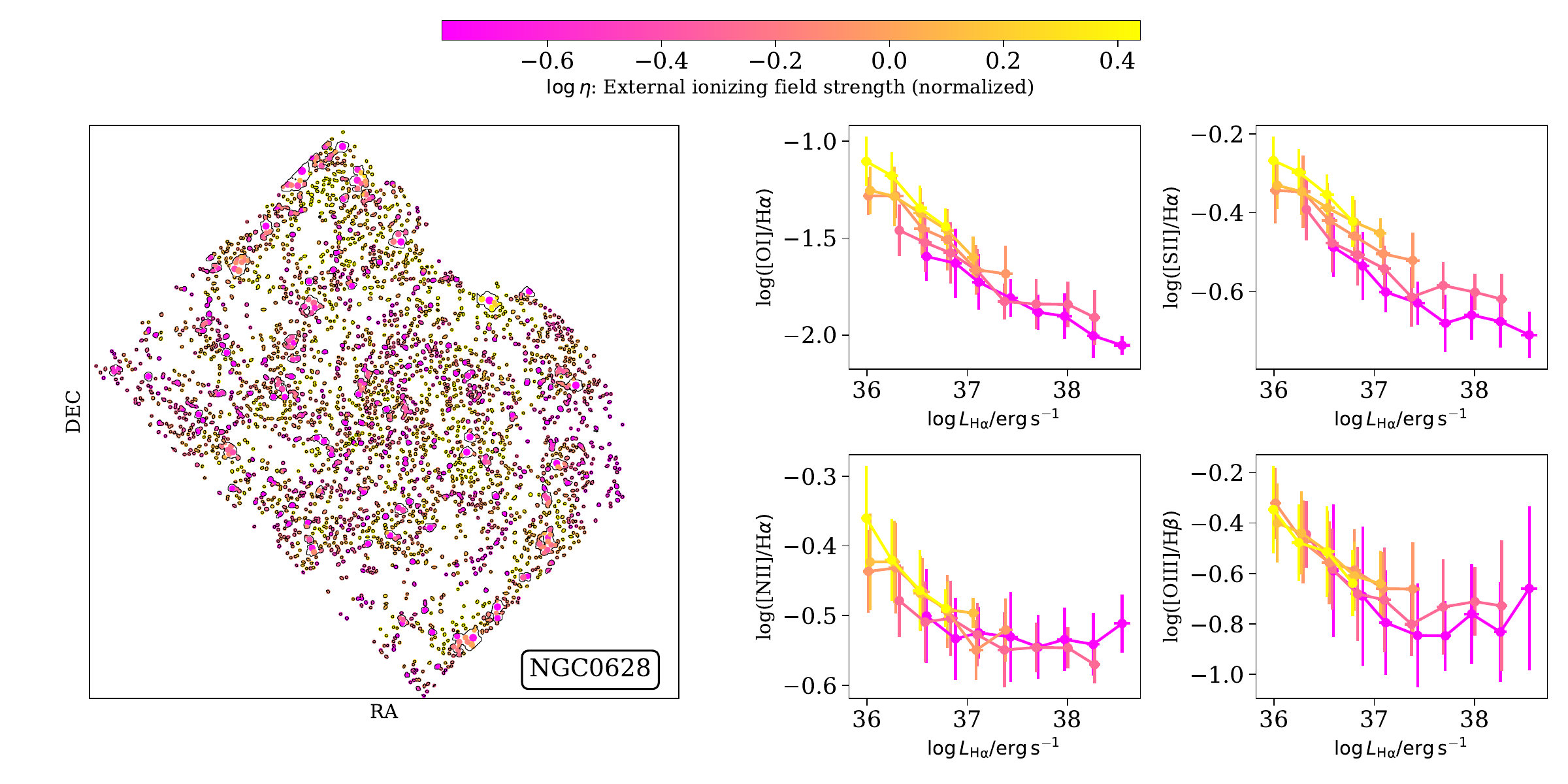}
\caption{\textbf{Dependence of line ratio--$L_{\mathrm{H}\alpha}$ relations on radiative environment in NGC~628.}
The left panel shows HII regions in NGC~628, color-coded by $\eta$, the normalized external ionizing field strength (Eq.~\ref{eq:eta}). $\eta$ quantifies the ionizing radiation field from surrounding HII regions, accounting for geometric dilution, attenuation, and weighting by source luminosities. Low values (magenta) indicate regions dominated by their own ionizing output, while high values (yellow) indicate regions embedded in more radiatively rich environments. The right panels show stacked \oi/\halpha, \sii/\halpha, \nii/\halpha, and \oiii/\hbeta\ as a function of $L_{\mathrm{H}\alpha}$ and $\eta$. 
\textbf{Summary:} HII regions show systematic variations with the large-scale radiative environment. At fixed $L_{\mathrm{H}\alpha}$, regions exposed to stronger ambient ionizing radiation exhibit enhanced low-ionization line ratios (\oi/\halpha\ and \sii/\halpha).}
\label{f:NGC0628_ratios_vs_eta_LHa}
\end{figure*}

In this picture, higher-sSFR systems may exhibit stronger enhancements either because a larger fraction of ionizing photons escapes through a more porous, feedback-processed ISM (e.g., \citealt{clarke02, oey07, kimm14, ma15}), or because, at fixed escape fraction, a larger absolute ionizing photon budget provides more radiation to power the DIG. We explore this scenario in detail in Section~\ref{sec:leaking}, where we model the leakage of ionizing radiation from HII regions, its spectral hardening during propagation through the ISM, and the resulting impact on both the DIG and faint HII regions. In Section~\ref{sec:discussion_leaking_vs_sSFR}, we show that this picture qualitatively reproduces the observed trends.

\subsection{Environmental dependence of HII region line ratio vs $\Sigma$(\halpha) sequences}\label{sec:HII_env_dependence}

In Sections~\ref{sec:global_HII_DIG_variations_line_ratios} and \ref{sec:global_HII_DIG_variations_line_ratios_with_sigma}, we showed that HII regions generally follow trends in optical line ratios and $\Sigma$(\halpha) that closely resemble those of the DIG. If the DIG is primarily powered by ionizing radiation leaking from HII regions, its properties should depend on the local ($\lesssim 1$ kpc) radiative environment, set by the number, spatial distribution, and luminosity of nearby HII regions. This picture was investigated by \citet{belfiore22}, who used the observed locations and luminosities of HII regions in PHANGS as inputs to a radiation propagation model, demonstrating that it can successfully reproduce the observed $\Sigma$(\halpha) distribution of the DIG. The continuity between HII regions and the DIG therefore raises the question of whether HII region properties themselves are influenced by this same environment. We investigate this possibility here.

To characterize the radiative environment of each HII region, we use the locations, angular sizes, and \halpha\ luminosities from the nebular catalog, following the formulation of \citet{belfiore22}. We define the normalized external ionizing field strength at HII region $i$, $\eta_i$, as the ratio between the ionizing radiation contributed by neighboring HII regions and the region's own ionizing output,
\begin{equation}\label{eq:eta}
    \eta_{i} = \frac{\sum_{i\neq j} K_{ij} L_{\mathrm{H}\alpha, j}}{L_{\mathrm{H}\alpha, i}},
\end{equation}
where $L_{\mathrm{H}\alpha,i}$ is the dust-corrected \halpha\ luminosity of region $i$, and $K_{ij}$ describes the attenuation and geometric coupling between regions $i$ and $j$. We use $\eta_i$ primarily to rank HII regions by the strength of their external radiation field rather than as an absolute measure. Consequently, if the escape fraction of ionizing photons is approximately constant across HII regions, it simply rescales all $\eta_i$ values by the same factor and does not affect their relative ordering.

For each pair of regions, $i$ and $j$, the radiative coupling kernel is
\begin{equation}\label{eq:Kij}
    K_{ij} = e^{-k_0 r_{ij}} \frac{\Delta \Omega_{ij}}{4 \pi},
\end{equation}
where $r_{ij}$ is the physical separation between the regions in $\mathrm{pc}$, $k_0$ is the effective absorption coefficient in the DIG (with corresponding mean free path $\lambda_0 = 1/k_0$), and $\Delta \Omega_{ij}$ is the solid angle subtended by region $i$ as seen from region $j$. We approximate $\Delta \Omega_{ij}$ as the solid angle of a spherical target (see \ref{app:solid_angles}), using the circularized nebular radius from the catalog and assuming that the HII regions lie within a common thin-disk geometry. Both $r_{ij}$ and $\Delta \Omega_{ij}$ are corrected for galaxy inclination.

\citet{belfiore22} found that $k_0 = 0.52\,\mathrm{kpc^{-1}}$ reproduces the observed DIG surface-brightness distribution in the PHANGS galaxies, and we adopt this value here. Varying $k_0$ by factors of 2--3 primarily rescales the overall normalization of $\eta$, but does not alter the relative ordering of HII regions (i.e., which regions reside in more or less radiatively rich environments). Our conclusions are therefore insensitive to the exact choice of $k_0$.

Using $\eta_i$ as a proxy for the local radiative environment relies on two additional assumptions. First, we assume that the escape fraction of ionizing photons does not vary strongly from one HII region to another. Second, we assume that $L_{\mathrm{H}\alpha,i}$ is a reliable tracer (up to a constant multiplicative factor) of the ionizing photon production rate of the stellar population powering region $i$. We adopt these simplifying assumptions here and revisit them in the full HII region + DIG radiation propagation model in Section~\ref{sec:leaking}, where we show that our conclusions are robust to more realistic treatments of ionizing-photon escape and the mapping between ionizing photon production and the observed \halpha\ emission.

In Figure~\ref{f:NGC0628_ratios_vs_eta_LHa}, we show the distribution of the normalized external ionizing field strength, $\eta$, across the HII regions in NGC~628. Low-$\eta$ HII regions are dominated by their own ionizing radiation, whereas high-$\eta$ regions reside in environments where ionizing photons from neighboring HII regions contribute a larger fraction of the local radiation field. The right-hand panels show stacked \oi/\halpha, \sii/\halpha, \nii/\halpha, and \oiii/\hbeta\ as a function of $L_{\mathrm{H}\alpha}$ and $\eta$.

In Figures~\ref{f:OIII_Hbeta_vs_Lhalpha_eta_all_gal}--\ref{f:OI_Halpha_vs_Lhalpha_eta_all_gal} in \ref{app:HII_line_ratios_versus_eta}, we present stacked line ratios as a function of $L_{\mathrm{H}\alpha}$ and $\eta$ for the full PHANGS-MUSE sample. For the low-ionization transitions, \oi/\halpha\ and \sii/\halpha, HII regions exhibit systematically higher line ratios at fixed $L_{\mathrm{H}\alpha}$ in more radiatively rich environments. The behavior of the higher-ionization and more metallicity-sensitive \nii/\halpha\ ratio is less uniform: some galaxies show a similar dependence on $\eta$, while others do not. For the high-ionization \oiii/\hbeta\ ratio, we find little to no dependence on $\eta$ at fixed $L_{\mathrm{H}\alpha}$. This behavior is observed for all but the brightest HII regions. For the brightest regions, $\eta$ remains uniformly low because their intrinsic luminosities dominate over the surrounding radiation field.

There are two plausible scenarios that could explain the variation of HII region line ratios with radiative environment. The first is line-of-sight mixing with the DIG. HII regions in more radiatively rich environments are also expected to be co-located with brighter DIG emission. Because DIG line ratios are typically higher than those of HII regions, such mixing would lead to elevated apparent line ratios, with the effect increasing in environments where the DIG is brighter. We explore this scenario in Section~\ref{sec:mixing}, where we show that mixing can reproduce the observed trends in \nii/\halpha\ and \oiii/\hbeta, but not in the lower-ionization transitions, which are also those that show dependence on environment here.

The second scenario is that fainter HII regions are increasingly influenced by ionizing radiation leaking from neighboring regions. As this radiation propagates through the ISM, it becomes progressively harder, resulting in an effectively harder ionizing spectrum incident on the gas. HII regions in more radiatively rich environments would therefore experience a larger contribution from this leaked radiation, leading to enhanced line ratios. We explore this scenario in Section~\ref{sec:leaking}, where we show that, under a range of plausible conditions, such contamination can be non-negligible for fainter regions, producing harder effective spectra and higher line ratios that quantitatively reproduce the observed trends.

The different behavior of the line ratios may also be understood in terms of the ionization structure of HII regions (\citealt{kewley19}). If leaking radiation contributes to the incident radiation field, its relative importance is expected to be greatest in the outer layers of an HII region, where it competes more effectively with the internally produced radiation. Consistent with this picture, \oi/\halpha\ and \sii/\halpha, which predominantly originate in partially ionized outer layers, show the strongest dependence on radiative environment. In contrast, \nii/\halpha\ and especially \oiii/\hbeta, which arise deeper within the HII region where the internal radiation field dominates, show little or no dependence on $\eta$.

\begin{figure}
	\centering
\includegraphics[width=0.65\columnwidth]{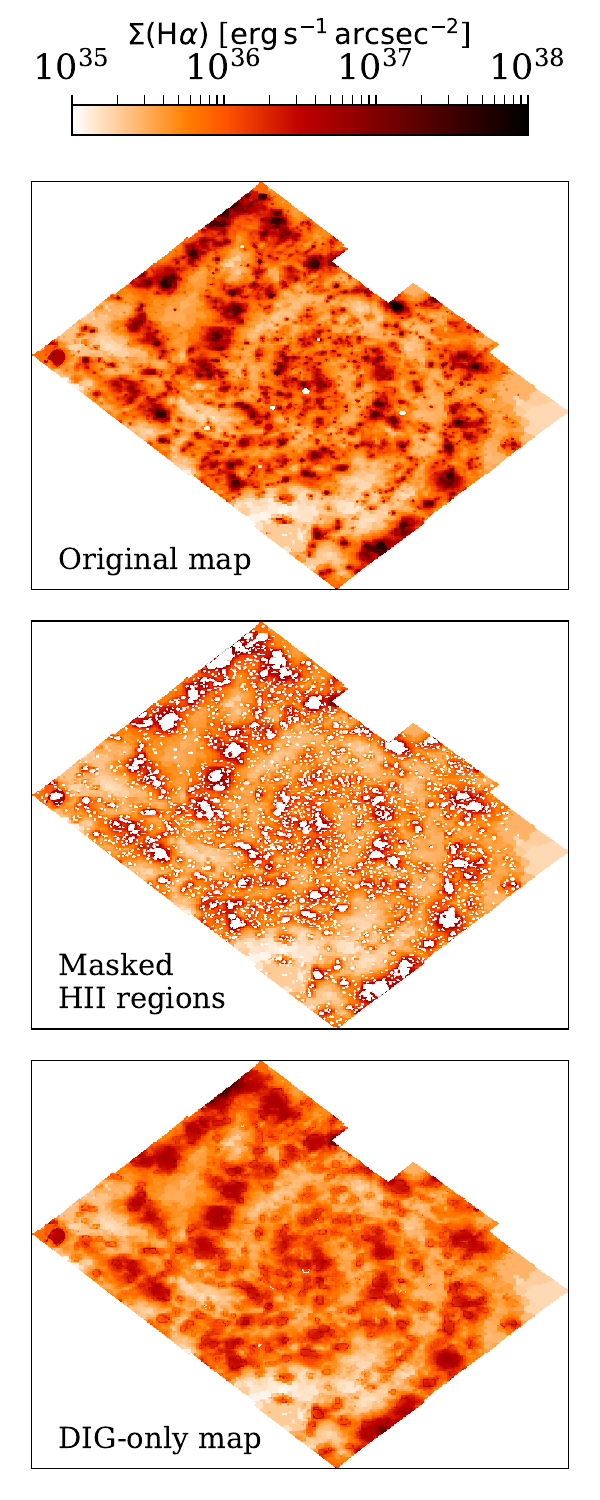}
\caption{\textbf{Estimating DIG-only $\Sigma$(\halpha) emission in NGC~628.} The top panel shows the original MUSE $\Sigma$(\halpha) map, measured in individual 0.2$''$ spaxels within HII regions and in Voronoi bins covering the DIG. The middle panel shows the same map with HII regions from the nebular catalog masked out. The bottom panel shows the Gaussian-smoothed version of the masked map, which we use as an approximation for the DIG-only emission across the galaxy.}
\label{f:DIG_smooth_maps_NGC0628}
\end{figure}

\section{HII region-DIG mixing along the line of sight}\label{sec:mixing}

At the angular resolution of MUSE for the moderately inclined PHANGS galaxies, the $\sim$100 pc-scale observations enable the separation of individual HII regions from the surrounding DIG in the plane of the disk. However, the star-forming disk and the DIG differ significantly in their vertical extent. The star-forming disk, which hosts molecular gas and HII regions, is relatively thin, with a characteristic scale height of $\sim$50--300 pc (e.g., \citealt{scoville93, yim14, vanDerKruit11}). In contrast, the DIG is substantially more extended, with scale heights of $\sim$0.5--1.5 kpc inferred for both the Milky Way and external galaxies (e.g., \citealt{reynolds91, ferriere01, gaensler08, levy19}, see however \citealt{miller03}). As a result, even when HII regions are well separated from the DIG in projection, emission from the vertically extended DIG component is unavoidably included along the line of sight (see e.g., \citealt{congiu23, mcclymont24}).

\begin{figure*}
	\centering
\includegraphics[width=1\textwidth]{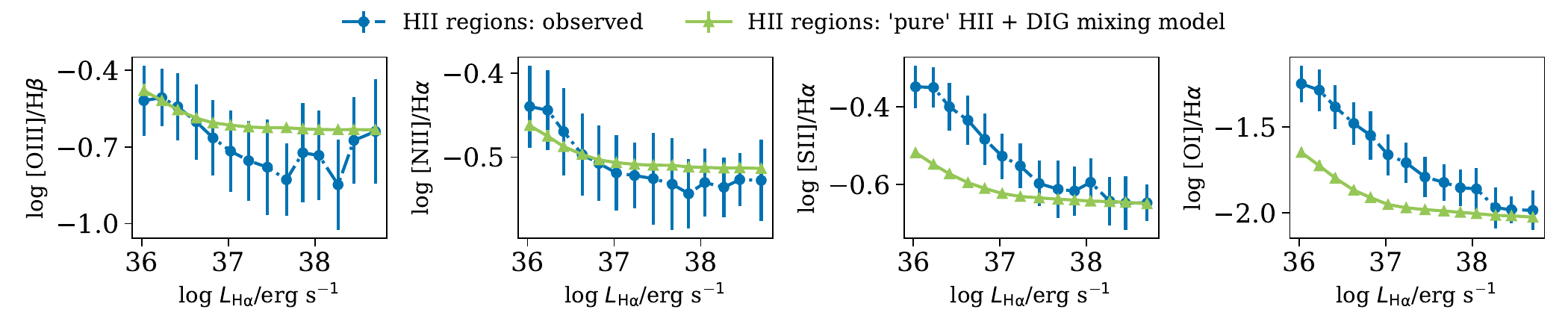}
\caption{\textbf{Observed versus predicted line ratios from the simple HII region + DIG mixing model for HII regions in NGC~628.} The panels show stacked \oiii/\hbeta, \nii/\halpha, \sii/\halpha, and \oi/\halpha\ line ratios in bins of HII region $L_{\mathrm{H}\alpha}$. Each panel shows the median and median absolute deviation, with observed HII region line ratios shown as blue circles and the predictions of the mixing model shown as green triangles. 
\textbf{Summary:} The mixing model reproduces the observed range of variation in the higher-ionization ratios, \oiii/\hbeta\ and \nii/\halpha, but underpredicts the observed low-ionization ratios, \oi/\halpha\ and \sii/\halpha, particularly for the faintest HII regions. }
\label{f:mixing_vs_observed_line_ratios}
\end{figure*}

The continuity between HII regions and the DIG, both in BPT space (Section~\ref{sec:global_HII_DIG_variations_line_ratios}) and in the relations between line ratios and $\Sigma$(\halpha) (Section~\ref{sec:global_HII_DIG_variations_line_ratios_with_sigma}), raises the question of whether these trends arise from line-of-sight mixing. We find that the brightest HII regions exhibit the lowest line ratios, while progressively fainter regions follow the same sequence toward higher line ratios, smoothly connecting to the DIG. In addition, HII region line ratios depend on their radiative environment: at fixed luminosity, regions in more radiatively rich environments, where the DIG is also expected to be brighter, show systematically elevated line ratios compared to more isolated regions (Section~\ref{sec:HII_env_dependence}). 

In this section, we test whether these highly regular trends can be explained by mixing with DIG emission along the line of sight. Given that DIG line ratios are typically higher than those of HII regions (e.g., \citealt{reynolds85, hoopes96, belfiore16, belfiore22}), an increasing contribution from DIG, particularly for fainter HII regions, could naturally produce the observed progression toward higher line ratios.

To mitigate contamination by projected DIG emission, \citet{congiu23} estimated and subtracted the local DIG contribution from each HII region in the PHANGS-MUSE galaxies. For every HII region, they measured the DIG spectrum within a surrounding annulus, excluding any overlapping HII regions, rescaled this spectrum by the ratio of the HII-region and annulus areas, and subtracted the resulting DIG contribution from the integrated HII-region spectrum. As shown in their Figure~2, this correction modifies the distribution of HII regions in BPT space. However, the diagonal sequences remain, indicating that projected DIG contamination alone cannot account for the observed HII-region trends.

We use the DIG-subtracted nebular catalog of \citet{congiu23} and repeat the analysis presented in Sections~\ref{sec:global_HII_DIG_variations_line_ratios} and \ref{sec:global_HII_DIG_variations_line_ratios_with_sigma}. We include all nebulae classified as HII regions with S/N$>3$ in \oiii, \hbeta, \oi, \nii, \halpha, and \sii\ after the DIG subtraction. We fit the slopes in the \nii/\halpha, \sii/\halpha, and \oi/\halpha\ versus \oiii/\hbeta\ diagrams for the PHANGS galaxies and find that they are consistent, within uncertainties, with those derived when fitting the relations in Figures~\ref{f:BPT_tracks_all_gals_NII}--\ref{f:BPT_tracks_all_gals_OI}. 

We then examine the relation between line ratios and $\Sigma$(\halpha). Using the DIG-corrected catalog, the relations involving the higher-ionization ratios, \oiii/\hbeta\ and \nii/\halpha, become noticeably flatter, whereas those involving the low-ionization ratios, \oi/\halpha\ and \sii/\halpha, largely persist (see Figures~\ref{f:OIIIHbeta_vs_LHa_DIGcorr}--\ref{f:OIHalpha_vs_LHa_DIGcorr} in \ref{app:DIGsub_trends}). In other words, even after correcting for line-of-sight DIG contamination, HII regions continue to show increasing \oi/\halpha\ and \sii/\halpha\ ratios toward lower \halpha\ luminosities.

The DIG-subtracted catalog of \citet{congiu23} provides an important consistency check, but it is based on a different HII region catalog than the one used throughout this work. Specifically, \citet{groves23} and \citet{congiu23} identify HII regions using different segmentation algorithms ({\sc HIIphot} and {\sc clumpfind}, respectively), resulting in different HII region boundaries and numbers of identified regions. We therefore perform a complementary test using the \citet{groves23} HII region catalog adopted throughout this paper.

We construct a simple HII region + DIG mixing model in which a ``pure'' HII region spectrum, characterized by relatively low emission-line ratios, is combined with a DIG spectrum exhibiting higher line ratios. The DIG contribution is scaled according to the expected DIG $\Sigma$(\halpha) at the location of each HII region. 

We define the ``pure HII region'' spectrum using the median line ratios of the brightest 10\% of HII regions within each galaxy. To estimate the DIG surface brightness across the galaxy, we begin with the observed \halpha\ map, consisting of individual 0.2$''$ spaxels within HII regions and Voronoi bins in the DIG. We then mask all identified HII regions from the nebular catalog and apply Gaussian smoothing to the masked map, interpolating across the masked regions and producing a continuous estimate of the DIG $\Sigma$(\halpha) field. Figure~\ref{f:DIG_smooth_maps_NGC0628} illustrates this procedure for NGC~628.

For each HII region, we use the predicted DIG $\Sigma$(\halpha) at its location together with the observed DIG relations between line ratios and $\Sigma$(\halpha) (e.g., Figure~\ref{f:line_ratios_vs_LHa}) to estimate the corresponding DIG line ratios. We then combine the ``pure HII region'' and DIG spectra, weighting each component by its intrinsic \halpha\ surface brightness. Because the observed HII region $\Sigma$(\halpha) already includes projected DIG emission, we first subtract the estimated DIG \halpha\ contribution from the observed $\Sigma$(\halpha). This ensures that the HII region component represents only the nebular emission, while the DIG contribution is added explicitly, avoiding double counting. The correction has little effect on the brightest HII regions, but becomes progressively more important toward lower luminosities.

For NGC~628, for example, the inferred HII region \halpha\ luminosities decrease by approximately 10\%, 25\%, 50\%, and 70\% at $\log L_{\mathrm{H}\alpha}/\mathrm{erg\,s^{-1}} = 38$, $37$, $36.5$, and $36$, respectively. These correspond approximately to the 90th, 70th, 30th, and 5th percentiles of the HII region luminosity distribution in NGC~628. The median-luminosity HII region is corrected downward by $\sim$40\%.

In Figure~\ref{f:mixing_vs_observed_line_ratios} we compare the observed HII region line ratios in NGC~628 with the predictions of the simple mixing model. The model reproduces the observed variation of the higher-ionization ratios, \oiii/\hbeta\ and \nii/\halpha, with \halpha\ luminosity. In contrast, for the lower-ionization transitions, \oi/\halpha\ and \sii/\halpha, the model systematically underpredicts the observed line ratios for regions with $\log L_{\mathrm{H}\alpha}/\mathrm{erg\,s^{-1}} < 37.5$. These are also the line ratios that show systematic variation with radiative environment in Section~\ref{sec:HII_env_dependence}.

These results suggest that line-of-sight mixing with the DIG contributes to the diagonal sequences observed in BPT space, as well as to the relations between line ratios and $\Sigma$(\halpha), but is not sufficient to fully explain them. In particular, the low-ionization transitions, \oi/\halpha\ and \sii/\halpha, require an additional process that enhances the line ratios toward lower \halpha\ luminosities. In Section~\ref{sec:leaking}, we explore whether leaking ionizing radiation that becomes progressively harder as it propagates through the ISM can account for the observed line-ratio sequences of both HII regions and the DIG.

\section{Radiative coupling of HII regions and the DIG through ionizing photon leakage}\label{sec:leaking}

In this section, we investigate whether leaking ionizing radiation can radiatively couple HII regions and the DIG, and whether such coupling can explain the observed continuity in line ratios and $\Sigma$(\halpha) across the two populations.

We extend the thin-slab ``leaky HII region'' framework developed by \citet{belfiore22}, itself based on the formalism of \citet{seon09}, to model the origin of DIG emission across the disks of the PHANGS galaxies. This framework is also conceptually similar to that proposed by \citet{zurita02}.  In the \citet{belfiore22} framework, the observed DIG \halpha\ surface brightness in each spatial element is treated as the cumulative contribution of ionizing photons escaping from a catalog of discrete HII regions (from the PHANGS-MUSE nebular catalog) and propagating through the disk. During propagation, the radiation undergoes both geometric dilution and attenuation by the diffuse ISM, the latter parameterized by an effective absorption coefficient, $k_0$, or equivalently by a mean free path for ionizing photons, $\lambda_0 = 1/k_0$. Our main extensions are to account for spectral hardening of the escaping radiation and to allow leaked radiation to influence neighboring HII regions as well as the DIG, as described below.

The \citet{belfiore22} model is designed to reproduce the spatial distribution of DIG \halpha\ emission using a small set of physically motivated parameters, including: (1) a global leakage normalization that links the ionizing output of HII regions to DIG recombination emission, and (2) an attenuation scale that determines how far ionizing photons propagate through the diffuse medium. \citet{belfiore22} showed that this framework is able to reproduce the observed spatial distribution of DIG emission across the PHANGS galaxies remarkably well. Fitting the model to the observed DIG emission therefore constrains both the escape fraction of ionizing radiation from HII regions, $f_{\mathrm{esc}}$ (assuming a known DIG scale height, $\Delta h$), and the effective mean free path of ionizing photons in the diffuse ISM.

At its core, the model assumes that the observed \halpha\ luminosity of an HII region is proportional to the ionizing photon production rate of the stellar population powering it, up to a constant conversion factor between ionizing photon production and recombination luminosity. However, if the DIG is powered primarily by ionizing radiation leaking from HII regions, then the continuity in line ratios and $\Sigma$(\halpha) observed between HII regions and the DIG raises the possibility that the same leaking radiation may also contribute to the ionization of neighboring HII regions themselves.

Motivated by this possibility, we first extend the \citet{belfiore22} formalism to investigate the impact of leaking radiation on nearby HII regions in Section~\ref{sec:leaking_HII}. This step is important because if HII regions are significantly contaminated by ionizing photons escaping from neighboring regions, then their observed \halpha\ luminosities no longer trace only their internal ionizing sources. Instead, the measured luminosity becomes the sum of locally produced and externally supplied ionizing radiation. In that case, the DIG radiation propagation model must also be modified, since the inferred ionizing photon budget escaping into the DIG depends on the intrinsic ionizing output of the HII regions themselves.

We therefore next construct a coupled radiation propagation model for both HII regions and the DIG, effectively re-deriving the \citet{belfiore22} DIG framework while correcting the intrinsic ionizing photon production rates of HII regions for external contamination. We then search for self-consistent solutions in terms of $k_0$ and $f_{\mathrm{esc}}$ that simultaneously reproduce the observed properties of both HII regions and the DIG. This is presented in Section~\ref{sec:leaking_HII_and_DIG}.

The primary motivation for this more general framework is to test whether leaking ionizing radiation can explain the observed increase in optical line ratios toward lower \halpha\ luminosities across both HII regions and the DIG. For that, in Section~\ref{sec:leaking_hardening}, we further generalize the model to include spectral hardening during propagation through the ISM, and examine whether such hardening can naturally produce increasingly harder effective ionizing spectra in progressively fainter HII regions. We provide a final test of this framework with photoionization models in Section~\ref{sec:leaking_photoionization}, which quantify the expected line-ratio behavior under the leaking and spectrally hardened radiation scenario, and assess whether such models can simultaneously reproduce the observed trends across both HII regions and the DIG.

\subsection{Radiative coupling between neighboring HII regions}\label{sec:leaking_HII}

\subsubsection{Model assumptions}

The observed \halpha\ luminosity of an HII region $i$ can be written as the sum of recombination emission powered by its internal ionizing source and recombination emission induced by externally supplied ionizing radiation leaking from neighboring HII regions:
\begin{equation}\label{eq:HII_prop_eq}
\begin{split}
    C_{\alpha} L_{\mathrm{H}\alpha,i} &= (1 - f_{\mathrm{esc}, i})Q_{i} + \sum_{j \neq i} e^{-k_0 r_{ij}} \Big( \frac{\Delta \Omega_{ij}} {4 \pi} \Big) f_{\mathrm{esc}, j} Q_{j} \\
    &= (1 - f_{\mathrm{esc}, i})Q_{i}
    + \sum_{j \neq i} K_{ij} f_{\mathrm{esc}, j} Q_{j},
\end{split}
\end{equation}
where $Q_i$ is the intrinsic ionizing photon production rate of region $i$, $f_{\mathrm{esc},i}$ is its escape fraction of ionizing photons, $K_{ij}$ is the radiative coupling kernel defined in Equation~\ref{eq:Kij}, and $C_{\alpha}$ is the Case B recombination conversion factor between ionizing photon production rate and \halpha\ luminosity. We adopt $C_{\alpha} = 7.31 \times 10^{11}\,\mathrm{photons\,erg^{-1}}$, corresponding to an electron temperature of $10^{4}\,\mathrm{K}$ and density of $10^{2}\,\mathrm{cm^{-3}}$ ($C_{\alpha}$ is only weakly sensitive to these choices; e.g., \citealt{draine11}).

The first term on the right-hand side therefore represents the contribution from the internal ionizing source powering the HII region (e.g., a young stellar cluster). The second term accounts for ionizing radiation escaping from neighboring HII regions and subsequently propagating through the diffuse ISM before reaching region $i$. During propagation, the radiation experiences attenuation parameterized by the effective absorption coefficient $k_0$, geometric dilution described by the solid angle $\Delta \Omega_{ij}$ subtended by region $i$ as seen from region $j$, and scaling with the ionizing photon production rate and escape fraction of the source region, $Q_j$ and $f_{\mathrm{esc},j}$.

We assume that $k_0$ is constant across the disk. A first-order estimate of $k_0$ can be obtained by fitting the observed \halpha\ surface brightness distribution of the DIG, as performed by \citet{belfiore22}, who inferred a characteristic value of $k_0 = 0.52\,\mathrm{kpc^{-1}}$, corresponding to a mean free path of $\lambda_0 \approx 1.9$ kpc. However, as discussed above, this estimate relies on the assumption that the observed \halpha\ luminosity of an HII region directly traces its internally produced ionizing photon budget. In Section~\ref{sec:leaking_HII_and_DIG} below, where we self-consistently couple the HII regions and DIG through a unified radiation propagation model, we infer somewhat larger values of $k_0 \sim 1$--$1.7\,\mathrm{kpc^{-1}}$ ($\lambda_0 \approx 0.6$--$1$ kpc). Motivated by these results, we adopt $k_0 = 1\,\mathrm{kpc^{-1}}$ as our fiducial value throughout this section, while also exploring the sensitivity of the model predictions to variations in $k_0$.

\begin{figure*}
	\centering
\includegraphics[width=1\textwidth]{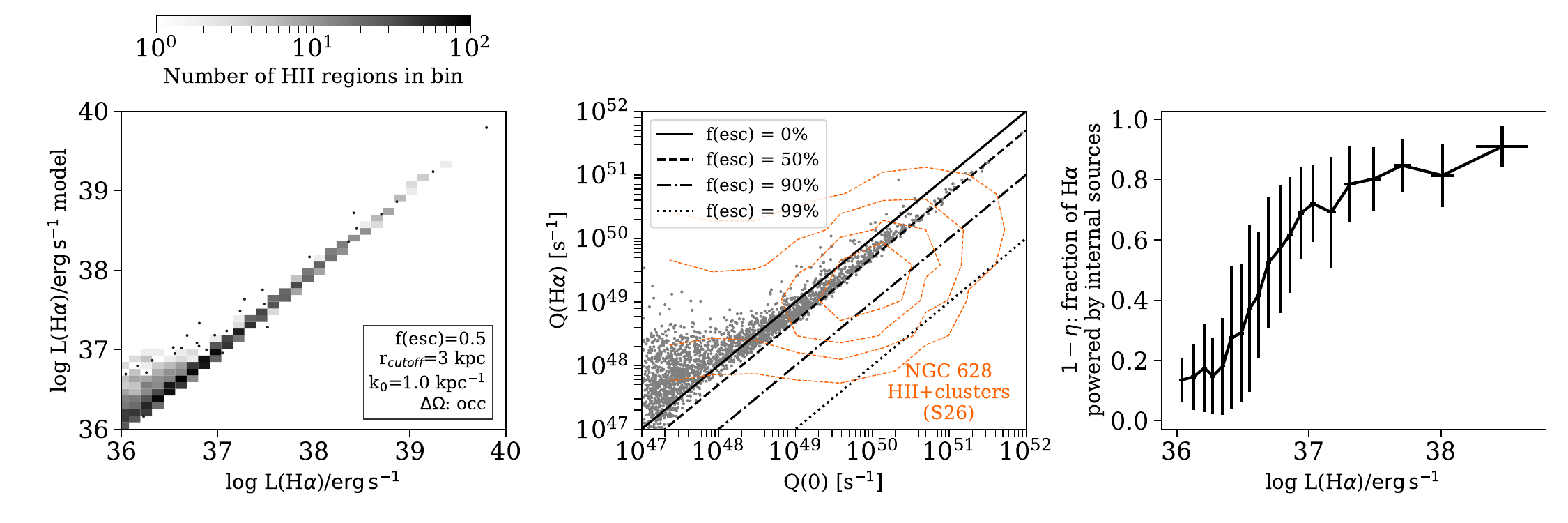}
\caption{\textbf{Fiducial model of radiative coupling between neighboring HII regions in NGC~628.} The panels show results from our fiducial model, which assumes $f_{\mathrm{esc}}=0.5$, $k_0 = 1\,\mathrm{kpc^{-1}}$, opaque HII regions (\texttt{occ}), and considers every HII region pair within a maximum coupling distance of $r_{\mathrm{cutoff}}=3\,\mathrm{kpc}$.
\textbf{Left panel:} observed versus model-predicted \halpha\ luminosities for HII regions, shown as 2D histograms with logarithmic color scaling; black points indicate bins containing a single source. 
\textbf{Middle panel:} intrinsic ionizing photon production rates derived from the model, $\mathrm{Q(0)}$, compared to $\mathrm{Q(H\alpha)}$ inferred directly from the observed \halpha\ luminosity assuming purely internal powering (as done in \citealt{scheuermann26}; S26). Orange contours show the measurements of star clusters+HII regions in PHANGS-MUSE from S26, and the black lines indicate constant apparent escape fractions, under the same assumption. 
\textbf{Right panel:} median and median absolute deviation of the internally powered fraction of the observed \halpha\ luminosity, $1-\eta$, as a function of HII region luminosity. $1-\eta=1$ corresponds to HII regions whose observed \halpha\ emission is entirely powered by internal sources, while $1-\eta=0.5$ corresponds to regions where only half of the observed \halpha\ emission is powered internally, with the remainder supplied by leaking radiation from neighboring HII regions.
\textbf{Summary: }The radiative coupling model qualitatively reproduces the two populations of cluster+HII region pairs observed in the data, including the seemingly ``impossible'' systems in which the ionizing radiation from the associated star cluster is insufficient to power the observed \halpha\ luminosity, through contamination by leaking ionizing radiation from neighboring HII regions (middle panel). The model predicts that bright HII regions are primarily powered by their own ionizing radiation, whereas progressively fainter HII regions receive an increasing contribution from leaked radiation originating in neighboring luminous regions (right panel).
} 
\label{f:HII_coupling_fiducial_model}
\end{figure*}

To estimate $\Delta \Omega_{ij}$, we assume that HII regions are spherical, with radii given by the circularized nebular radii from the nebular catalog, and that they lie within a common 2D disk plane (i.e., neglecting any vertical scatter in their spatial distribution). We consider two definitions for the solid angle $\Delta \Omega_{ij}$, whose derivations are presented in \ref{app:solid_angles}. The first definition, which we denote \texttt{simple}, approximates $\Delta \Omega_{ij}$ as the solid angle subtended by a spherical target of radius $R_i$ at a distance $r_{ij}$. In this case, HII regions are effectively assumed to be optically thin obstacles, such that intervening regions do not block radiation propagating between two other regions. While this approximation neglects absorption by intervening nebulae, it may partially mimic the effects of a finite vertical distribution of HII regions within the disk. The second definition, denoted \texttt{occ}, treats HII regions as optically thick spheres that can occult one another, such that radiation propagation depends on both the sizes and projected spatial distribution of all the nebulae between $i$ and $j$. This approximation likely overestimates the degree of blocking, since real HII regions should exhibit some vertical scatter that is not captured in our 2D geometry. Neither approximation fully represents the true three-dimensional structure of HII regions within galactic disks. Instead, they are intended to bracket a plausible range of geometries, with the real configuration expected to produce results lying between these two limiting cases.

\subsubsection{Fiducial model output for NGC~628}\label{sec:fiducial_model_NGC628}

We use the DIG-corrected \halpha\ luminosities of HII regions as inputs, where the DIG contribution is estimated as described in Section~\ref{sec:mixing}. We further correct $r_{ij}$ and $\Delta \Omega_{ij}$ (which itself depends on $r_{ij}$) for the galaxy inclination and position angle. All calculations are carried out for NGC~628, one of the closest galaxies in the PHANGS-MUSE sample; we discuss the implications for the broader PHANGS-MUSE sample in Section~\ref{sec:discussion}. This galaxy contains 2,182 identified HII regions, such that Equation~\ref{eq:HII_prop_eq} corresponds to a coupled system of 2,182 equations, where the unknowns are the intrinsic ionizing photon production rates, $Q_i$.

Using the radiative coupling kernel, $K_{ij}$, we can estimate the effective separation between HII regions in NGC~628 by computing the $K_{ij}$-weighted mean distance between regions. The resulting distribution has a mean and median of 320 pc and 200 pc, respectively, with 16th and 84th percentiles of 100 pc and 570 pc. At the same time, the HII regions in NGC~628 span approximately 2.5 orders of magnitude in DIG-corrected \halpha\ luminosity, from $L_{\mathrm{H}\alpha}/\mathrm{erg\,s^{-1}} \sim 10^{36}$ to $10^{38.5}$. The 16th, 50th, and 84th percentiles of the luminosity distribution are $2 \times 10^{36}\,\mathrm{erg\,s^{-1}}$, $5.4 \times 10^{36}\,\mathrm{erg\,s^{-1}}$, and $4 \times 10^{37}\,\mathrm{erg\,s^{-1}}$, respectively. Throughout this section, we will often use $L_{\mathrm{H}\alpha}=10^{37}\,\mathrm{erg\,s^{-1}}$ as a representative reference luminosity; this corresponds to approximately the 67th percentile of the HII region luminosity distribution.

In Figure~\ref{f:HII_coupling_fiducial_model} we present the solution to Equation~\ref{eq:HII_prop_eq} under our fiducial model assumptions: $f_{\mathrm{esc}}=0.5$, $k_{0}=1\,\mathrm{kpc^{-1}}$, and opaque HII regions (\texttt{occ}). The left panel compares the observed \halpha\ luminosities to those predicted by the model, and therefore serves as a measure of the model goodness of fit. The agreement is excellent for HII regions with $L_{\mathrm{H\alpha}} \geq 10^{37}\,\mathrm{erg\,s^{-1}}$. At lower luminosities, the agreement remains generally good, although the model overpredicts the emission for a subset of faint HII regions. Reducing this excess requires significantly lowering the escape fraction, to values that become inconsistent with observational constraints from the DIG (see \ref{app:HII_rad_coupling_app}).

The middle panel of Figure~\ref{f:HII_coupling_fiducial_model} compares the intrinsic ionizing photon production rates derived by the model, $Q(0)$, with the ionizing photon production rates inferred directly from the observed \halpha\ luminosities under the assumption that the emission is powered entirely by internal sources, $Q(\mathrm{H\alpha})$. We present these quantities because they are the closest analogs to the observational measurements of \citet{scheuermann26} for the PHANGS-MUSE galaxies. In their analysis, $Q(0)$ was estimated from UV--I SED fitting of young stellar clusters, while $Q(\mathrm{H\alpha})$ was inferred from the observed \halpha\ luminosities of the associated HII regions. Their distribution for NGC~628 is shown as orange contours.

The model qualitatively reproduces the two distinct populations identified by \citet{scheuermann26}: (1) bright HII regions, where the intrinsic ionizing photon production rate exceeds that inferred from \halpha, consistent with substantial leakage of ionizing radiation, and (2) faint HII regions, where the observed \halpha\ luminosity is larger than can be powered by the associated stellar cluster alone. In our framework, the latter systems arise naturally from contamination by ionizing radiation leaking from nearby HII regions, which supplements the locally produced ionizing emission.

\begin{figure*}
	\centering
\includegraphics[width=1\textwidth]{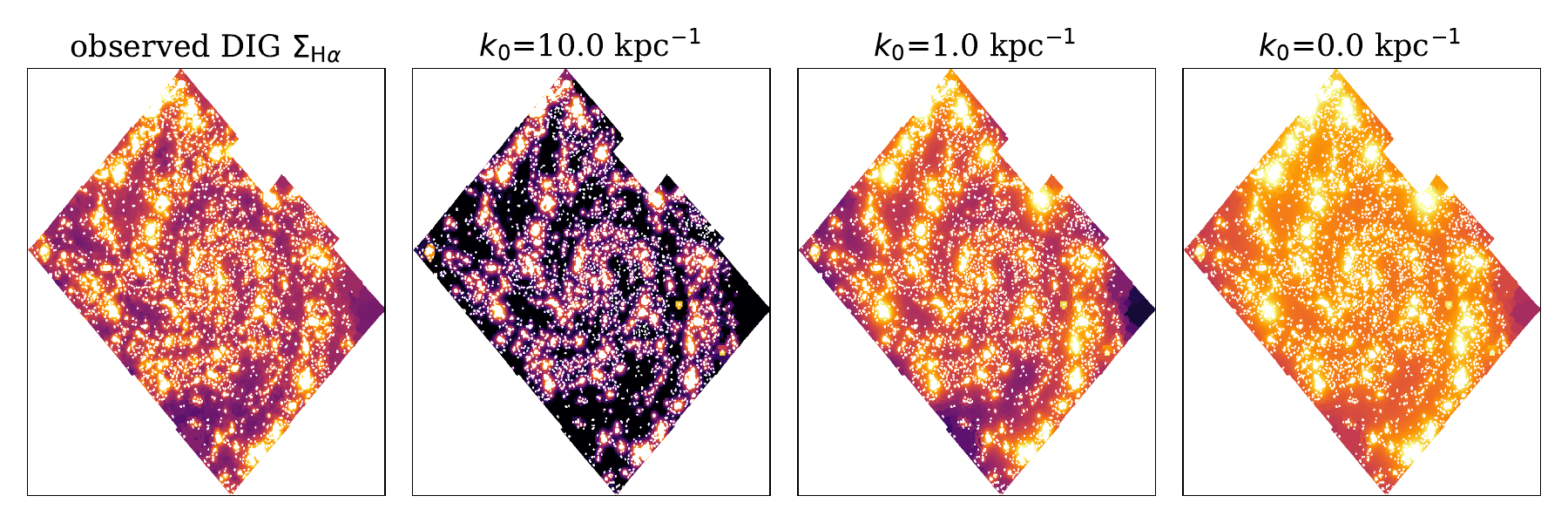}
\caption{\textbf{DIG surface brightness model illustration for NGC~628}. Each panel shows the \halpha\ surface brightness (logarithmic scale) across the DIG of NGC~628, with HII regions masked out. The leftmost panel shows the observed emission. The next three panels show the model-predicted DIG surface brightness obtained from Equation~\ref{eq:DIG_SB} for different values of the effective absorption coefficient $k_0$.
\textbf{Summary:} Variations in the assumed $k_0$ change the predicted DIG surface brightness distribution, allowing the DIG surface brightness to constrain $k_0$ under the leaking radiation scenario.
} 
\label{f:NGC628_different_k0_maps}
\end{figure*}

We do not attempt to tune the model to reproduce the \citet{scheuermann26} distribution more precisely for several reasons. First, \citet{scheuermann26} do not correct the observed \halpha\ luminosities for DIG contamination. Applying such a correction would shift the orange contours downward, particularly for faint HII regions and to a lesser extent for bright ones (see Section~\ref{sec:mixing} for summary statistics). Second, their analysis derives ionizing photon production rates from stellar population synthesis fits to broad-band photometry of young stellar clusters. These estimates depend on assumptions about the stellar initial mass function, metallicity, stellar evolutionary tracks, and the treatment of binary interactions. Third, our assumed escape fractions cannot be directly compared to the escape fractions inferred by \citet{scheuermann26}. Their analysis assumes that all ionizing photons not traced by \halpha\ recombination emission escape the HII region environment. In practice, this means that ionizing photons absorbed by dust in the vicinity of HII regions are not explicitly accounted for. In contrast, our framework defines $f_{\mathrm{esc}}$ as the fraction of ionizing radiation that leaves the HII region environment altogether, and therefore implicitly includes losses due to both gas and dust absorption. Dust absorption of ionizing photons may itself be significant (e.g., \citealt{choi20}). Given the combination of observational and modeling uncertainties in both approaches, we do not consider a more precise fit to the detailed distribution to be physically justified.

The right panel of Figure~\ref{f:HII_coupling_fiducial_model} shows the fractional contribution of internal sources to the observed \halpha\ emission as a function of HII region \halpha\ luminosity. We define $\eta$ as:
\begin{equation}\label{eq:eta_model}
\eta_{i} = \frac{\sum_{j \neq i} K_{ij} f_{\mathrm{esc}, j} Q_{j}}{(1 - f_{\mathrm{esc}, i})Q_{i}
+ \sum_{j \neq i} K_{ij} f_{\mathrm{esc}, j} Q_{j}},
\end{equation}
where the numerator represents the contribution to the observed \halpha\ emission from external sources, while the denominator represents the total ionizing photon budget, including both internal and external contributions. Thus, $1 - \eta$ corresponds to the fraction of the observed \halpha\ emission powered by internal sources within a given HII region. This definition is the model-dependent analog of Equation~\ref{eq:eta}, introduced in Section~\ref{sec:HII_env_dependence} to characterize the strength of the external ionizing radiation field.

We show the median and median absolute deviation of $1-\eta$ in bins of HII region \halpha\ luminosity. The resulting distribution exhibits a characteristic trend, which persists across the full range of model assumptions we explored: bright HII regions are predicted to be dominated by internal ionizing sources, with contamination from leaking radiation typically limited to $\lesssim10\%$, while progressively fainter HII regions experience increasingly large contributions from externally supplied ionizing radiation. In particular, HII regions with $L_{\mathrm{H}\alpha}\sim10^{37}\,\mathrm{erg\,s^{-1}}$ are predicted to have roughly 40\% (median) of their observed \halpha\ emission powered by external sources.

The large error bars around the median values reflect the diversity of radiative environments experienced by HII regions. At fixed \halpha\ luminosity, the degree of contamination depends on the local environment, with HII regions located near bright neighboring sources experiencing on-average larger contamination from leaking radiation.

\subsubsection{General conclusions from a broad parameter exploration}

In \ref{app:HII_rad_coupling_app} we explore the sensitivity of our results to the various model parameters. In particular, we examine the impact of the adopted geometry ($\Delta \Omega_{ij}$) and mean free path, as well as the dependence on the escape fraction, both when assumed to be constant across HII regions and when allowed to depend on the unknown $Q_{i}$.

Overall, for all models that are not extreme or ruled out by other observational constraints, the emerging picture is that for bright HII regions, leaking radiation is negligible relative to the intrinsic radiation, owing to their substantial internal ionizing photon production. At the same time, the existence of these bright HII regions, together with the clustering of fainter HII regions around them, leads to a robust trend across all model assumptions: progressively fainter HII regions experience increasing contamination from leaking radiation.

\begin{figure*}
	\centering
\includegraphics[width=0.8\textwidth]{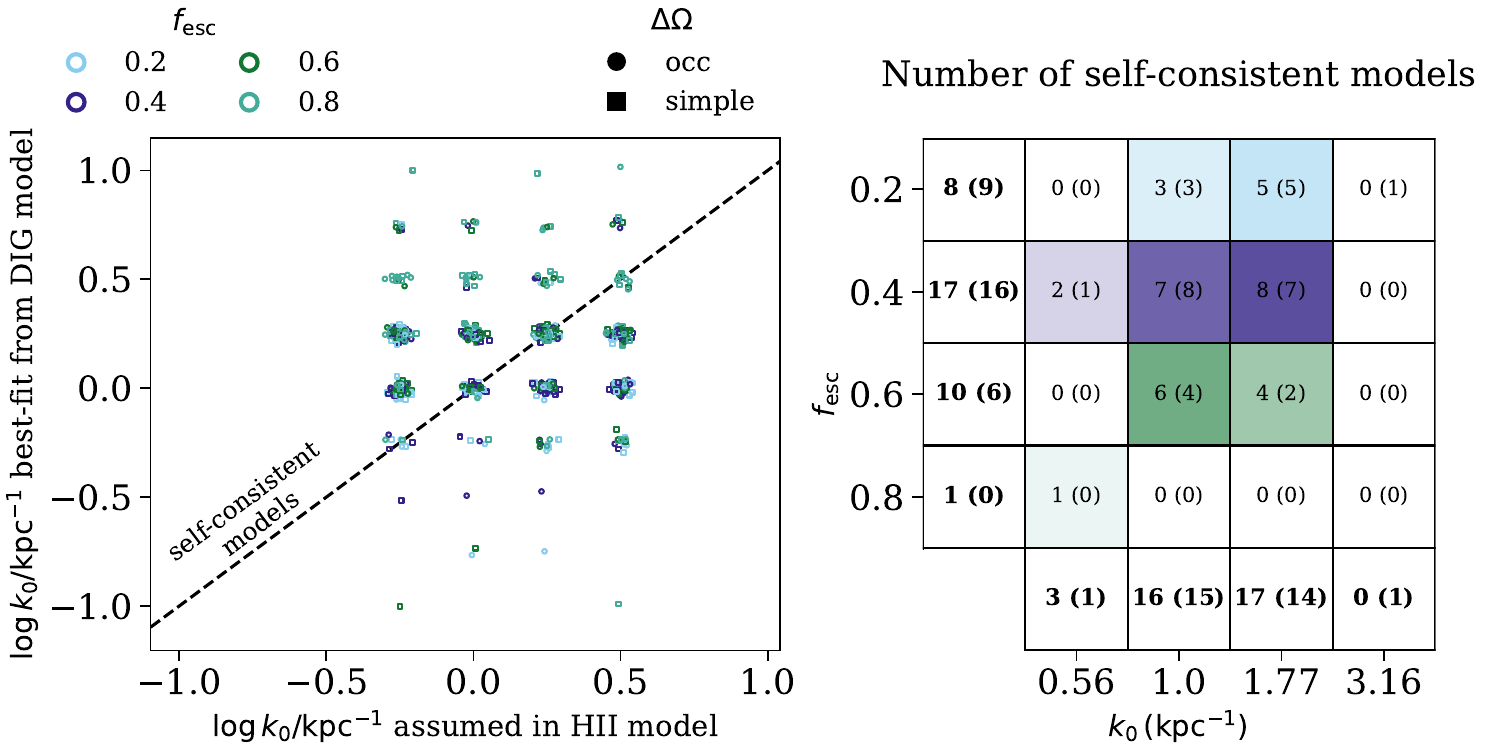}
\caption{\textbf{Self-consistency of HII region + DIG radiation propagation models across the PHANGS-MUSE galaxies.} \textbf{The left panel} shows all coupled HII region + DIG models for the 19 PHANGS-MUSE galaxies, where the x-axis represents the $k_0$ assumed in the HII region model and the y-axis shows the best-fitting $k_0$ inferred from the DIG model. Marker colors and shapes indicate the assumed $f_{\mathrm{esc}}$ and $\Delta \Omega$ in the HII region model (see legend in the upper left panel). For each of the 19 galaxies, we explored 32 realizations of the model parameters ($k_0$, $f_{\mathrm{esc}}$, $\Delta \Omega$), resulting in 608 model runs across the full sample.
\textbf{The right panel} shows the number of self-consistent models, defined as cases where the assumed $k_0$ and $f_{\mathrm{esc}}$ in the HII region model are also recovered by the DIG fit. The first number corresponds to optically thin HII region models (\texttt{simple}), while the number in parentheses corresponds to optically thick models (\texttt{occ}). \textbf{Summary:} Across the full sample, self-consistent solutions are primarily found for escape fractions $f_{\mathrm{esc}} = 0.4$--$0.6$ and effective absorption coefficients $k_0 = 1$--$1.77\,\mathrm{kpc^{-1}}$ (mean free path $\lambda=0.5$--$1$ kpc). Figure~\ref{f:chi2_dof_improvement} further shows that the coupled models improve the overall fit to the data.} 
\label{f:k0_comparison_all_gals}
\end{figure*}

Across HII regions of all luminosities, the inferred degree of contamination is higher when HII regions are assumed to be optically thin (\texttt{simple} $\Delta \Omega$), when the escape fraction is higher, or when the mean free path is longer. Synthesizing results from all model variants, we find that for an HII region with $L(\mathrm{H\alpha})\sim10^{37}\,\mathrm{erg\,s^{-1}}$, the inferred contamination fraction ranges from 20 to 60\%, depending on the adopted assumptions. This is clearly not a negligible effect and may significantly influence the ionization conditions within HII regions.

\subsection{A self-consistent HII region + DIG radiation propagation model}\label{sec:leaking_HII_and_DIG}

\subsubsection{DIG \halpha surface brightness model}

Following the framework of \citet{belfiore22}, the \halpha\ surface brightness of the DIG in spatial bin $k$ can be modeled as:
\begin{equation}\label{eq:DIG_SB}
C_{\alpha} \Sigma_{\mathrm{H\alpha},k} = a_0 + a_1 \sum_{j} e^{-k_0 r_{kj}} \frac{Q_j}{4 \pi r_{kj}^2},
\end{equation}
where $C_{\alpha}$ is the conversion factor between ionizing photon production rate and \halpha\ luminosity, $Q_j$ is the ionizing photon production rate of HII region $j$, $k_0$ is the effective absorption coefficient, and $r_{kj}$ is the inclination-corrected distance between HII region $j$ and DIG bin $k$. For a fixed value of $k_0$, this expression can be fitted to the observed DIG surface brightness distribution to determine the best-fitting values of $a_0$ and $a_1$. If the DIG emission is fully accounted for by leaking radiation from HII regions, the best-fitting background term $a_0$ should be close to zero (as found both by \citealt{belfiore22} and in this work). According to Equation~1 of \citet{belfiore22}, $a_1$ is proportional to $[f_{\mathrm{esc}} / (1-f_{\mathrm{esc}})]\Delta h$, where $\Delta h$ is the DIG scale height.

The key difference between the \citet{belfiore22} framework and the approach adopted here is that we use the intrinsic ionizing photon production rates derived from the HII region coupling model, $Q_j$, rather than the observed \halpha\ luminosities of the HII regions themselves. In our framework, the observed \halpha\ luminosities generally include some contribution from ionizing radiation leaking from neighboring HII regions. Using the observed \halpha\ luminosities directly would therefore effectively double count some of the ionizing photons in the DIG radiation budget.

In Figure~\ref{f:NGC628_different_k0_maps}, we illustrate the model output for NGC~628 for three values of $k_0/\mathrm{kpc^{-1}}$: 10, 1, and 0. For each PHANGS-MUSE galaxy, \citet{belfiore22} computed DIG models for nine logarithmically spaced values of $k_0$ between $0.1$ and $10\,\mathrm{kpc^{-1}}$, identified the value that provides the best agreement with the observed DIG surface brightness (in terms of $\chi^2_{\mathrm{dof}}$), and used the corresponding best-fitting $a_{1}$, together with an assumed DIG scale height of $\Delta h = 1\,\mathrm{kpc}$, to estimate the escape fraction. We follow a similar approach here.

\begin{figure*}
	\centering
\includegraphics[width=1\textwidth]{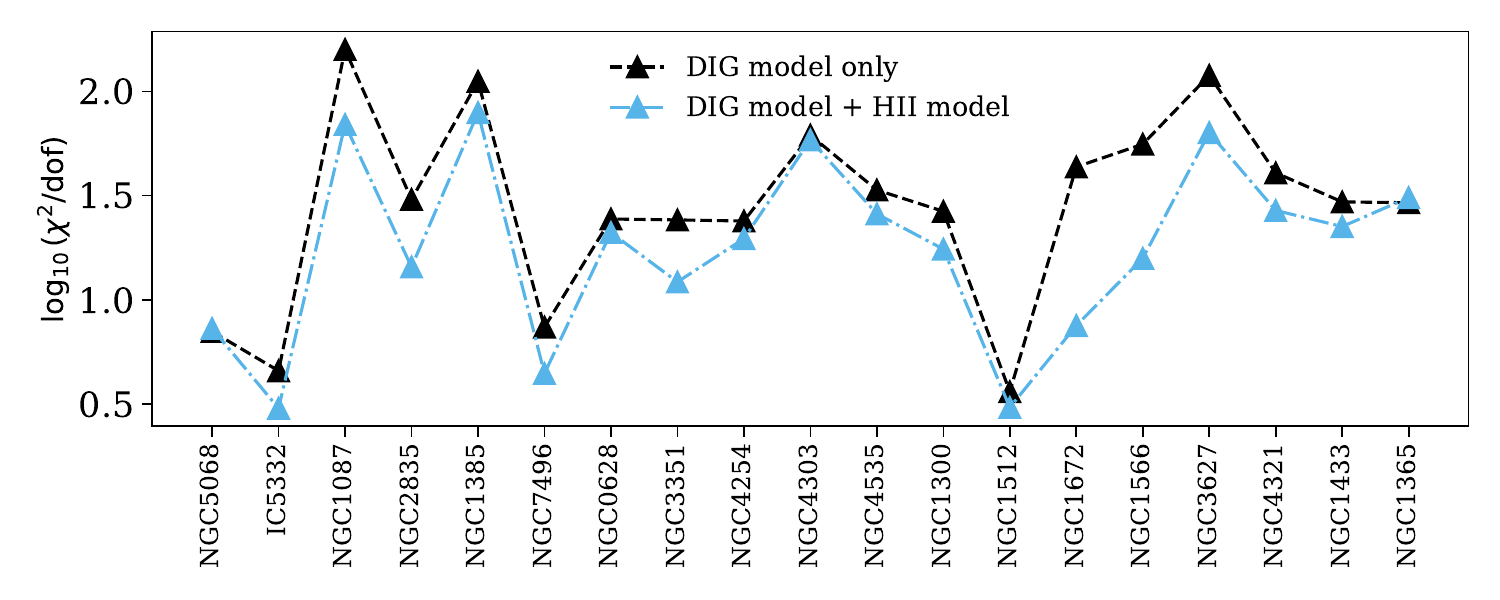}
\caption{\textbf{Comparison of DIG surface brightness goodness of fit between the DIG-only and coupled HII region+DIG models.} For each galaxy, we compare the goodness of fit of the DIG radiation propagation model when using the observed \halpha\ luminosities of HII regions directly as source terms (DIG-only model) with the goodness of fit obtained when using the intrinsic ionizing photon production rates, $Q$, derived from the HII region coupling model (coupled HII region+DIG model). \textbf{Summary:} In 13 of the 19 galaxies, the coupled model provides a better fit to the observed DIG surface brightness, while in the remaining 6 galaxies the two approaches yield comparable goodness of fit. Both approaches have the same number of free parameters for the DIG.} 
\label{f:chi2_dof_improvement}
\end{figure*}

\subsubsection{Coupling the HII regions and DIG propagation models}

Our HII region radiative coupling model presented in Section~\ref{sec:leaking_HII} requires assumptions for $f_{\mathrm{esc}}$, $k_0$, and the geometry ($\Delta \Omega$), and returns the best-fitting ionizing photon production rates for each HII region, $Q_i$. The DIG propagation model, in contrast, takes $Q_i$ as an input and, by fitting the observed DIG surface brightness, returns best-fitting values of $k_0$ and $f_{\mathrm{esc}}$. The two components therefore form a coupled system, for which a fully self-consistent solution can, in principle, be obtained through iteration until convergence.

In practice, however, each combined HII region–DIG model evaluation is computationally expensive, and given the simplified nature of both models relative to the complexity of real galaxies, we adopt a more pragmatic approach. Our goals are: (1) to test whether values of $k_0$ and $f_{\mathrm{esc}}$ exist that yield self-consistent solutions between the two frameworks (i.e., where the assumed parameters in the HII region model are consistent with those recovered by the DIG fit), and (2) to assess whether this coupled treatment improves the overall fit to the DIG surface brightness relative to the approach of \citet{belfiore22}, which uses observed \halpha\ luminosities directly.

To achieve this, we examine the following input grid of parameters for the HII region model: $f_{\mathrm{esc}} \in [0.2, 0.4, 0.6, 0.8]$, $k_{0} \in [0.56, 1, 1.77, 3.16]$, and $\Delta \Omega_{ij} \in [\texttt{simple}, \texttt{occ}]$. The chosen $k_0$ values are drawn from the set of nine logarithmically spaced bins adopted by \citet{belfiore22}, restricted here based on preliminary tests indicating that the best-fitting values lie within this range. For each set of HII region model parameters, we run the coupling model to derive the best-fitting $Q_i$. We then use these $Q_i$ as input to the DIG propagation model, and consider the nine logarithmically spaced $k_0$ bins between $0.1$ and $10\,\mathrm{kpc^{-1}}$. We identify the $k_0$ that provides the best overall fit, and, using the corresponding best-fitting $a_1$, we derive the escape fraction $f_{\mathrm{esc}}$.

We consider a model self-consistent if the assumed $k_0$ in the HII region model matches the best-fitting $k_0$ obtained from the DIG propagation model. The escape fraction inferred from the DIG model depends on the assumed DIG scale height, which is not fixed to $1\,\mathrm{kpc}$ but was adopted by \citet{belfiore22} as representative of median scale heights measured in the Milky Way and in edge-on galaxies. We therefore adopt the range $\Delta h \in [0.5, 1.5]\,\mathrm{kpc}$ from \citet{levy19} as a plausible interval. A model is then considered self-consistent if, within this range of scale heights, the escape fraction inferred from the DIG model is consistent with the escape fraction assumed in the HII region model.

Finally, requiring the same $k_0$ in both the HII region and DIG models may not be strictly valid. In the immediate vicinity of HII regions, particularly in clustered environments, one may expect higher gas densities and a greater presence of neutral and molecular clouds, whereas in the more extended DIG the gas is likely more diffuse and less neutral/molecular (e.g., \citealt{haffner09}). Given the simplified nature of our framework, we nevertheless enforce this equality as a working assumption. More realistic treatments, such as full radiative transfer (e.g., \citealt{seon09, pellegrini20}) or galaxy evolution simulations (e.g., \citealt{kim17}), would be better suited to capture these environmental variations in a self-consistent way.

\begin{figure*}
	\centering
\includegraphics[width=1\textwidth]{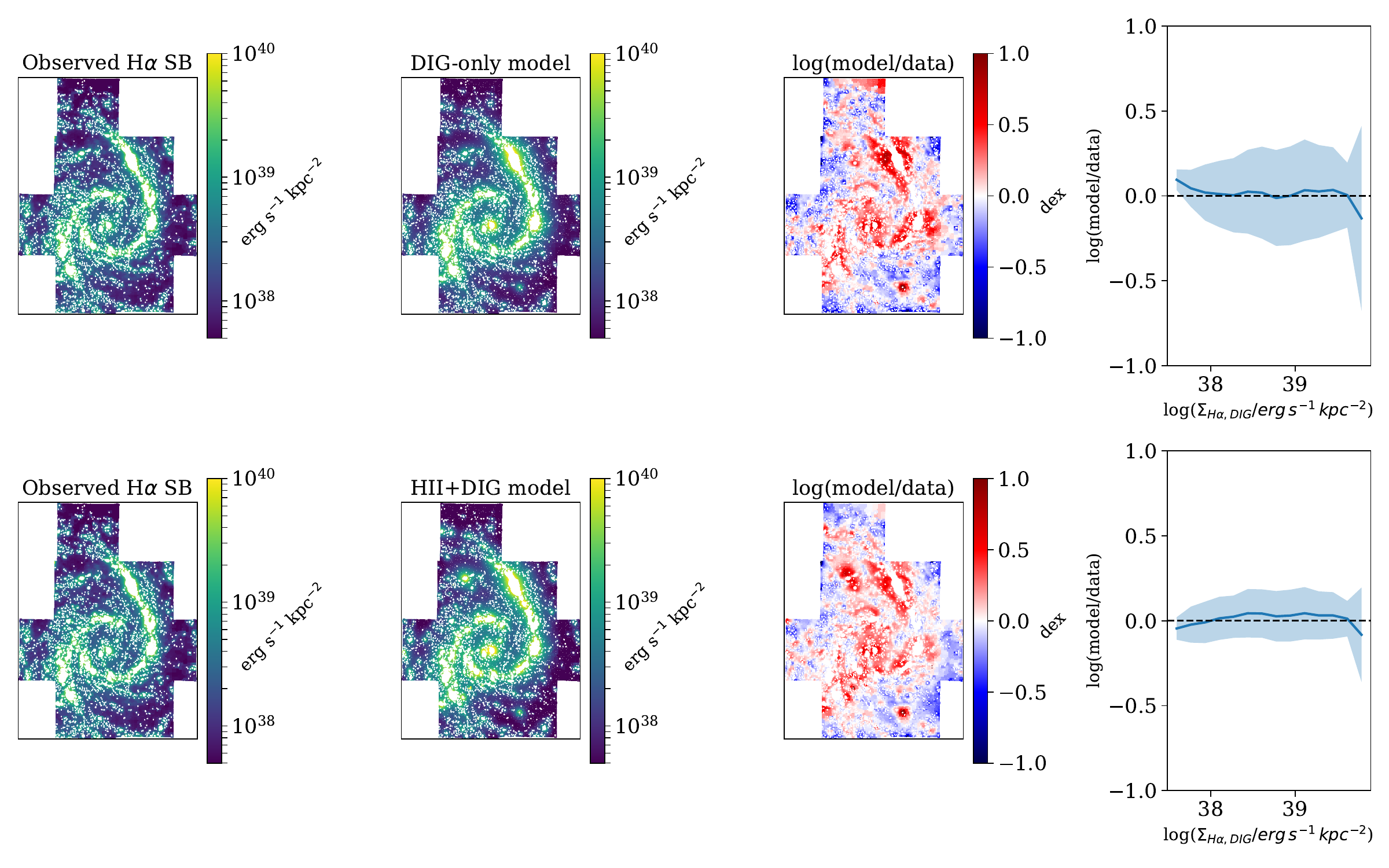}
\caption{\textbf{Comparison of DIG surface brightness models for NGC~1566: DIG-only versus coupled HII region+DIG framework.} The top row shows the original DIG-only model, while the bottom row shows the self-consistent coupled HII region+DIG model. The leftmost panel, identical in both rows, presents the observed DIG \halpha\ surface brightness in NGC~1566, with HII regions masked out. The second panel from the left shows the predicted DIG \halpha\ surface brightness from the best-fitting models (DIG-only in the top row and coupled HII region+DIG in the bottom row), while the third panel shows the residuals between the observations and the model. The rightmost panels show the median model-to-data ratio in bins of DIG \halpha\ surface brightness. \textbf{Summary:} The coupled HII region+DIG model provides a better overall fit, with smaller residuals particularly along spiral arms and improved agreement across the full range of DIG surface brightness. Small-scale residuals remain in both models, suggesting that a spatially varying $k_0$ may be required.} 
\label{f:DIG_SB_comparison_NGC1566}
\end{figure*}

\vspace{1in}
\subsubsection{Self-consistent models across the PHANGS-MUSE sample}

We run the coupled HII region + DIG radiation propagation model for each of the 19 PHANGS-MUSE galaxies, exploring 32 realizations of the model parameters ($f_{\mathrm{esc}}$, $k_0$, $\Delta \Omega$). In total, this results in 608 model runs across the full sample. Figure~\ref{f:k0_comparison_all_gals} summarizes the results. The left panel compares the $k_0$ assumed in the HII region model with the best-fitting $k_0$ derived from the DIG model, while the right panel shows the number of self-consistent models as a function of ($f_{\mathrm{esc}}$, $k_0$, and $\Delta \Omega$).

Out of the 608 models, only 67 are self-consistent, suggesting that the combination of the HII region and DIG constraints significantly restricts the allowed parameter space. Every galaxy has at least one self-consistent solution. From the figure, we see that self-consistent models are primarily found for escape fractions $f_{\mathrm{esc}} = 0.4$--$0.6$ and effective absorption coefficients $k_0 = 1$--$1.77\,\mathrm{kpc^{-1}}$. There is only a marginal excess of self-consistent solutions for optically thin HII region models (\texttt{simple}) relative to optically thick models (\texttt{occ}), not at a level that would allow us to strongly prefer one geometry over the other.

\subsubsection{Improved DIG surface brightness modeling with the coupled models}

In Figure~\ref{f:chi2_dof_improvement} we compare the goodness of fit to the DIG \halpha\ surface brightness for two classes of models. The first is the best-fitting DIG-only model following the \citet{belfiore22} framework, where the HII region source terms are taken directly from the observed \halpha\ luminosities, without correcting for contamination by leaking radiation. The second is the coupled and self-consistent HII region+DIG model, where the source terms are the intrinsic ionizing photon production rates, $Q$, derived after accounting for the contribution of leaking radiation to the HII regions themselves. 

For 13 out of the 19 galaxies, the coupled HII region+DIG model provides an improved fit to the DIG surface brightness distribution. For the remaining 6 galaxies, the goodness of fit is comparable between the two approaches.

In Figure~\ref{f:DIG_SB_comparison_NGC1566} we compare the predicted DIG \halpha\ surface brightness from the DIG-only model with that from the coupled and self-consistent HII region+DIG model for NGC~1566, one of the galaxies showing the largest improvement in $\chi_{dof}^2$. The coupled model produces overall smaller residuals between the observed and predicted surface brightness distributions, reducing the over-prediction near the spiral arms while also alleviating the under-prediction around fainter clustered HII regions. 

However, it is important to note that even the coupled model still tends to over-predict the DIG surface brightness near spiral arms, where bright and clustered HII regions are concentrated, while under-predicting the emission in fainter regions. Since the contribution from leaking radiation has been accounted for, this residual pattern may instead reflect our assumption of a single constant $k_0$ across the entire disk. The effective mean free path of ionizing photons could depend on the local ISM environment, becoming shorter in dense regions associated with neutral and molecular gas and longer in more diffuse environments (e.g., \citealt{zurita02, haffner09}). Allowing $k_0$ to vary spatially may therefore reduce these systematic residuals. In principle, molecular gas maps from PHANGS-ALMA \citep{leroy21a}, or H\,\textsc{i} maps available for a subset of the PHANGS galaxies \citep{eibensteiner24}, could be used to construct a spatially varying mean-free-path model. Exploring such models is beyond the scope of the present work.

\subsection{Spectral hardening of leaking radiation in the diffuse ISM}\label{sec:leaking_hardening}

In Sections~\ref{sec:leaking_HII} and \ref{sec:leaking_HII_and_DIG}, we treated the escaping ionizing radiation field in an integrated sense, considering only the total Lyman continuum photon budget available to power the observed \halpha\ emission, without explicitly following the spectral dependence of the radiation field. In that framework, the effective absorption coefficient, $k_0$, parametrizes the attenuation of the ionizing continuum as a whole.

In reality, however, the spectral shape of the escaping radiation is expected to evolve as it propagates through the ISM. In particular, photoelectric absorption by neutral hydrogen preferentially removes photons near the Lyman edge, where the \hi photoionization cross section is largest, while higher-energy photons propagate more efficiently. As a result, the transmitted radiation field becomes progressively harder with increasing propagation distance. In this section, we extend the model to include this spectral hardening of the leaking ionizing radiation field.

In Section~\ref{sec:attenuation_model}, we describe our adopted wavelength-dependent attenuation model. In Section~\ref{sec:wavelength_dependent_K}, we extend the radiative coupling kernel, $K_{ij}$, to include this wavelength dependence in the attenuation. In Section~\ref{sec:spectral_hardness_in_HII}, we examine the resulting effective spectral hardness of the ionizing radiation field as experienced by HII regions of different luminosities.

\begin{figure}
	\centering
\includegraphics[width=1\columnwidth]{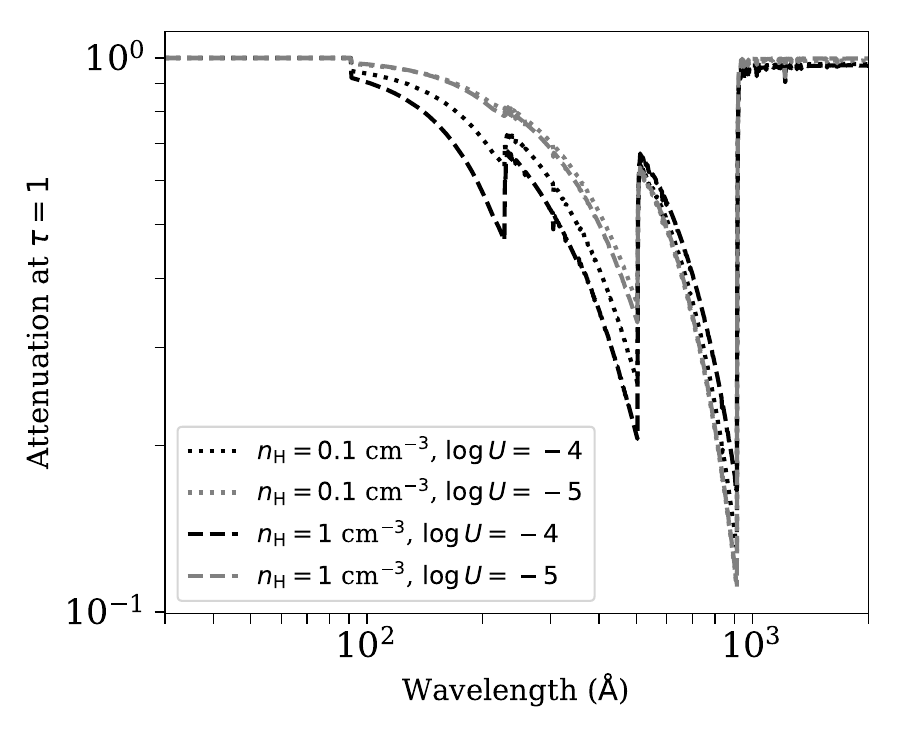}
\caption{\textbf{Comparison of wavelength-dependent attenuation models.} The curves show the transmitted fraction of the ionizing spectrum, defined as $f_{\lambda,\,\mathrm{obs}}(\tau_{\mathrm{LyC}}=1)/f_{\lambda,\,\mathrm{emitted}}$, where $\tau_{\mathrm{LyC}}=1$ corresponds to the optical depth at which a fraction $e^{-1}$ of the Lyman continuum radiation remains unabsorbed. The attenuation models were computed with {\sc cloudy}, assuming $n_{\mathrm{H}}=0.1\,\mathrm{cm^{-3}}$ and $n_{\mathrm{H}}=1\,\mathrm{cm^{-3}}$, and ionization parameters $\log U=-4$ and $-5$.}
\label{f:attenuation_comparison}
\end{figure}

\subsubsection{Wavelength-dependent attenuation of the ionizing continuum}\label{sec:attenuation_model}

We use the {\sc cloudy} photoionization code \citep{ferland17, chatzikos23} to construct a physically motivated wavelength-dependent attenuation model for the propagation of ionizing radiation through the DIG. We consider hydrogen number density of $n_{\mathrm{H}}=0.1\,\mathrm{cm^{-3}}$ and $n_{\mathrm{H}}=1\,\mathrm{cm^{-3}}$, roughly consistent with typical electron densities inferred for DIG clouds in the Milky Way \citep{reynolds91, ferriere01}.

\begin{figure}
	\centering
\includegraphics[width=1\columnwidth]{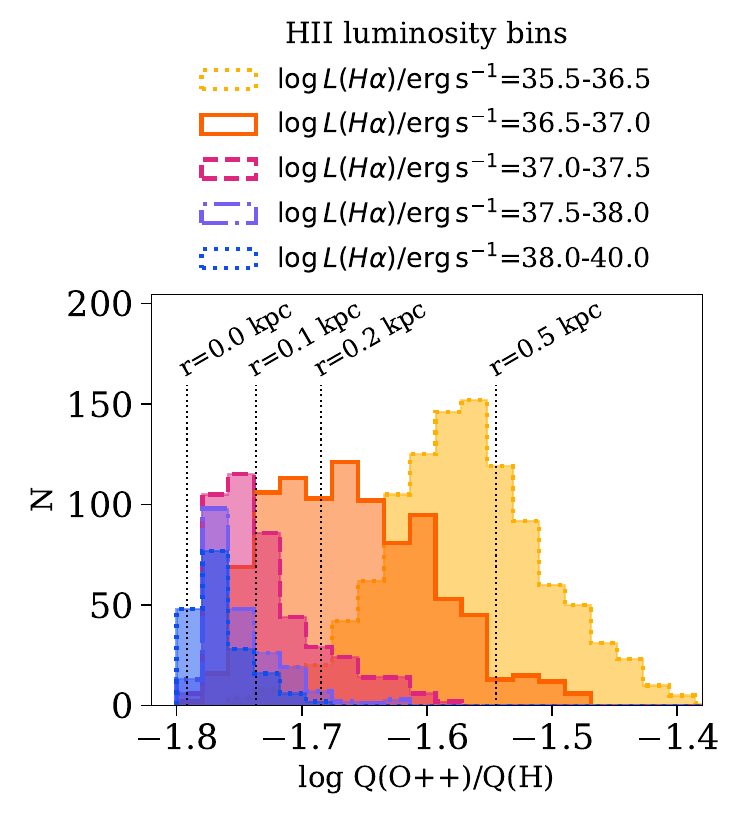}
\caption{\textbf{Effective spectral hardness of radiatively coupled HII regions of different luminosities in NGC~628.} The histograms show the distribution of effective spectral hardness, characterized by $Q(\mathrm{O^{++}})/Q(\mathrm{H^{+}})$, for HII regions in NGC~628. The effective spectrum of each HII region combines its locally produced ionizing spectrum, modeled as a 2 Myr FSPS single stellar population, with the external ionizing field from neighboring HII regions, propagated using the wavelength-dependent radiative coupling kernel that accounts for attenuation, geometric dilution, and spectral hardening. The five histograms correspond to different \halpha\ luminosity bins and show that progressively fainter HII regions exhibit harder effective spectra due to the increasing contribution of leaked and hardened radiation. The vertical dashed lines indicate the spectral hardness of a 2 Myr FSPS single stellar population after propagating a distance $r$ through the ISM.
\textbf{Summary:} Within the leaking radiation picture, fainter HII regions experience larger contributions from leaked, spectrally hardened radiation, leading to harder effective ionizing spectra on average.} 
\label{f:hardening_as_a_function_of_L_halpha_k0_1p5_DIGcorr}
\end{figure}

\citet{belfiore22} argued that if the electron density is approximately uniform throughout the DIG, then $\Sigma$(\halpha), which in our framework traces the leaked ionizing radiation field from HII regions, should scale approximately with the ionization parameter. Based on the observed \halpha\ surface brightness distributions of HII regions and the DIG in NGC~628 (Figure~\ref{f:DIG_smooth_maps_NGC0628}), we therefore estimate that the ionization parameter in the DIG is roughly 1--2 dex lower than in HII regions. Since typical HII regions in NGC~628 have ionization parameters of $\log U \sim -3$, we adopt representative DIG values of $\log U \sim -4$ to $-5$.

The remaining photoionization model parameters are identical to those adopted for Figure~\ref{f:BPT_tracks} and are described in detail in \ref{app:cloudy_grids}, where the gas-phase abundances were tuned to simultaneously reproduce the observed \oiii/\hbeta, \nii/\halpha, \sii/\halpha, and \oi/\halpha\ line ratios of HII regions in NGC~628. As the incident radiation field, we adopt a 2~Myr FSPS single stellar population spectrum.

In Figure~\ref{f:attenuation_comparison}, we show the wavelength-dependent attenuation curves for the four {\sc cloudy} models. The curves represent the transmitted fraction of the incident radiation field at $\tau_{\mathrm{LyC}}=1$, corresponding to the optical depth at which a fraction $e^{-1}$ of the Lyman continuum radiation remains unabsorbed. We find that the attenuation curves are nearly unchanged when the calculations are repeated without dust, demonstrating that dust has a negligible effect on the attenuation in these models.

Figure~\ref{f:attenuation_comparison} shows that the $\log U=-5$ models produce a harder transmitted radiation field, as they preferentially allow more high-energy photons to escape. In particular, the transmission remains closer to unity at short wavelengths (e.g., $\lambda=300\,\mathrm{\AA}$) compared to the lower-energy ionizing photons near the Lyman limit ($\lambda=912\,\mathrm{\AA}$), indicating weaker absorption of the harder radiation. Throughout the remainder of this section, we adopt the $\log U=-4$ and $n_{\mathrm{H}}=1\,\mathrm{cm^{-3}}$ model, which produces an intermediate degree of spectral hardening, as our fiducial attenuation prescription. Repeating the analysis using the other models yields qualitatively unchanged conclusions.

\subsubsection{Generalizing the radiative coupling kernel to include spectral dependence}\label{sec:wavelength_dependent_K}

We generalize the radiative coupling kernel, $K_{ij}$, used in both the HII region coupling and HII region+DIG models to include wavelength-dependent attenuation:
\begin{equation}\label{eq:Kij_lambda}
K_{ij}(\lambda) = e^{-\alpha_{*}\, k_0\, g(\lambda)\, r_{ij}} \frac{\Delta \Omega_{ij}}{4 \pi},
\end{equation}
where $g(\lambda)$ is the attenuation curve introduced above, and $\alpha_*$ is a normalization factor chosen to preserve consistency with the wavelength-independent treatment adopted in Sections~\ref{sec:leaking_HII} and \ref{sec:leaking_HII_and_DIG}. Since spectral hardening causes the effective attenuation of the integrated Lyman continuum to evolve with distance, a single wavelength-independent $k_0$ cannot reproduce the attenuation at all radii simultaneously.

We therefore calibrate $\alpha_*$ at a fiducial propagation scale $r_*$ by requiring that the transmitted fraction of ionizing photons matches that of the gray model:
\begin{equation}\label{eq:alpha_star_def}
\frac{\int f_\lambda\, e^{-\alpha_*\, k_0\, r_* g(\lambda)}\, d\lambda}{\int f_\lambda\, d\lambda}
= e^{-k_0 r_*},
\end{equation}
where $f_\lambda$ is the incident ionizing photon spectrum. We solve Equation~\ref{eq:alpha_star_def} numerically using the adopted FSPS ionizing spectrum and set $r_* = 2\,\mathrm{kpc}$. This normalization preserves the effective mean free path implied by $k_0$ for the integrated ionizing photon budget at $r_*$, while still allowing the spectrum to harden with distance.

For $r < 2\,\mathrm{kpc}$, the wavelength-dependent and gray attenuation models differ in their integrated attenuation by less than 10\%, whereas the discrepancy grows substantially above $2\,\mathrm{kpc}$, to $\sim250\%$ by $r=5\,\mathrm{kpc}$. We therefore restrict the hardening model to $r \leq 2\,\mathrm{kpc}$. This comfortably encompasses the physically relevant propagation scales in our analysis: the $K_{ij}$-weighted distances between coupled HII regions have 16th, 50th, and 84th percentiles of 100, 200, and 570 pc, respectively, while the corresponding HII-to-DIG weighted distances are 400, 700, and 1000 pc, with a 99.9th percentile of 1.6 kpc.

\begin{figure*}
	\centering
\includegraphics[width=1\textwidth]{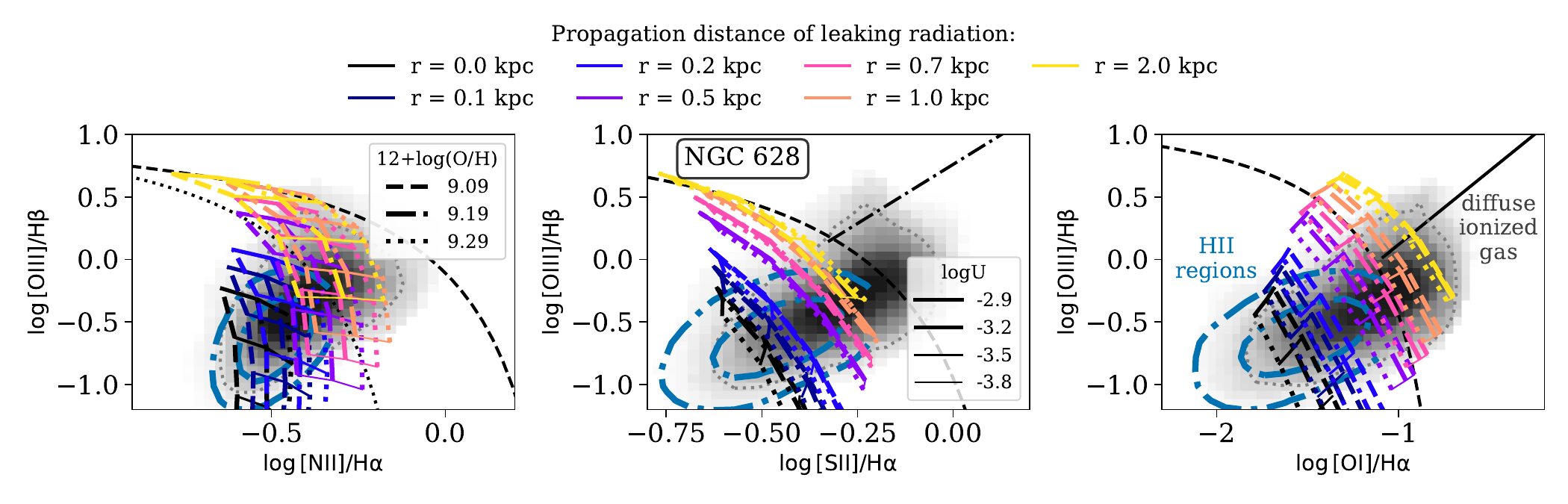}
\caption{\textbf{Photoionization model grid for leaking radiation progressively hardened during propagation through the ISM.} \textbf{Data:} BPT diagrams of \nii/\halpha\ (left), \sii/\halpha\ (middle), and \oi/\halpha\ (right) versus \oiii/\hbeta\ for NGC~628. Blue contours show the 50th and 95th percentiles of the HII region distribution, while the DIG is shown as a logarithmically scaled grayscale density map, with the grey dotted contour marking its 95th percentile. \textbf{Models:} Colored curves show {\sc cloudy} predictions for a 2 Myr FSPS stellar population filtered by propagation through the ISM using our adopted wavelength-dependent attenuation model, for different propagation distances (different colors). For each propagation distance, the grids span a range of gas-phase (and stellar) metallicities and ionization parameters, as indicated in the legends.
\textbf{Summary:} Spectral hardening of the ionizing radiation field with propagation distance increases both low- and high-ionization line ratios, broadly reproducing the observed trends across the PHANGS galaxies. Allowing the ionization parameter to decrease with distance to account for geometric dilution produces shallower trends that are in even better agreement with the observations.
} 
\label{f:filtered_radiation_cloudy_models}
\end{figure*}

\subsubsection{Emergent spectral hardness of radiatively coupled HII regions}\label{sec:spectral_hardness_in_HII}

With the wavelength-dependent radiative coupling kernel, Equation~\ref{eq:HII_prop_eq} can be generalized to describe the full ionizing spectrum incident on each HII region. For the fiducial NGC~628 model presented in Section~\ref{sec:fiducial_model_NGC628}, we estimate the effective ionizing spectrum seen by every HII region. This spectrum consists of two components: (1) the locally produced radiation field, modeled as a 2 Myr FSPS single stellar population, and (2) externally supplied radiation leaking from neighboring HII regions. The external component is also assumed to originate from a 2 Myr FSPS spectrum, but is progressively hardened as it propagates through the ISM according to the adopted attenuation model. The effective spectrum for each HII region is therefore obtained by summing the internal and external contributions, accounting for the propagation distance and attenuation associated with each neighboring HII region.

In Figure~\ref{f:hardening_as_a_function_of_L_halpha_k0_1p5_DIGcorr} we show the distribution of spectral hardness, quantified as $Q(\mathrm{O^{++}})/Q(\mathrm{H^{+}})$, for HII regions in NGC~628. The sample is divided into bins of \halpha\ luminosity, and the hardness distribution is shown separately in each bin. As shown in the figure, progressively fainter HII regions, which have a larger contribution from leaked ionizing radiation, exhibit systematically harder effective spectra.

This behavior indicates that the combination of radiation leakage and spectral hardening reproduces the qualitative trends seen in the observations, namely increasing optical line ratios (\oiii/\hbeta, \nii/\halpha, \sii/\halpha, and \oi/\halpha) toward lower \halpha\ luminosities across HII regions. It is also consistent with the observed trend that, at fixed \halpha\ luminosity, HII regions in more radiatively rich environments (higher $\eta$) exhibit higher optical line ratios, as these regions are expected to receive a larger contribution from leaked ionizing radiation, which is spectrally harder than the original emitted 2 Myr stellar population.

\subsection{Photoionization models of leaking and spectrally hardened radiation}\label{sec:leaking_photoionization}

To illustrate the impact of spectral hardening on the predicted emission-line ratios, we run a grid of {\sc cloudy} photoionization models in which the incident radiation field is progressively filtered using our adopted wavelength-dependent attenuation model. Figure~\ref{f:filtered_radiation_cloudy_models} shows the resulting model predictions for a 2 Myr FSPS single stellar population propagated through the ISM over distances ranging from 0 to 2 kpc. For each propagation distance, we consider a range of gas-phase (and stellar) metallicities, $12 + \log (\mathrm{O/H}) = 9.09$--$9.29$, and ionization parameters, $\log U = -2.9$ to $-3.8$. The ionization parameter range was chosen to span the regime between typical HII-region values ($\log U \approx -3$) and the ionization parameter adopted for our fiducial DIG attenuation prescription ($\log U=-4$; Section~\ref{sec:attenuation_model}).

As the propagation distance increases, the hardening of the radiation field produces a simultaneous increase in the optical line ratios \oiii/\hbeta, \nii/\halpha, \sii/\halpha, and \oi/\halpha. At the same time, the predicted \halpha\ equivalent width decreases by a factor of $\sim$15-30 between propagation distances of 0 and 2 kpc. Overall, the leaking and spectrally hardened radiation model qualitatively reproduces the observed trends of increasing line ratios toward lower \halpha\ luminosities and lower \halpha\ EWs.

The predicted trends across the BPT diagrams are somewhat steeper than those observed in NGC~628 when the ionization parameter is held fixed across all propagation distances. However, it is reasonable to expect the ionization parameter to decrease along the sequence, particularly if it scales with the observed $\Sigma$(\halpha), which itself decreases toward larger line ratios. Allowing the ionization parameter to decrease with propagation distance produces shallower tracks that better match the observed distributions, since decreasing $\log U$ generally lowers \oiii/\hbeta\ while increasing \nii/\halpha, \sii/\halpha, and \oi/\halpha (see e.g., Figure~\ref{f:BPT_tracks}). Given the simplified nature of the radiation propagation and attenuation model, we do not attempt to fine tune the model further.

Building a more quantitative prediction of line ratio variations across the BPT diagrams for the DIG in PHANGS-MUSE depends sensitively on the adopted radiation propagation model, including the treatment of spectral hardening and the structure of the multiphase medium through which the radiation propagates (e.g., \citealt{wood04, haffner09, mcclymont24}). Additional constraints on the hardening may come from the \hei\ $\lambda5875$ recombination line, which provides a direct probe of the propagation of higher-energy photons ($24.6\,\mathrm{eV}$) from HII regions into the DIG and could therefore be used to empirically constrain attenuation models. Constraining the clumpy structure of the diffuse medium outside HII regions is less straightforward, but in principle information may be obtained from cold molecular gas maps from PHANGS-ALMA at $\sim 150$ pc resolution \citep{leroy21a}, or \hi\ maps at a coarser resolution \citep{eibensteiner24}. Furthermore, recent PHANGS-JWST results suggest that mid-infrared PAH (Polycyclic Aromatic Hydrocarbon; \citealt{tielens08, li20}) emission closely traces gas column density (e.g., \citealt{leroy23, chown25}), and may therefore serve as a high-resolution ($\sim10$ pc) tracer of the clumpy medium through which ionizing photons propagate.\vspace{0.1in}

\section{Discussion}\label{sec:discussion}

\begin{figure}
    \centering
    \includegraphics[width=1\linewidth]{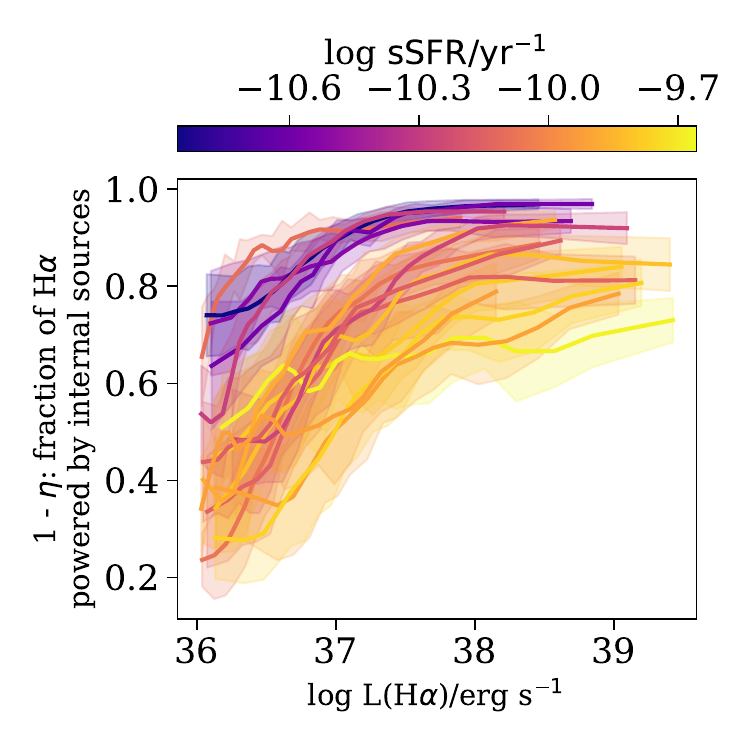}
    \caption{\textbf{The fraction of HII region \halpha\ luminosity powered by internal sources across the PHANGS-MUSE galaxies.} Each curve shows the median $1-\eta$ as a function of \halpha\ luminosity for one PHANGS-MUSE galaxy, color coded by the galaxy-integrated sSFR. \textbf{Summary:} At fixed \halpha\ luminosity, HII regions in galaxies with higher sSFR generally experience larger contributions from leaked ionizing radiation originating in neighboring HII regions. This provides a qualitative explanation for the higher low- and high-ionization line ratios observed in galaxies with higher sSFR (Figure~\ref{f:BPT_tracks_partial_correlation_all_gals_control_all}).}
    \label{fig:leakage_as_a_function_of_sSFR_all_gals}
\end{figure}

\subsection{Dependence of leaking radiation on global sSFR}\label{sec:discussion_leaking_vs_sSFR}

In Section~\ref{sec:global_HII_DIG_variations_line_ratios_with_sigma} we showed that, at fixed \halpha\ luminosity, the optical line ratios increase systematically with galaxy sSFR (Figure~\ref{f:BPT_tracks_partial_correlation_all_gals_control_all}). Within the framework of leaking and spectrally hardened ionizing radiation, this trend can be interpreted as indicating that galaxies with higher star formation activity produce larger amounts of leaking ionizing radiation that propagate through the ISM and contribute to the ionization of neighboring HII regions and the DIG. Because the leaked radiation field becomes progressively harder as it propagates through the ISM (Sections~\ref{sec:leaking_hardening} and \ref{sec:leaking_photoionization}), regions receiving a larger contribution from this non-local radiation are expected to exhibit elevated optical line ratios, naturally producing the observed correlation with sSFR.

Figure~\ref{fig:leakage_as_a_function_of_sSFR_all_gals} shows the fraction of \halpha\ luminosity powered by internal sources, $1-\eta$, across the PHANGS-MUSE sample. We exclude the two AGN hosts NGC~1365 and NGC~7496, since some of their HII regions may receive additional ionizing contributions from the AGN that are not included in our model. The curves are computed using our fiducial HII region radiative coupling model described in Section~\ref{sec:leaking_HII}, assuming a constant ionizing photon escape fraction of $f_{\mathrm{esc}}=0.5$, and are color coded by the global sSFR of each galaxy.

Overall, the figure shows that HII regions in higher-sSFR galaxies experience systematically larger contamination from leaking radiation, qualitatively supporting the idea that the observed dependence of optical line ratios on sSFR is driven by the amount of leaked ionizing radiation. Notably, this trend emerges even though we assume a constant escape fraction across all galaxies. In reality, stronger star formation activity may produce a more porous ISM through stellar feedback and supernova-driven clearing, potentially increasing the escape fraction of ionizing radiation and further strengthening the dependence on sSFR \citep{clarke02, oey07, kimm14, ma15}. Our results, however, show that an sSFR dependence in the amount of leaking radiation arises naturally even with a fixed escape fraction, simply from propagating radiation through the different spatial and luminosity distributions of HII regions across galaxies.

In our fiducial model, the stronger contamination in high-sSFR systems is driven primarily by differences in the HII region luminosity distribution. Across the PHANGS-MUSE galaxies, the effective separation between neighboring HII regions remains relatively similar, typically around 200--300 pc\footnote{This is likely due to the limited angular resolution of the MUSE observations. When observed at higher spatial resolution with narrow-band HST imaging, individual nebulae are often resolved into more closely separated structures \citep{barnes25}.}. In contrast, higher-sSFR galaxies exhibit broader \halpha\ luminosity distributions, such that the average HII region is surrounded by comparatively brighter neighbors, leading to a larger contribution from leaked ionizing radiation even when the escape fraction is held fixed.

Interestingly, Figure~\ref{fig:leakage_as_a_function_of_sSFR_all_gals} shows that in high-sSFR galaxies, even the brightest HII regions exhibit non-negligible contributions from externally ionizing radiation. We will explore this effect in more detail in a follow-up study.

\subsection{Metallicity and ionization parameter estimation in the presence of DIG mixing and leaking radiation}\label{sec:disc_metallicity_and_logU} 

The need to account for DIG contamination in nebular diagnostics has been recognized for kpc-scale IFU surveys, where each spaxel blends emission from HII regions and an extended DIG component. Empirically, several works proposed to correct HII region line ratios by assuming a characteristic DIG spectrum and subtracting its contribution \citep[e.g.,][]{sanders17, valeAsari19, espinosaPonce20}. A central limitation of these approaches is that, at kpc resolution, ``clean'' DIG measurements are typically restricted to the lowest-$\Sigma$(\halpha) regions far from star formation, while the DIG component most relevant for contaminating HII regions is the higher-$\Sigma$(\halpha) DIG in the vicinity of HII regions, whose line ratios vary systematically with $\Sigma$(\halpha) \citep[e.g.,][]{zhang17, belfiore22}. The move to $\sim$50 pc resolution with PHANGS-MUSE was expected to reduce these systematics by enabling a more direct separation of HII regions and DIG.

Results from PHANGS-MUSE, discussed by \citet{belfiore22}, \citet{congiu23}, and throughout this work, show that although the $\sim$50 pc spatial resolution substantially alleviates beam smearing, a non-negligible line-of-sight contribution from the DIG remains because the DIG has a larger scale height than the star-forming disk. This mixing systematically enhances the observed \oiii/\hbeta, \nii/\halpha, \sii/\halpha, and \oi/\halpha\ line ratios of HII regions (Section~\ref{sec:mixing}; \citealt{congiu23}), with the strongest effects expected for faint HII regions and for regions embedded in radiatively rich environments (high $\eta$; Section~\ref{sec:HII_env_dependence}), where the DIG surface brightness is elevated .

Because strong-line metallicity calibrations rely on the same line ratios affected by DIG mixing (e.g., \citealt{kreckel19, groves23}), the inferred metallicities may be systematically biased. While DIG contamination has often been discussed in the context of biases to radial metallicity gradients in kpc-scale studies, at PHANGS-MUSE resolution its impact may instead appear primarily as a dependence on local environment, producing correlated metallicity residuals within spiral arms or clustered star-forming complexes where the DIG contribution is enhanced. Approaches for mitigating this include inferring the DIG contribution from EW(\halpha) \citep{espinosaPonce20}, modeling the morphology of the DIG distribution \citep{Lugo-Aranda22, Lugo-Aranda24}, or estimating and subtracting the DIG contribution to each HII region using a surrounding DIG annulus \citet{congiu23}.

In addition to line-of-sight mixing with the DIG, we suggested that leaking radiation from neighboring HII regions may contribute a non-negligible fraction of the H-ionizing photons powering an HII region. For example, for an HII region with $L_{\mathrm{H\alpha}}=10^{37}\,\mathrm{erg\,s^{-1}}$, corresponding to the 67th percentile of the \halpha\ luminosity distribution in NGC~628, the inferred contamination fraction ranges from 20\% to 60\% depending on the adopted model assumptions (Section~\ref{sec:leaking_HII}). Observationally, we found that, at fixed \halpha\ luminosity, HII regions embedded in more radiatively rich environments exhibit enhanced \sii/\halpha\ and \oi/\halpha\ ratios (Section~\ref{sec:HII_env_dependence}) that cannot be explained by DIG mixing alone (Section~\ref{sec:mixing}), but could arise from contamination by leaking radiation that becomes spectrally hardened during propagation through the ISM. Such hardened radiation can preferentially enhance extended low-ionization and partially ionized zones at the edges of HII regions, increasing \sii/\halpha\ and \oi/\halpha. 

The radiative coupling scenario explored in this work remains tentative. As discussed in Section~\ref{sec:disc_HII_coupling_uncertainties}, a number of uncertainties remain, and additional observational and theoretical tests will be needed to determine whether this interpretation is correct. If this interpretation is confirmed, it may have implications for the interpretation of nebular emission-line diagnostics. In particular, metallicity estimates based on calibrations involving \sii\ (e.g., the PHANGS $S$-cal metallicities; \citealt{pilyugin16, groves23}), as well as ionization parameter estimates based on line ratios such as \siii/\sii, could be systematically biased in environments where external ionizing radiation makes a significant contribution to the observed emission, including faint HII regions and HII regions located near bright neighboring star-forming regions. 

As a simple illustrative exercise, we assume that the ``pure'' HII region sequence corresponds to the lower-left edge of the HII region distribution in the BPT diagrams of NGC~628, and then shift the line ratios along the observed sequences associated with increasing contamination. Under this toy model, the inferred $S$-cal metallicities can be underestimated by up to $\sim0.04$ dex, with the magnitude of the bias increasing systematically with the inferred contamination level. This behavior arises because the $S$-cal metallicity calibration is inversely correlated with \sii/\halpha.

These considerations have implications for recent results that explored azimuthal abundance variations, where the magnitude of the effect of radiative coupling we estimate is similar to the variations observed \citep{kreckel19, Sanchez-Menguiano20, Bresolin25}. It may also have implications for the inferred metallicity mixing scales, where homogeneity observed on sub-kpc scales is impacted by the kpc-scale DIG radiation \citep{kreckel20, metha21, williams22, li23}. 

\begin{figure*}
	\centering
\includegraphics[width=1\textwidth]{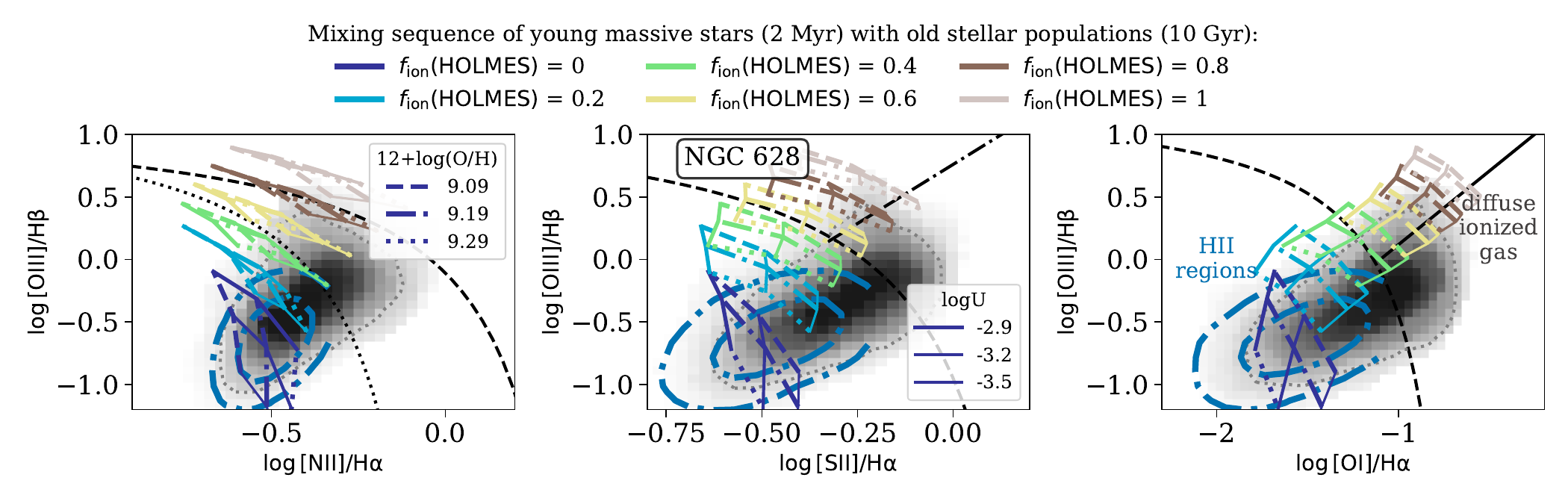}
\caption{\textbf{Photoionization model grid for a HOLMES mixing sequence.} \textbf{Data:} BPT diagrams of \nii/\halpha\ (left), \sii/\halpha\ (middle), and \oi/\halpha\ (right) versus \oiii/\hbeta\ for NGC~628. Blue contours show the 50th and 95th percentiles of the HII region distribution, while the DIG is shown as a logarithmically scaled grayscale density map, with the gray dotted contour marking its 95th percentile. \textbf{Models:} Colored curves show {\sc cloudy} predictions for varying levels of spectral mixing between a young stellar population (2 Myr FSPS SSP) and an old stellar population (10 Gyr FSPS SSP). The mixing is parameterized by $f_{\mathrm{ion}}(\mathrm{HOLMES})$, the fractional contribution of the old stellar population to the total ionizing photon budget (different colors). For each value of $f_{\mathrm{ion}}(\mathrm{HOLMES})$, the grids span a range of gas-phase metallicities and ionization parameters, as indicated in the legends.
Summary: Producing line ratios above the standard star-forming demarcation line requires a HOLMES contribution of $>$40\% to the ionizing photon budget, larger than the contribution expected across most DIG sightlines in PHANGS.} 
\label{f:HOLMES_mixing_cloudy_models}
\end{figure*}

\subsection{The role of HOLMES in powering DIG emission in the PHANGS-MUSE galaxies}\label{sec:disc_holmes}

The systematic analysis of the DIG across the PHANGS-MUSE galaxies by \citet{belfiore22} showed that the observed \halpha\ surface brightness distribution of the DIG can generally be reproduced under the assumption that it is powered by ionizing radiation leaking from HII regions. \citet{belfiore22}, however, needed to invoke a non-negligible contribution from HOLMES to explain the simultaneous increase in both high-ionization (\oiii/\hbeta) and low-ionization (\sii/\halpha\ and \oi/\halpha) line ratios across the DIG. This arises because, in models where leaking radiation experiences only geometric dilution (as considered by \citealt{belfiore22}), the resulting decrease in ionization parameter increases \sii/\halpha\ and \oi/\halpha, but decreases \oiii/\hbeta. Reproducing the observed behavior therefore required a progressively harder ionizing radiation field, which \citet{belfiore22} attributed to HOLMES. 

In the present work, however, we show that such hardening can arise within the leaking-radiation framework itself, as the escaping radiation field is spectrally reprocessed during propagation through the ISM (Section~\ref{sec:leaking_hardening}), thereby alleviating the need to invoke HOLMES. Photoionization models incorporating this propagation-induced hardening reproduce the simultaneous increase in both high- and low-ionization line ratios (Section~\ref{sec:leaking_photoionization}). In addition, Section~\ref{sec:global_HII_DIG_variations_line_ratios_with_sigma} shows that, at fixed \halpha\ surface brightness, the line ratios \oiii/\hbeta, \nii/\halpha, \sii/\halpha, and \oi/\halpha\ increase with sSFR, opposite to the trend expected if HOLMES were the primary driver of the observed line-ratio variations.

HOLMES are capable of producing hard ionizing spectra that generate LINER-like line ratios in BPT diagrams. However, the ionizing emissivity of 1--10 Gyr stellar populations (in units of $L_{\odot}/M_{\odot}$) is roughly six orders of magnitude lower than that of young massive stars (e.g., Figure 12 in \citealt{mcclymont24}), implying characteristic HOLMES-powered $\mathrm{EW}(\mathrm{H}\alpha)\sim1.1\,\mathrm{\AA}$ (e.g., \citealt{stasinska08, cidFernandes11, flores-fajardo11, byler19}). Across the PHANGS-MUSE galaxies, the observed DIG $\mathrm{EW}$(\halpha) typically spans $3$--$14\,\mathrm{\AA}$ (16th--84th percentiles), implying that HOLMES generally contribute $\sim$7--30\% to the ionizing photon budget across most sightlines. The main exception is the population of ``star-formation desert'' regions identified in several galaxies (e.g., NGC~1300, NGC~1433, NGC~1512, and NGC~3351), where $\mathrm{EW}$(\halpha) approaches the theoretical expectation for HOLMES-dominated ionization and where HOLMES may therefore play an important role \citep{belfiore22}.

In Figure~\ref{f:HOLMES_mixing_cloudy_models}, we show {\sc cloudy} predictions for a HOLMES mixing sequence, where the incident radiation field combines contributions from young massive stars (represented by a 2 Myr FSPS SSP) and old stellar populations (represented by a 10 Gyr FSPS SSP). The relative contribution of HOLMES is parameterized by $f_{\mathrm{ion}}(\mathrm{HOLMES})$, which describes the fraction of the total ionizing photon budget contributed by the old stellar population. The figure shows that a substantial HOLMES contribution can indeed push line ratios into the LINER regime, as suggested by numerous previous studies (e.g., \citealt{stasinska08, cidFernandes11, flores-fajardo11, zhang17, byler19, belfiore22}). However, moving line ratios beyond the standard demarcation curves that define the star-forming sequence in BPT space requires $f_{\mathrm{ion}}(\mathrm{HOLMES})\gtrsim0.4$, larger than the few-percent contribution expected across most of the DIG in PHANGS-MUSE. 

The threshold of $f_{\mathrm{ion}}(\mathrm{HOLMES})\gtrsim0.4$ inferred here is higher than the $f_{\mathrm{ion}}(\mathrm{HOLMES})\gtrsim0.1$ found by \citet{belfiore22}. This difference arises because our photoionization models are anchored to the observed HII-region line ratios in PHANGS under the assumption that HII regions and the DIG have approximately the same gas-phase metallicity, such that models with $f_{\mathrm{ion}}(\mathrm{HOLMES})=0$ reproduce the HII-region locus. In contrast, the \citet{belfiore22} models do not simultaneously reproduce both the HII-region and DIG line-ratio distributions under this assumption. Instead, matching the HII-region locus with $f_{\mathrm{ion}}(\mathrm{HOLMES})=0$ would require HII regions to have gas-phase metallicities that are $\sim0.4$ dex higher than those of the surrounding DIG.

We therefore suggest that leaking ionizing radiation is likely the dominant source of DIG optical line emission, driving the observed increase in line ratios with decreasing \halpha\ surface brightness. If HOLMES contribute, their contribution to the \halpha\ budget appears to be small (consistent with \citealt{belfiore22}), and they are not required to explain the higher-ionization emission, such as \oiii.

Finally, it is worth considering how these results connect to systems where the role of HOLMES is firmly established: elliptical, or ``retired,'' galaxies, which generally lack ongoing star formation and therefore should not contain a significant contribution from ionizing radiation leaking from young HII regions. These systems exhibit LINER-like line ratios and \halpha\ EWs approaching $\sim1\,\mathrm{\AA}$ \citep{stasinska08, yan12, singh13, cidfernandes10, belfiore17b}. In contrast, the PHANGS-MUSE galaxies are actively star-forming systems with sufficient ionizing photon production to plausibly power a large fraction of the DIG through leaking radiation. Together, these populations represent opposite ends of the galaxy star-formation activity sequence, suggesting that both mechanisms may contribute across the broader galaxy population. 

This naturally raises the question of how to distinguish observationally between ionization powered by HOLMES and ionization powered by leaking, spectrally hardened radiation in systems with intermediate levels of star formation. Standard optical line-ratio diagnostic diagrams alone may not be sufficient to separate these scenarios, as both mechanisms can produce LINER-like line ratios. However, the two processes predict different distributions of $\mathrm{EW}$(\halpha), making the 
WHAN diagram proposed by \citet{cidfernandes10} or similar diagrams proposed by others \citep[e.g.][]{Sanchez25}  potential discriminators. Another statistical discriminator may be their expected correlations with galaxy properties: the contribution from HOLMES should scale primarily with stellar mass surface density, whereas the contribution from leaking radiation should be more closely tied to recent star formation activity. The early type galaxies MUSE sample presented by \citet{williams25} or the MAUVE-MUSE sample \citep{catinella25, brown26} may be particularly valuable to explore in tandem with the PHANGS galaxies, given their comparable spatial resolution while extending to generally less star-forming systems.

Another potentially informative diagnostic is the spatially resolved relation between PAH band ratios and optical emission-line ratios. In \citet{baron24} and \citet{baron25}, we identified a remarkably tight, nearly universal relation between the PAH 11.3 \mic/7.7 \mic band ratio and optical line ratios across the PHANGS-MUSE galaxy sample, which we interpreted as being driven by spatial variations in the spectral shape of the radiation field. Because PAHs predominantly reside in neutral regions shielded from direct ionizing radiation \citep{chastenet19, chastenet23a, egorov23, sutter24, egorov25}, the 11.3 \mic/7.7 \mic ratio is primarily sensitive to changes in the non-ionizing far-ultraviolet (FUV) radiation field. In this context, different physical scenarios may imprint distinct FUV spectral shapes: an increasing contribution from HOLMES, or alternatively progressive spectral filtering as radiation propagates through the ISM, could produce measurable differences in the FUV continuum shape, and hence in PAH band ratios, even at fixed optical line ratios. This may therefore provide an additional lever arm for distinguishing between these scenarios. We defer a detailed exploration of this diagnostic, and its dependence on intrinsic PAH properties, to future work.

\subsection{Caveats of the HII-region radiative coupling scenario}\label{sec:disc_HII_coupling_uncertainties}

The radiative coupling scenario proposed here should be regarded as a tentative interpretation. While the combination of observations and photoionization modeling provides evidence that ionizing radiation leaking from neighboring HII regions may contribute significantly to the ionization of a given HII region, both the observational analysis and the theoretical framework are subject to uncertainties that could alter the inferred significance of this effect. Below we discuss the main caveats and outline several tests that could help establish whether radiative coupling is indeed a significant effect.

The dominant observational uncertainty arises from the limited spatial resolution of the PHANGS-MUSE data. At the $\sim$50 pc resolution of the survey, a substantial fraction of the HII-region population is unresolved \citep{chandar25, barnes25}, making the identification of HII regions and the placement of their boundaries uncertain. Different segmentation schemes directly affect the measured H$\alpha$ luminosities \citep{groves23, congiu23}, propagating into estimates of the DIG fraction \citep{belfiore22} and local escape fractions derived from star cluster--HII region pairs \citep{scheuermann26}. They are also expected to alter the measured optical line ratios, particularly \sii/\halpha\ and \oi/\halpha, since these transitions are strongest near HII-region boundaries. Because the observational evidence for radiative coupling relies heavily on systematic variations in these line ratios, the significance of the inferred environmental trends may depend on how HII regions are segmented.

Related to this is the possibility that the faintest objects identified as HII regions may, in some cases, not be classical HII regions at all but local enhancements within the diffuse ionized medium. Using hydrodynamic simulations of a Milky Way-like galaxy, \citet{mcclymont24} showed that applying standard observational segmentation techniques can identify bright DIG clumps as HII regions. Such misclassifications are particularly important because they occur in the luminosity regime where our analysis predicts the largest contribution from external ionizing radiation, and where segmentation and line-ratio uncertainties are also expected to be greatest.

A natural observational test of the radiative coupling scenario is therefore to repeat this analysis in nearby galaxies where individual HII regions are spatially resolved. Ideally, these observations should combine high spatial resolution ($\sim$10 pc), integral-field spectroscopy covering multiple diagnostic emission lines, sufficient depth to map the DIG, and a field of view large enough to characterize HII regions' surrounding radiative environment. Nearby, relatively face-on systems such as NGC~300 \citep{micheva22}, M33 \citep{feltre26}, and NGC~253 \citep{congiu25, mcClain26}, all observed with MUSE, together with Local Volume Mapper observations of the LMC and SMC \citep{drory24}, provide particularly promising datasets for such tests.

An additional observational test concerns the angular dependence of the contamination. If HII regions are ionized in part by radiation escaping from neighboring regions, then enhanced low-ionization line ratios might preferentially occur on the sides of HII regions facing their brightest neighbors, rather than being distributed isotropically. Detecting such directional signatures would provide an independent test of the radiative coupling hypothesis.

The theoretical framework presented here also relies on several assumptions that remain uncertain. Most importantly, the predicted degree of radiative coupling depends sensitively on the adopted escape fraction of ionizing photons from individual HII regions. While recent PHANGS analyses suggest typical escape fractions of $\sim$80\% \citep{scheuermann26}, previous studies that account for stochastic sampling of the stellar initial mass function infer substantially lower values of $\sim$10\% \citep{teh23}. Adopting such low escape fractions reduces the predicted radiative coupling to negligible levels for most HII regions. Resolving this discrepancy will require improved constraints on the stellar populations powering HII regions and the fraction of their ionizing radiation that escapes into the surrounding ISM.

The results are also sensitive to the adopted HII-region sizes, which determine the solid angle subtended by each region and therefore the amount of external ionizing radiation intercepted. Throughout this work, we adopt sizes measured from the PHANGS-MUSE nebular catalog \citep{groves23}. However, HST observations indicate that many of these nebulae remain unresolved by MUSE, leading to systematically overestimated sizes \citep{barnes25}. Smaller intrinsic sizes reduce the effective cross section of HII regions and thus the amount of ambient ionizing radiation they intercept. To estimate the impact of this effect, we converted the PHANGS-MUSE sizes into approximate HST-equivalent sizes using the empirical relation presented by \citet[][their Figure D.2]{barnes25}. This reduces the externally powered H$\alpha$ fraction ($\eta$; e.g., Figures~\ref{f:HII_coupling_fiducial_model} and \ref{fig:leakage_as_a_function_of_sSFR_all_gals}) by approximately 0.1--0.15 across the full luminosity range. Although our fiducial NGC~628 model still predicts non-negligible radiative coupling under this assumption, the exercise illustrates the importance of accurately resolving HII-region sizes.

Finally, our propagation model assumes a single effective absorption coefficient, $k_0$, throughout the multiphase ISM. The residuals shown in Section~\ref{sec:leaking_HII_and_DIG} suggest that this approximation is overly simplistic and that the clumpy structure of the ISM should be treated more explicitly. In particular, HII regions that remain embedded within dense neutral or molecular material are likely to be more effectively shielded from external ionizing radiation, reducing the degree of radiative coupling.

Taken together, these considerations suggest that the radiative coupling scenario remains tentative. Resolving individual HII regions, obtaining improved constraints on escape fractions and nebular sizes, and incorporating a more realistic treatment of the multiphase ISM will be essential for determining whether radiative coupling represents a significant contribution to the ionization of HII regions.

\section{Summary}\label{sec:summary}

By mapping nearby galaxies at high angular resolution, PHANGS provides a unique laboratory for connecting star formation and feedback on cloud scales to the structure of entire galactic disks \citep{leroy21a, emsellem22, lee22, lee23}. In this paper, we investigate ionizing radiation from massive stars as a mechanism linking compact HII regions to the DIG, and to neighboring HII regions. Using 50--100 pc-scale observations from the PHANGS-MUSE survey for 19 nearby star-forming galaxies, we jointly analyze HII regions and the faint, extended DIG to investigate how ionizing-photon leakage radiatively couples these components and shapes their observed emission-line properties. Specifically, we find:\vspace{-0.07in}
\begin{enumerate}

\item \textbf{Continuous emission-line sequences}: HII regions and the DIG populate common, continuous tracks across all three standard BPT diagrams (\niihalpha, \siihalpha, and \oihalpha\ versus \oiiihbeta). Bright HII regions lie on the classical star-forming locus, progressively fainter regions extend smoothly to higher line ratios, and the DIG continues these sequences into the composite and LINER regimes, with no clear separation in line-ratio space (Section~\ref{sec:global_HII_DIG_variations_line_ratios}; Figure~\ref{f:BPT_tracks}).\vspace{-0.07in}

\item \textbf{A common dependence on \halpha\ surface brightness}: All four optical line ratios (\oiii/\hbeta, \nii/\halpha, \sii/\halpha, and \oi/\halpha) increase systematically with decreasing $\Sigma$(\halpha) across both HII regions and the DIG, defining nearly continuous relations over $\sim$2--3 orders of magnitude. The DIG extends trends already present within the HII region population (Section~\ref{sec:global_HII_DIG_variations_line_ratios_with_sigma}; Figure~\ref{f:line_ratios_vs_LHa}).\vspace{-0.07in}

\item \textbf{Correlation with global galaxy properties}: At fixed \halpha\ surface brightness, optical line ratios show strong positive correlations with sSFR and the expected dependence on gas-phase metallicity. These trends extend across the combined HII region--DIG sequence. The simultaneous increase of both high-ionization (\oiii/\hbeta) and low-ionization (\nii/\halpha, \sii/\halpha, \oi/\halpha) line ratios with sSFR suggests a progressively harder ionizing radiation field with increasing star-formation activity. This trend is opposite to that expected if HOLMES were driving the observed line-ratio variations (Section~\ref{sec:global_HII_DIG_variations_line_ratios_with_sigma}; Figure~\ref{f:BPT_tracks_partial_correlation_all_gals_control_all}).\vspace{-0.07in}

\item \textbf{Environmental dependence within HII regions}: At fixed \halpha\ luminosity, HII regions located near other luminous regions and concentrations of recent star formation show systematically enhanced low-ionization line ratios (\oihalpha\ and \siihalpha), indicating a dependence of their emission properties on local environment (Section~\ref{sec:HII_env_dependence}; Figure~\ref{f:NGC0628_ratios_vs_eta_LHa}).\vspace{-0.05in}

\end{enumerate}

\noindent We explored two scenarios to explain these trends. \textbf{(I)~Line-of-sight mixing (Section~\ref{sec:mixing}; Figure~\ref{f:mixing_vs_observed_line_ratios}):} We test whether projection effects can explain the continuity between HII regions and the DIG, given the DIG’s larger scale height ($\sim$0.5–1.5 kpc) compared to the star-forming disk ($\sim$50–300 pc). Using a DIG-subtracted HII region catalog, we recover roughly similar trends across the BPT diagrams. Furthermore, a simple spectral mixing model combining ``pure'' HII region spectra with local DIG emission reproduces the trends in \oiiihbeta\ and \niihalpha, but fails for the low-ionization ratios \oihalpha\ and \siihalpha. While DIG contamination contributes to the observed sequences, it cannot fully account for the behavior of the low-ionization lines.\vspace{0.07in}

\noindent \textbf{(II) Radiative coupling through leaking radiation (Section~\ref{sec:leaking}):} We examine whether ionizing photons escaping from HII regions can radiatively couple neighboring HII regions and the DIG. We extend the radiation propagation model of \citet{belfiore22} to include spectral hardening during ISM propagation. The model uses the locations and luminosities of HII regions to compute ionizing radiation propagation across the ISM, accounting for geometric dilution, attenuation in a diffuse medium with an effective absorption coefficient, and wavelength-dependent filtering that progressively hardens the spectrum. We find:\vspace{-0.07in}
\begin{enumerate}
\item \textbf{HII region radiative coupling}: If ionizing photons escape from HII regions, they can contribute to the ionization of neighboring regions. We solve for the intrinsic ionizing photon production rates of individual HII regions while accounting for contamination from leaked radiation produced by surrounding sources. Across all model assumptions, we find a strong luminosity dependence: bright HII regions experience minimal external contamination, while progressively fainter regions are increasingly affected by leaked ionizing radiation. For NGC~628, HII regions with $L_{\mathrm{H}\alpha}\sim10^{37}\,\mathrm{erg\,s^{-1}}$ (approximately the 67th percentile of the luminosity distribution) have 20–60\% of their observed H$\alpha$ emission powered by external ionizing radiation, depending on model assumptions (Section~\ref{sec:leaking_HII}; Figure~\ref{f:HII_coupling_fiducial_model}).\vspace{-0.07in}

\item \textbf{Coupled HII region--DIG solutions}: Joint modeling of HII regions and DIG yields global solutions with escape fractions $f_{\mathrm{esc}} = 0.4$–0.6 and effective absorption coefficients $k_0 = 1$–$1.77\,\mathrm{kpc^{-1}}$ (mean free path $\sim$0.5–1 kpc; Figure~\ref{f:k0_comparison_all_gals}). These models improve fits to DIG surface brightness in 13 of 19 galaxies relative to models without coupling (Figure~\ref{f:chi2_dof_improvement}; Section~\ref{sec:leaking_HII_and_DIG}).\vspace{-0.07in}

\item \textbf{Spectral hardening and photoionization predictions}: We use {\sc cloudy} to model wavelength-dependent attenuation through the diffuse medium. As ionizing radiation propagates through the ISM, preferential absorption of lower-energy photons progressively hardens the transmitted spectrum. Because fainter HII regions are more strongly affected by leaked radiation, this produces systematically harder effective ionizing fields in progressively fainter HII regions (Section~\ref{sec:leaking_hardening}; Figure~\ref{f:hardening_as_a_function_of_L_halpha_k0_1p5_DIGcorr}). {\sc cloudy} calculations show that spectra filtered over propagation distances of 0–2 kpc produce diagonal sequencess in BPT space, simultaneously increasing \oiiihbeta, \niihalpha, \siihalpha, and \oihalpha\ while decreasing \halpha\ equivalent width, in qualitative agreement with observations (Section~\ref{sec:leaking_photoionization}; Figure~\ref{f:filtered_radiation_cloudy_models}).\vspace{-0.07in}

\item \textbf{Connection to sSFR trends}: Applying the radiative coupling model across the full PHANGS-MUSE sample, we find that higher-sSFR galaxies exhibit systematically larger levels of radiative contamination of HII regions by neighboring sources. Because this contaminating radiation is, on average, harder due to propagation effects, this results in a global trend in which higher-sSFR systems experience harder effective ionizing fields, providing a qualitative explanation for the observed dependence of line ratios on sSFR (Section~\ref{sec:discussion_leaking_vs_sSFR}; Figure~\ref{fig:leakage_as_a_function_of_sSFR_all_gals}).\vspace{-0.07in}

\end{enumerate}

\noindent Overall, these results add to the growing evidence that in star-forming galaxies, massive-star feedback operates across a wide range of scales, from individual HII regions to entire galactic disks. They support a picture in which leaking, progressively hardening ionizing radiation is the dominant source of the optical line emission observed in the DIG of star-forming galaxies, while HOLMES, which are known to dominate the ionization budget of elliptical and other ``retired'' galaxies, contribute only a minor fraction. They also raise the possibility that HII regions are not isolated nebulae, but are instead radiatively coupled through leaked ionizing radiation. At present, this interpretation should be regarded as tentative (see Section~\ref{sec:disc_HII_coupling_uncertainties}), and further observational and theoretical work is needed to test its validity.

If this interpretation is confirmed, then the properties and evolution of HII regions may depend not only on their local conditions but also on the surrounding radiation field established by neighboring star-forming regions. Quantities commonly used to characterize stellar feedback, including ionizing-photon escape fractions and the timescales over which natal gas is dispersed, could therefore exhibit an environmental dependence. Likewise, because optical emission-line diagnostics are sensitive to non-local radiation, estimates of ionization parameters, gas-phase metallicities, and metal-mixing length scales may require explicit accounting of the ambient radiation field. More fundamentally, the surrounding radiation field may be used to redefine the boundaries of HII regions, with their outer edge marking the transition where locally produced ionizing radiation no longer dominates over the ambient radiation field. 

\acknowledgments{The data used in this work is based on observations collected at the European Southern Observatory under ESO programmes 094.C-0623 (PI: Kreckel), 095.C-0473,  098.C-0484 (PI: Blanc), 1100.B-0651 (PHANGS-MUSE; PI: Schinnerer), as well as 094.B-0321 (MAGNUM; PI: Marconi), 099.B-0242, 0100.B-0116, 098.B-0551 (MAD; PI: Carollo) and 097.B-0640 (TIMER; PI: Gadotti).

KK gratefully acknowledges funding from the European Research Council’s starting grant ERC StG-101077573 (“ISM-METALS”).
ES acknowledges support from the European Research Council (ERC) under the Horizon Europe framework programme. This work is supported by an ERC Advanced Grant (Acronym: eCMZs, Number 101201230). Funded by the European Union. Views and opinions expressed are however those of the author(s) only and do not necessarily reflect those of the European Union or the European Research Council Executive Agency. Neither the European Union nor the granting authority can be held responsible for them. 
ES and JN acknowledge funding from the European Research Council (ERC) under the European Union's Horizon 2020 research and innovation program (grant agreement No. 694343). RSK acknowledges financial support from the ERC via Synergy Grant ``ECOGAL'' (project ID 855130) and from the German Excellence Strategy via the Heidelberg Cluster ``STRUCTURES'' (EXC 2181 - 390900948). In addition RSK is grateful for funding from the German Ministry for Economy and Energy (BMWE) in project ``MAINN'' (funding ID 50OO2206), and from DFG and ANR for project ``STARCLUSTERS'' (funding ID KL 1358/22-1). J.E.M.-D. gratefully acknowledges support from the Secretar\'ia de Ciencia, Humanidades, Tecnolog\'ia e Innovaci\'on (SECIHTI) project CBF-2025-I-2048, ``Resolviendo la F\'isica Interna de las Galaxias: De las Escalas Locales a la Estructura Global con el SDSS-V Local Volume Mapper'', and from the UNAM/DGAPA/PAPIIT IA103326 project ``DESIRED (DEep Spectra of ionised Regions Database): de las emisiones m\'as sutiles a la f\'isica fundamental del universo''. TGW gratefully acknowledges support from the UK ALMA Regional Centre (ARC) Node, which is supported by the Science and Technology Facilities Council grant number ST/Y004108/1. OE acknowledges funding from the Deutsche Forschungsgemeinschaft (DFG, German Research Foundation) -- project-ID 541068876.}

\software{Astropy \citep{astropy13, astropy18, astropy22},
		  IPython \citep{perez07},
          scikit-learn \citep{pedregosa11}, 
          SciPy \citep{scipy01},
		  matplotlib \citep{hunter07},
		  reproject \citep{reproject_2020}}\vspace{2cm}

\clearpage

\bibliography{ref.bib}

\appendix

\onecolumngrid

\section{Line ratio sequences from HII regions to the DIG for the full PHANGS sample}\label{app:bpt_trends_all_galaxies}
\setcounter{figure}{0}

\begin{figure*}[h!]
	\centering
\includegraphics[width=0.9\textwidth]{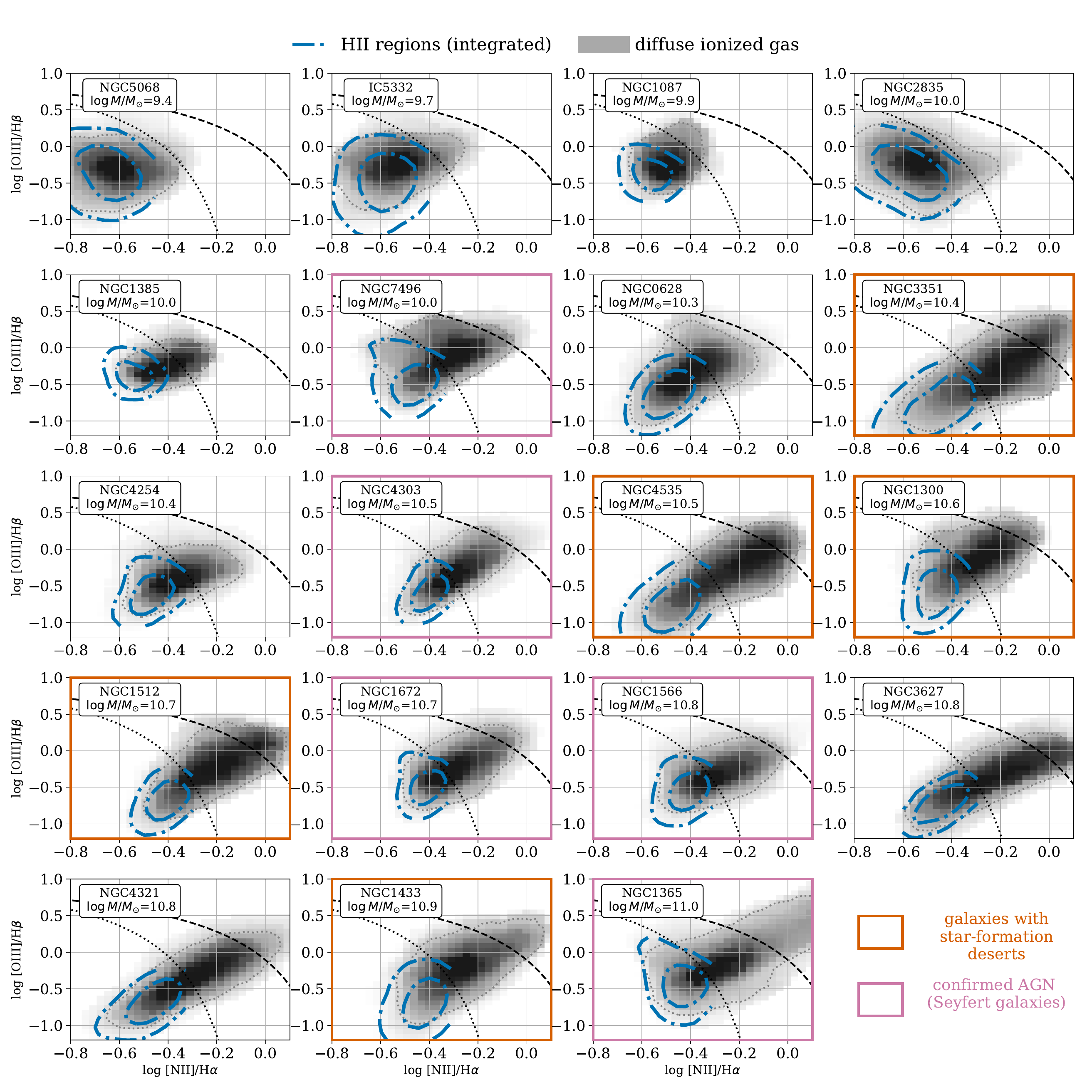}
\caption{\textbf{Distribution of \nii/\halpha versus \oiii/\hbeta across HII regions and the DIG in the PHANGS-MUSE sample.} Each panel shows one galaxy, ordered by stellar mass from low to high. Pink and orange borders indicate galaxies hosting an AGN or extended star-formation desert, respectively. Blue contours show the 50th and 95th percentiles of the HII region distribution, while the DIG is shown as a logarithmic grayscale density with the 95th percentile indicated by a gray dotted contour. Black demarcation curves from the literature are shown as dashed \citep{kewley01}, dotted \citep{kauff03a}, dash-dotted \citep{kewley06}, and solid \citep{cidfernandes10}. }
\label{f:BPT_tracks_all_gals_NII}
\end{figure*}

\begin{figure*}
	\centering
\includegraphics[width=0.9\textwidth]{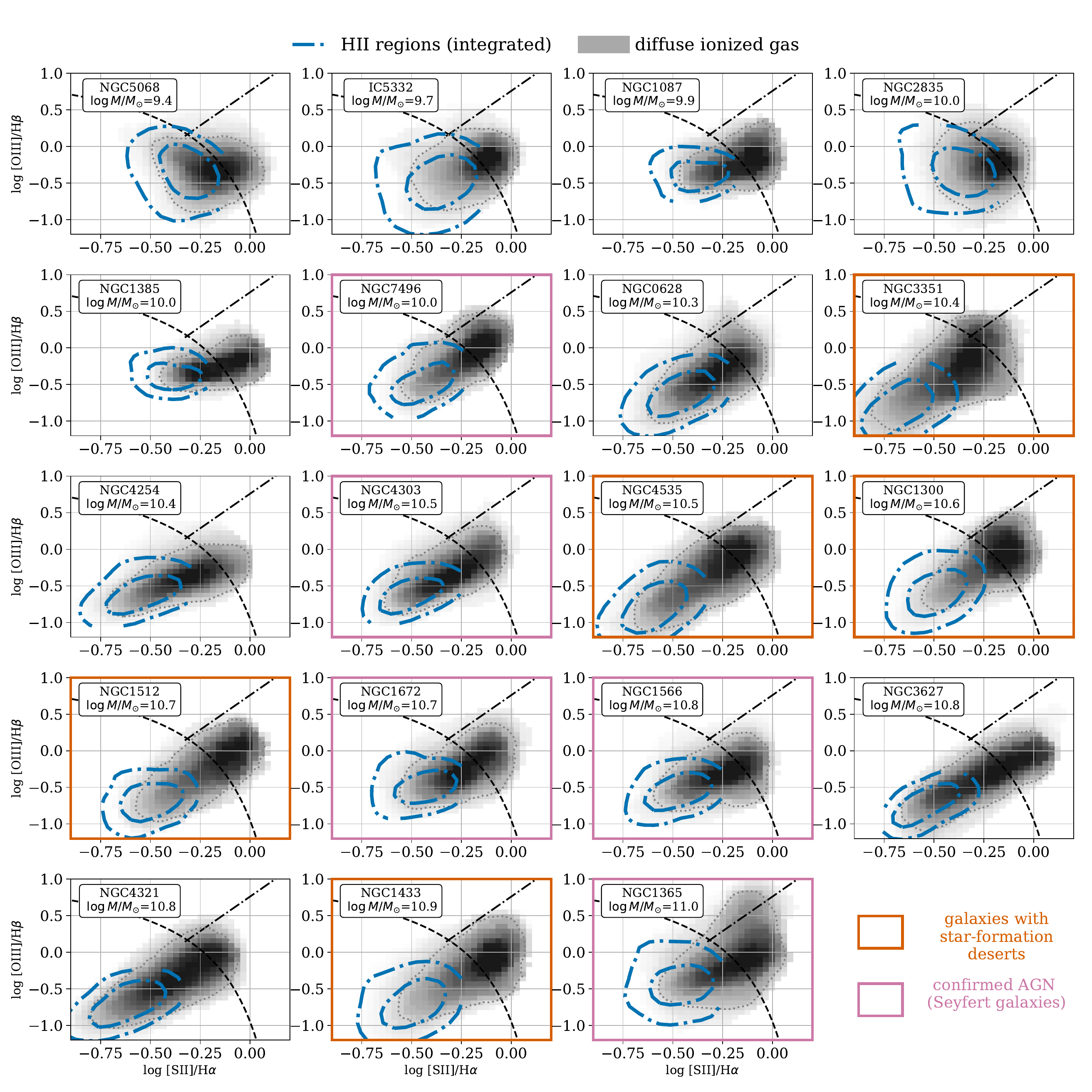}
\caption{\textbf{Distribution of \sii/\halpha versus \oiii/\hbeta across HII regions and the DIG in the PHANGS-MUSE sample.} Each panel shows one galaxy, ordered by stellar mass from low to high. Pink and orange borders indicate galaxies hosting an AGN or extended star-formation desert, respectively. Blue contours show the 50th and 95th percentiles of the HII region distribution, while the DIG is shown as a logarithmic grayscale density with the 95th percentile indicated by a gray dotted contour. Black demarcation curves from the literature are shown as dashed \citep{kewley01}, dotted \citep{kauff03a}, dash-dotted \citep{kewley06}, and solid \citep{cidfernandes10}.}
\label{f:BPT_tracks_all_gals_SII}
\end{figure*}

\begin{figure*}
	\centering
\includegraphics[width=0.9\textwidth]{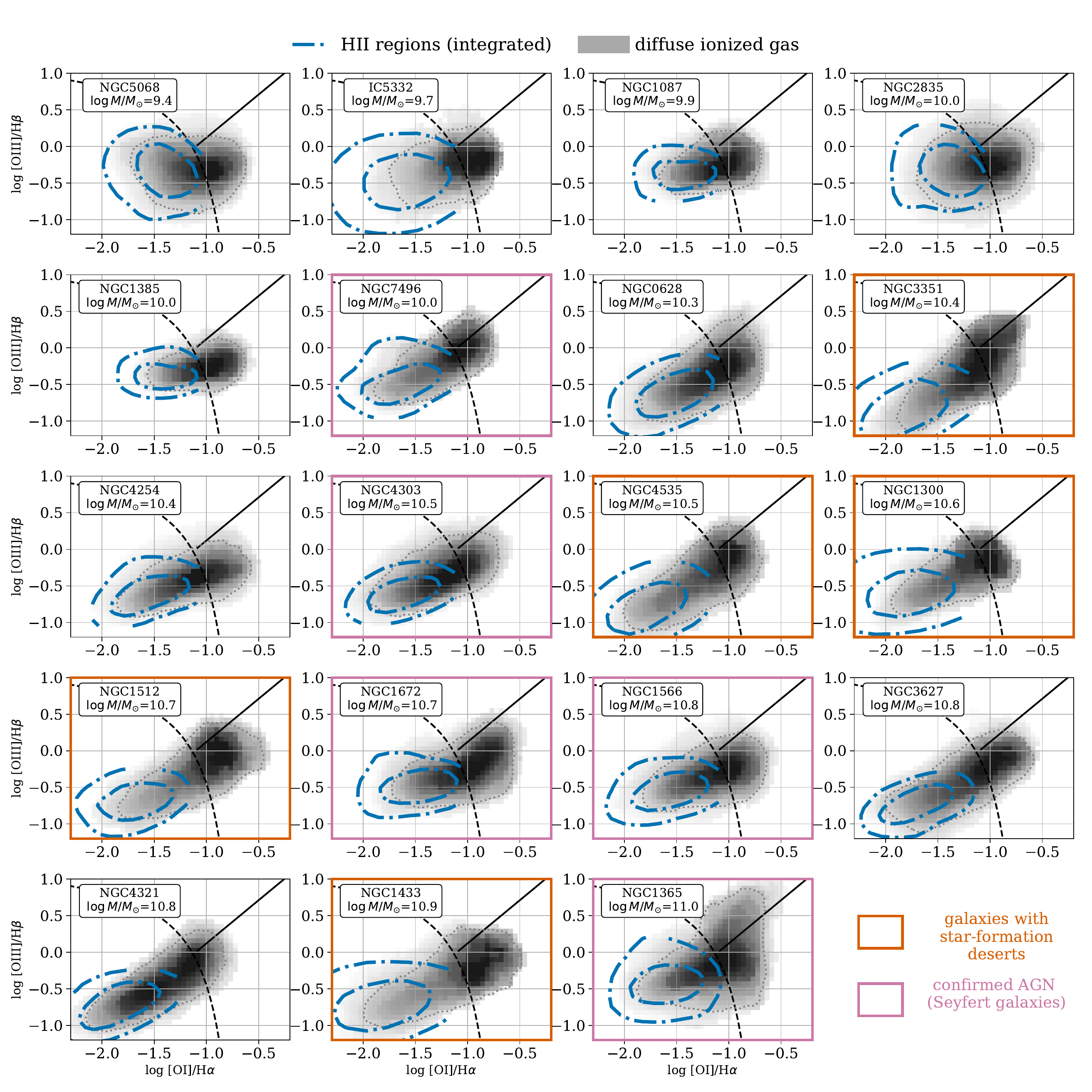}
\caption{\textbf{Distribution of \oi/\halpha versus \oiii/\hbeta across HII regions and the DIG in the PHANGS-MUSE sample.} Each panel shows one galaxy, ordered by stellar mass from low to high. Pink and orange borders indicate galaxies hosting an AGN or extended star-formation desert, respectively. Blue contours show the 50th and 95th percentiles of the HII region distribution, while the DIG is shown as a logarithmic grayscale density with the 95th percentile indicated by a gray dotted contour. Black demarcation curves from the literature are shown as dashed \citep{kewley01}, dotted \citep{kauff03a}, dash-dotted \citep{kewley06}, and solid \citep{cidfernandes10}. }
\label{f:BPT_tracks_all_gals_OI}
\end{figure*}

Figures~\ref{f:BPT_tracks_all_gals_NII}--\ref{f:BPT_tracks_all_gals_OI} show the distributions of line ratios for HII regions and the DIG on the \niihalpha, \siihalpha, and \oihalpha\ BPT diagrams, respectively, across the full PHANGS-MUSE sample.

\clearpage

\section{Photoionization modeling}\label{app:cloudy_grids}
\setcounter{figure}{0}

\subsection{General model assumptions}

The photoionization models in this work are computed with {\sc cloudy} v23.01 \citep{chatzikos23}, following and extending the approach of \citet{belfiore22}. We use the {\sc python} package {\sc cloudy-fsps}\footnote{\url{https://github.com/nell-byler/cloudyfsps}} to generate stellar SEDs in a {\sc cloudy} format, construct and run the {\sc cloudy} input files, and analyze their outputs. 

For the incident radiation field, we use the \texttt{FSPS} stellar population synthesis (SPS) library \citep{conroy09, conroy10} to generate the input ionizing spectra. Specifically, we adopt single stellar population models with ages of 2 Myr and 10 Gyr, representing young massive stars and evolved stellar populations, respectively. The models assume a \citet{chabrier03} IMF with an upper mass cutoff of $100\,\mathrm{M_{\odot}}$. We consider stellar metallicities in the range $-0.2 \leq \log (Z/Z_{\odot}) \leq 0.6$, adopting $Z_{\odot}=0.014$ and $12+\log(\mathrm{O/H})_{\odot}=8.69$ \citep{asplund09}.

Our baseline gas-phase abundance model follows \citet{byler18}. We adopt the solar abundance pattern of \citet{asplund09} with dust depletion factors from \citet{dopita03}. Carbon and nitrogen abundances are scaled with metallicity using the fitting functions of \citet{byler18}, while all other heavy elements scale linearly with metallicity. We include ISM grains in our modeling, which are scaled with the gas-phase metallicity.

The stellar and gas-phase metallicities are tied as follows. For the young stellar population models, we tie the SPS metallicity to the gas-phase oxygen abundance, motivated by both the strong metallicity dependence of the ionizing spectrum \citep{byler17} and the expectation that young stars and their surrounding HII regions have similar metallicities. For the old stellar population models, we instead fix the stellar metallicity to solar. This choice is motivated by the lack of a direct physical connection between the metallicity of evolved stars and that of the DIG, as well as the weak dependence of post-AGB ionizing spectra on metallicity \citep{byler19}. Our results are unchanged if the stellar and gas-phase metallicities are tied in the old-population models as well.

We assume a closed geometry using the \texttt{sphere} command. We adopt two stopping criteria for the {\sc cloudy} computation: electron temperature below 100 K and ionized fraction below 0.01. These criteria ensure that we reach the ionization front, where all the ionizing radiation has been absorbed, modeling radiation-bounded clouds. The predicted line ratios do not change if instead we use 1000 K stopping temperature.

\subsection{Chemical abundances for the baseline model}

Our goal is to identify photoionization models that simultaneously reproduce the line ratios of PHANGS HII regions across the three BPT diagrams: \niihalpha, \siihalpha, and \oihalpha\ versus \oiiihbeta.

The gas-phase abundance prescription of \citet{byler18} reproduces the mean N/O and C/O trends observed in HII regions, but both direct abundance measurements and observations reveal significant scatter around these relations (e.g., \citealt{berg15, brazzini24}). Moreover, this baseline model produces grids that are offset by $\sim0.5$ dex toward higher \siihalpha\ compared to PHANGS HII regions (see Figure~15 of \citealt{belfiore22}), suggesting that lower S/H abundances may be required. Because the location of model grids in BPT space depends sensitively on O/H, N/H, and S/H, we explore a range of nitrogen and sulfur abundances to identify the combinations that best reproduce the observed distribution of PHANGS HII regions.

We tie O/H to the SPS metallicity and compute the corresponding baseline N/O and S/O ratios using the \citet{byler18} prescription. We then vary N/O and S/O independently in steps of $0.2\,\mathrm{dex}$ over the range $[-0.6,0.2]\,\mathrm{dex}$ relative to the baseline values. The C/O ratio is tied to N/O, although fixing C/O to the baseline prescription produces comparable predictions in the BPT diagrams over the parameter space relevant to our observations.

The model grid spans $12+\log(\mathrm{O/H}) \in [8.4,9.2]$, $\log(\mathrm{N/O}) \in [-1.9,0.3]$, and $\log(\mathrm{S/O}) \in [-2.3,-1.7]$, with ionization parameters in the range $\log U \in [-3.5,-2.5]$, typical of HII regions (e.g., \citealt{kewley19}). To identify the models that best reproduce the PHANGS HII regions, we compare each model to the median line ratios of the brightest 10\% of HII regions in NGC~628, which occupy the lower-left portion of the BPT diagrams. Because the PHANGS galaxies exhibit broadly similar BPT sequences, we adopt NGC~628 as a representative reference galaxy.

Models consistent within $\pm3\sigma$ of the median line ratios have $12+\log(\mathrm{O/H}) = 9.12 \pm 0.1$, $\log(\mathrm{N/O}) = -1.2 \pm 0.1$, $\log(\mathrm{S/O}) = -2.0 \pm 0.1$, and $\log U = -3.2 \pm 0.15$. The derived N/O abundance is broadly consistent with auroral-line measurements in NGC~628 \citep{berg15, brazzini24}, while the preferred S/O lies toward the lower end of the observed distribution \citep{brazzini24}. The inferred oxygen abundance is higher than auroral-line estimates by $\sim0.5$ dex, reflecting the well-known offset between photoionization-model and direct-method metallicities (e.g., \citealt{kewley02, stasinska05, blanc15, dopita16}).

These parameters should not be interpreted as precise abundance measurements. Our photoionization models adopt a simplified single-cloud, constant-density, radiation-bounded geometry. In practice, reproducing the full set of observed strong-line ratios is likely to require a more complex description of the ionized gas. For example, comparable agreement can be achieved using combinations of multiple cloud components (e.g., {\sc homerun}; \citealt{marconi24}) or mixtures of radiation-bounded and matter-bounded clouds (Ramambason et al., in prep.). The abundance values derived here should therefore be viewed as defining a reference model rather than a unique physical solution.

Once this reference model is established, we fix the chemical abundances and focus on how changes in the ionizing radiation field alter the predicted line ratios. Because the subsequent analysis relies primarily on variations in the radiation field rather than on the absolute abundance scale, the qualitative conclusions regarding spectral hardening and radiation propagation should be largely insensitive to the specific choice of baseline model.

\clearpage

\section{Dependence of $\Sigma$(\halpha)--line-ratio relations on global galaxy properties}\label{app:sigma_global_appendix}
\setcounter{figure}{0}

This appendix provides supporting material for Section~\ref{sec:global_HII_DIG_variations_line_ratios_with_sigma}. Section~\ref{app:sigma_galaxy_by_galaxy} describes the galaxy-to-galaxy variations of line ratios with $\Sigma$(\halpha), and Section~\ref{app:line_ratio_vs_sigma_global_properties} describes their dependence on global galaxy properties. 

\subsection{Galaxy-by-galaxy line-ratio trends with $\Sigma$(\halpha)}\label{app:sigma_galaxy_by_galaxy}

\begin{figure*}[h!]
	\centering
\includegraphics[width=1\textwidth]{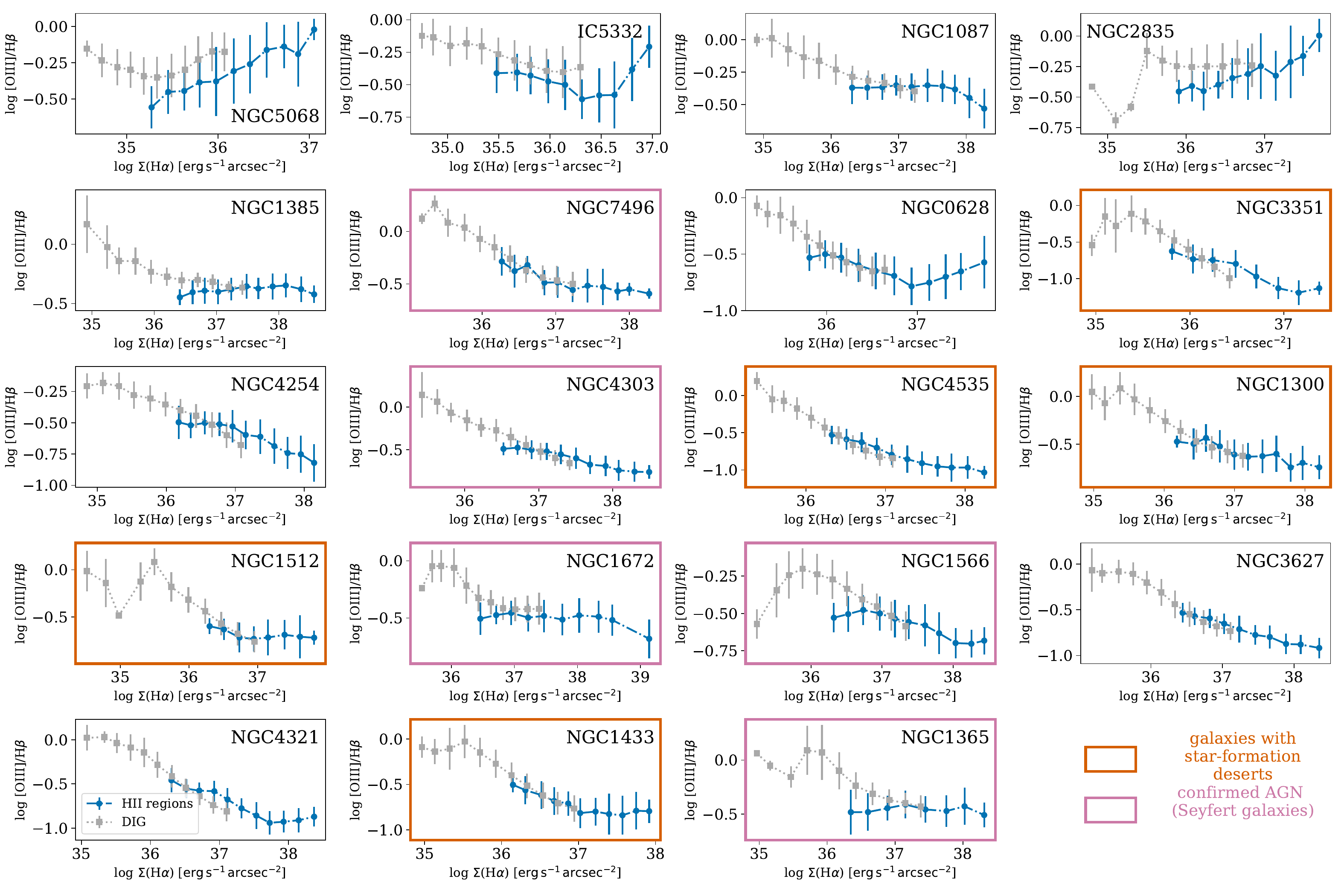}
\caption{\textbf{Variation of \oiii/\hbeta\ with \halpha\ surface density for HII regions and the DIG across the PHANGS-MUSE sample.} Each panel shows one galaxy, ordered by increasing stellar mass. Pink and orange borders indicate galaxies hosting an AGN or extended star-formation desert, respectively. Line ratios are shown as the median and median absolute deviation in bins of \halpha\ surface density, $\Sigma$(\halpha). Blue circles and dash-dotted lines show HII region stacks, while gray squares and dotted lines show DIG stacks.}
\label{f:OIIIHbeta_vs_L_Ha_all_galaxies}
\end{figure*}

\begin{figure*}[h!]
	\centering
\includegraphics[width=1\textwidth]{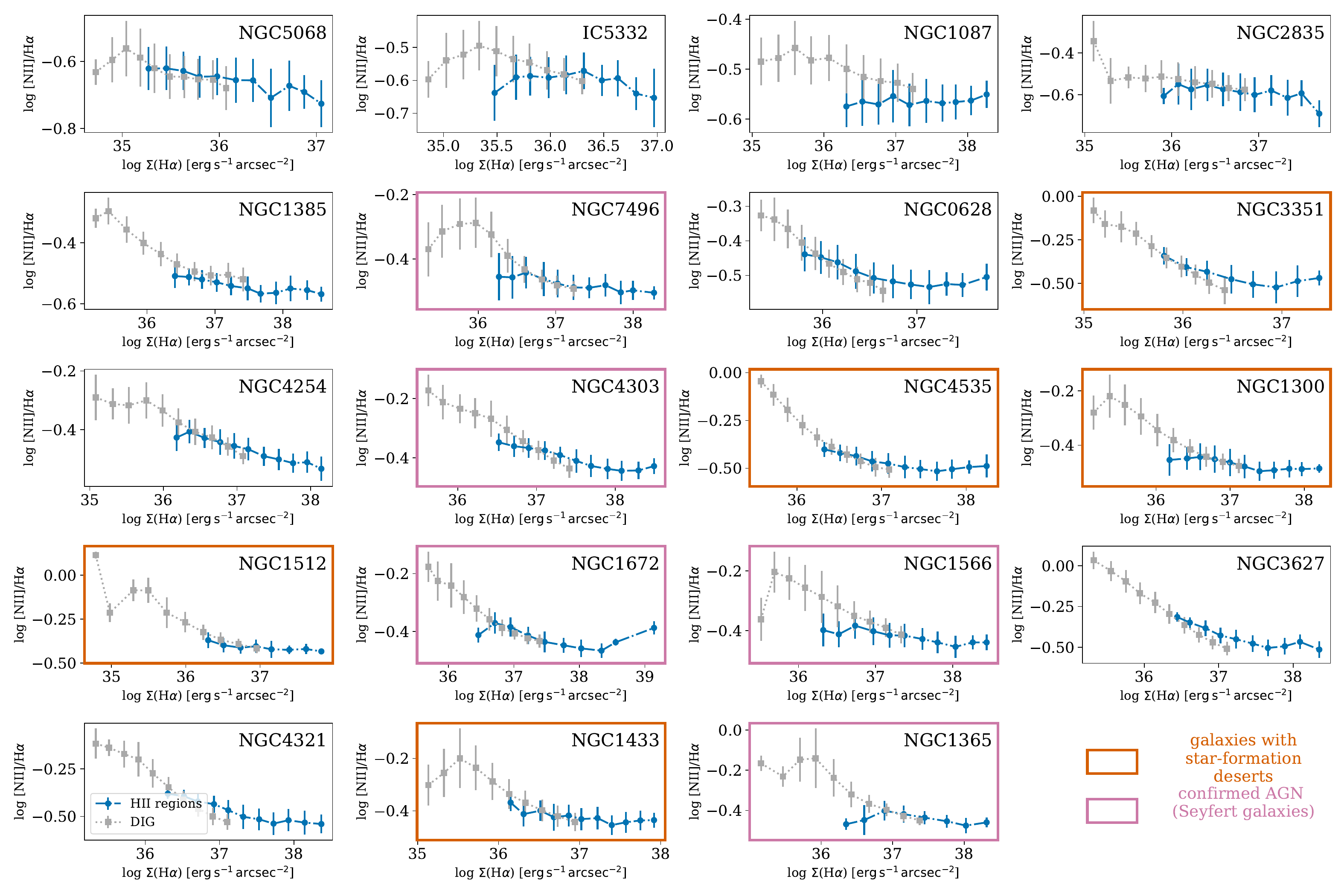}
\caption{\textbf{Variation of \nii/\halpha\ with \halpha\ surface density for HII regions and the DIG across the PHANGS-MUSE sample.} Each panel shows one galaxy, ordered by increasing stellar mass. Pink and orange borders indicate galaxies hosting an AGN or extended star-formation desert, respectively. Line ratios are shown as the median and median absolute deviation in bins of \halpha\ surface density, $\Sigma$(\halpha). Blue circles and dash-dotted lines show HII region stacks, while gray squares and dotted lines show DIG stacks.}
\label{f:NIIHalpha_vs_L_Ha_all_galaxies}
\end{figure*}

\begin{figure*}
	\centering
\includegraphics[width=1\textwidth]{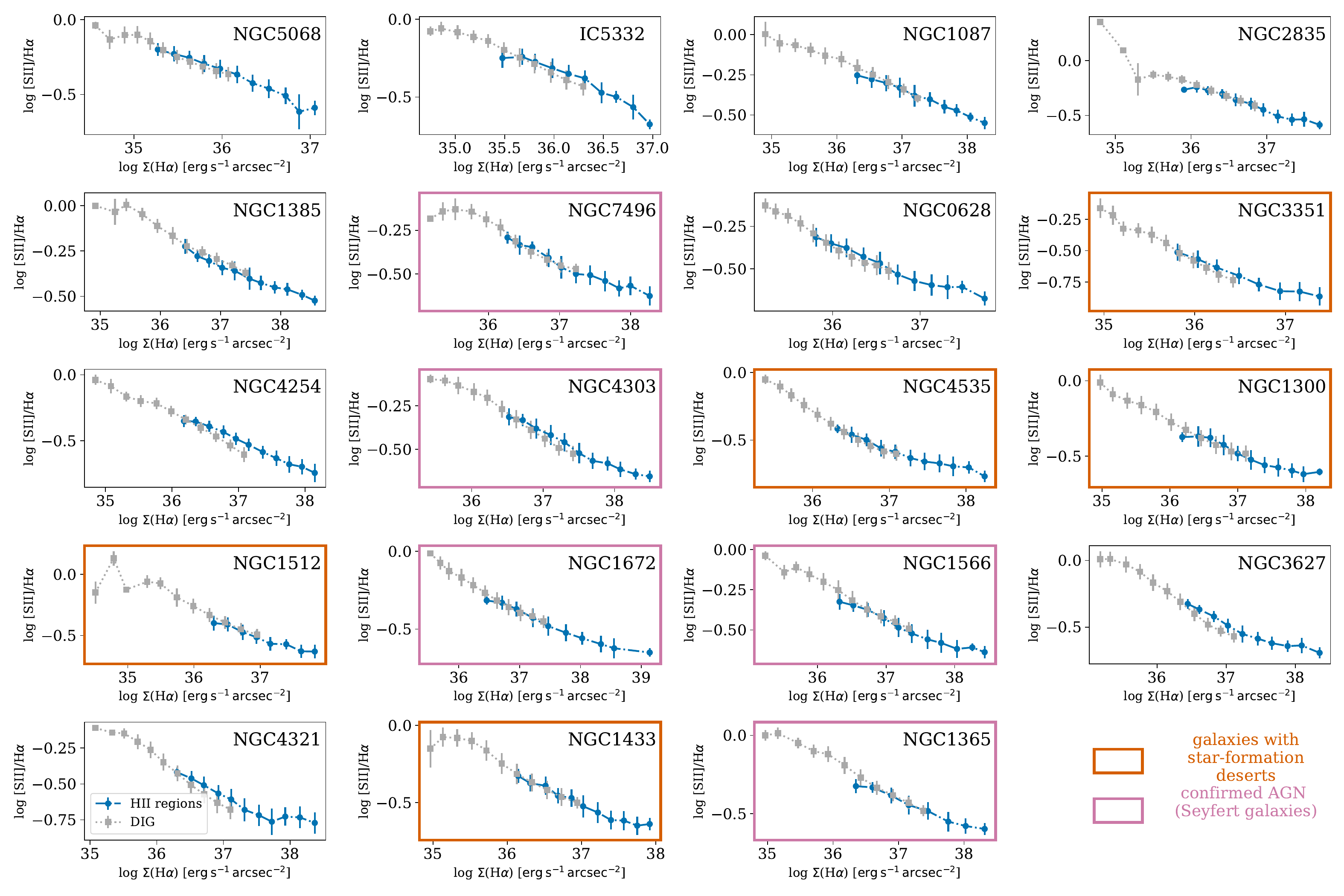}
\caption{\textbf{Variation of \sii/\halpha\ with \halpha\ surface density for HII regions and the DIG across the PHANGS-MUSE sample.} Each panel shows one galaxy, ordered by increasing stellar mass. Pink and orange borders indicate galaxies hosting an AGN or extended star-formation desert, respectively. Line ratios are shown as the median and median absolute deviation in bins of \halpha\ surface density, $\Sigma$(\halpha). Blue circles and dash-dotted lines show HII region stacks, while gray squares and dotted lines show DIG stacks.}
\label{f:SIIHalpha_vs_L_Ha_all_galaxies}
\end{figure*}

\begin{figure*}
	\centering
\includegraphics[width=1\textwidth]{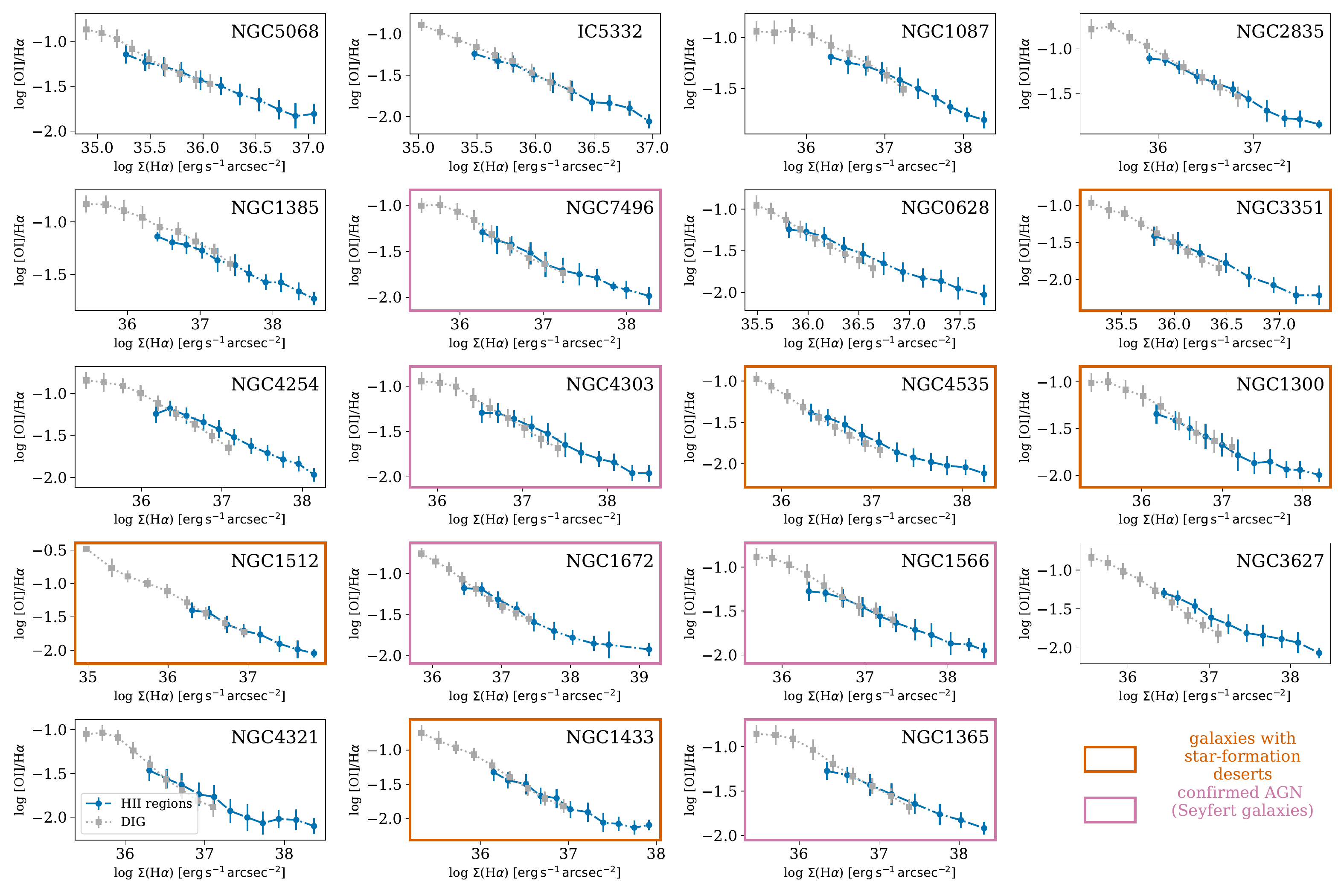}
\caption{\textbf{Variation of \oi/\halpha\ with \halpha\ surface density for HII regions and the DIG across the PHANGS-MUSE sample.} Each panel shows one galaxy, ordered by increasing stellar mass. Pink and orange borders indicate galaxies hosting an AGN or extended star-formation desert, respectively. Line ratios are shown as the median and median absolute deviation in bins of \halpha\ surface density, $\Sigma$(\halpha). Blue circles and dash-dotted lines show HII region stacks, while gray squares and dotted lines show DIG stacks.}
\label{f:OIHalpha_vs_L_Ha_all_galaxies}
\end{figure*}

Figures~\ref{f:OIIIHbeta_vs_L_Ha_all_galaxies}--\ref{f:OIHalpha_vs_L_Ha_all_galaxies} show the median and median absolute deviation of line ratios (\oiii/\hbeta, \nii/\halpha, \sii/\halpha, and \oi/\halpha) in bins of $\Sigma$(\halpha) for HII regions and the DIG across the full PHANGS-MUSE sample. 

Similarly to \citet{belfiore22}, all four line ratios in the DIG decrease with increasing $\Sigma$(\halpha), with the exception of \oiii/\hbeta for the lowest mass galaxies, which we further discuss in Section~\ref{app:line_ratio_vs_sigma_global_properties} below. HII regions, however, exhibit a more nuanced behavior that depends on the ionization level traced by each ratio. The higher-ionization diagnostics, \oiii/\hbeta\ (IP$_{\mathrm{O^{++}}}=35.12$ eV) and \nii/\halpha\ (IP$_{\mathrm{N^{+}}}=14.53$ eV), show comparable or slightly shallower trends over the range in $\Sigma$(\halpha) where both HII regions and DIG are present, followed by a flattening at higher surface brightness levels. In contrast, the lower-ionization ratios, \sii/\halpha\ (IP$_{\mathrm{S^{+}}}=10.36$ eV) and \oi/\halpha\ (with \oi arising from neutral oxygen in extended low-ionization zones), closely track the DIG trends across most of the $\Sigma$(\halpha) range, with flattening apparent only in some galaxies and only at the highest surface brightnesses. 

\clearpage

\subsection{Dependence on global galaxy properties}\label{app:line_ratio_vs_sigma_global_properties}

Throughout this section, we occasionally refer to luminosity ranges in which HII regions and DIG ``co-exist''. This terminology does not imply spatial overlap, but instead denotes luminosity regimes in which both components are present in our sample. The DIG surrounding HII regions typically exhibits \halpha equivalent widths (EW) approximately an order of magnitude lower than those of the associated HII regions (e.g., \citealt{belfiore22}). Therefore, this ``co-existence'' regime corresponds to the brightest DIG in our sample (surrounding the brightest HII regions) and the faintest HII regions (embedded in the faintest DIG).

\begin{figure*}
	\centering
\includegraphics[width=1\textwidth]{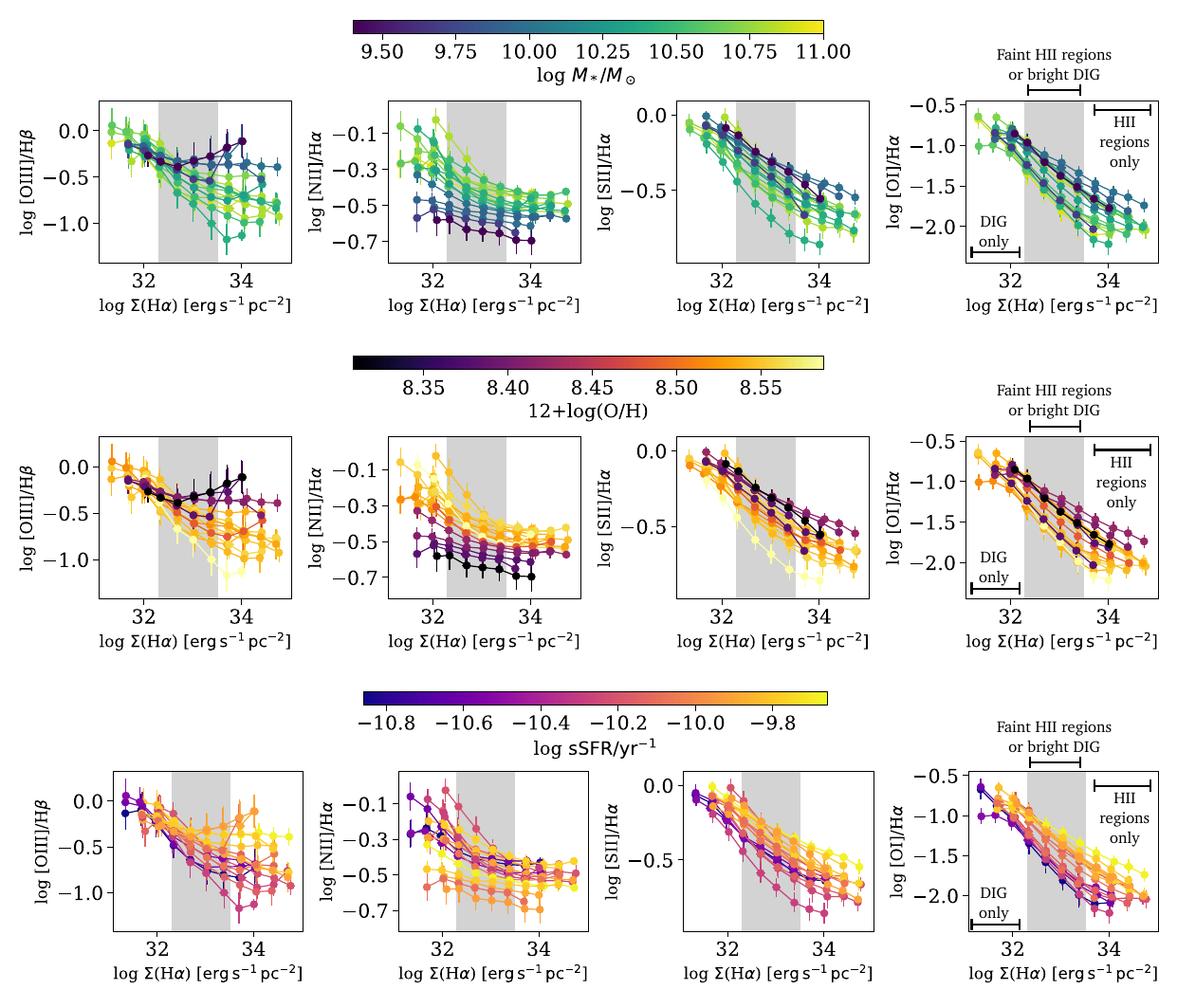}
\caption{\textbf{Line-ratio variation with $\Sigma$(\halpha) and global galaxy properties across the PHANGS-MUSE galaxies.} Each row shows, from left to right, the median \oiii/\hbeta, \nii/\halpha, \sii/\halpha, and \oi/\halpha\ as a function of $\Sigma$(\halpha), combining HII regions and DIG emission for individual galaxies. The markers are color-coded by the stellar mass (top), median gas-phase metallicity (middle), and global sSFR (bottom). The gray band indicates the range in $\Sigma$(\halpha) where both
bright DIG and faint HII regions are present. To the left of the dark gray band, the emission is dominated by faint DIG sightlines. Beyond the gray band, the emission is dominated by the brightest HII regions.}
\label{f:line_ratios_vs_Lha_vs_galaxy_properties}
\end{figure*}

Figure~\ref{f:line_ratios_vs_Lha_vs_galaxy_properties} shows stacked line ratios as a function of $\Sigma$(\halpha) for individual galaxies, color-coded by stellar mass, median gas-phase metallicity from the nebular catalog, and the global sSFR. 

We first examine the dependence of the line-ratio trends on stellar mass. Previous studies have used the variation of line ratios with $\Sigma$(\halpha) to constrain the ionizing sources of the DIG. In particular, \citet{zhang17} and \citet{belfiore22} found that the relation between \oiii/\hbeta\ and $\Sigma$(\halpha) depends on stellar mass: in lower-mass galaxies, \oiii/\hbeta\ remains approximately constant with $\Sigma$(\halpha), while in higher-mass systems it increases toward lower $\Sigma$(\halpha), i.e., deeper into the DIG. Because such an increase cannot be reproduced with ionizing photons from young massive stars that escape HII regions and are geometrically diluted (lower $\log U$), \citet{belfiore22} interpreted the trend as evidence for an increasing contribution from a harder radiation field, attributed to HOLMES. This interpretation is qualitatively consistent with the expectation that more massive galaxies host larger populations of evolved stars.

In the top-left panel of Figure~\ref{f:line_ratios_vs_Lha_vs_galaxy_properties}, we recover a similar trend when considering individual galaxies. However, when all galaxies are examined together, a different picture emerges. In the faint DIG regime (low $\Sigma$(\halpha)), \oiii/\hbeta\ is roughly uniform across the sample and shows little dependence on stellar mass. Differences between galaxies become apparent only at higher $\Sigma$(\halpha), where DIG and HII regions coexist. This suggests that the apparent flattening of the \oiii/\hbeta--$\Sigma$(\halpha) relation in lower-mass galaxies is not driven by changes in the faintest DIG, but rather by variations in brighter regions closer to HII regions. Further insight comes from the dependence of line ratios on sSFR, discussed in Section~\ref{sec:global_HII_DIG_variations_line_ratios_with_sigma}. We find that line ratios increase with sSFR, arguing against HOLMES as the primary driver of the harder ionizing radiation, since their relative contribution is expected to increase as star formation activity declines.

In addition to stellar mass, we color-code the trends by gas-phase metallicity and sSFR. The trends with stellar mass and metallicity closely mirror each other, reflecting the strong correlation between these quantities across our sample. Given the well-established sensitivity of optical line ratios to gas-phase (and stellar) metallicity (see e.g., Figure~\ref{f:BPT_tracks}, or review by \citealt{kewley19}), it is more instructive to interpret the trends primarily in terms of metallicity.

Focusing first on \oiii/\hbeta\ as a function of $\Sigma$(\halpha), we find a strong increase in \oiii/\hbeta\ with decreasing metallicity at fixed $\Sigma$(\halpha), both in bright HII regions and in the intermediate-$\Sigma$(\halpha) regime where faint HII regions and bright DIG coexist. This behavior is consistent with expectations from photoionization models: lower metallicity leads to harder ionizing spectra and reduced metal-line cooling (e.g., \citealt{byler17}), both of which enhance collisionally excited emission such as \oiii. At the lowest $\Sigma$(\halpha), corresponding to faint DIG sightlines, the relation flattens. This can be understood in the regime of very low ionization parameter, where most oxygen resides in O$^{+}$ rather than O$^{++}$, so that the \oiii\ emission is set primarily by the limited abundance of O$^{++}$ and responds only weakly to changes in metallicity.

For \nii/\halpha, we observe a clear positive correlation with metallicity, such that \nii/\halpha\ increases with increasing metallicity. This trend reflects the super-linear increase of N/H with metallicity (e.g., \citealt{kewley08, dopita13}). The dependence on metallicity becomes even stronger in the faint DIG regime, likely again due to the low ionization parameter, where most nitrogen is in the form of N$^{+}$ and abundance variations dominate the line ratios. At higher $\Sigma$(\halpha), the relation is comparatively shallower, consistent with an additional dependence on the ionization parameter in regions closer to HII regions. 

The trends of \sii/\halpha\ and \oi/\halpha\ with metallicity at fixed $\Sigma$(\halpha) are broadly consistent with photoionization model expectations, following the same physical interpretation discussed for \oiii/\hbeta.

\section{Solid Angles}\label{app:solid_angles}
\setcounter{figure}{0}

In our radiative coupling model for HII regions, each region $i$ receives ionizing radiation from neighboring regions $j$ according to a coupling matrix $K_{ij}$ that depends on geometric factors and attenuation. Two key geometric quantities are computed: the solid angle $\Delta\Omega_{ij}$ subtended by region $j$ as seen from region $i$ (\texttt{simple}), and an occlusion-corrected version that accounts for intervening regions blocking the line of sight (\texttt{occ}).

\subsection{Simple solid angle (\texttt{simple})}
For the unoccluded case, we model each HII region as a circular aperture of angular radius $\theta_j$ at projected separation $r_{ij}$ on the sky. We first convert celestial coordinates (RA, Dec) to a tangent-plane projection centered on the galaxy, yielding physical separations in kpc using the galaxy distance $D$.
For a circular source of physical radius $R_j$ at distance $r_{ij}$ (where $r_{ij} > R_j$), the solid angle subtended is
\begin{equation}
\Delta\Omega_{ij} = 2\pi\left(1 - \sqrt{1 - \left(\frac{R_j}{r_{ij}}\right)^2}\right).
\end{equation}
When two regions spatially overlap ($r_{ij} \leq R_j$), we set $\Delta\Omega_{ij} = 2\pi$ by default, representing nearly complete sky coverage from the target's perspective. Diagonal elements are set to zero ($\Delta\Omega_{ii} = 0$), and pairs beyond an optional cutoff radius $r_{\rm cutoff}$ are also zeroed.

\subsection{Occlusion-corrected solid angle (\texttt{occ})}

To account for foreground regions that partially or fully obscure the line of sight between target $i$ and source $j$, we employ a Monte Carlo occlusion model as follows. For each $(i, j)$ pair, we find all regions $k$ that lie within the cone defined by the separation $r_{ij}$ from source $j$. We retain only those blockers $k$ closer to $j$ than the target ($r_{jk} < r_{ij}$). We then project all blockers onto an angular disk of radius $\alpha_i = \arcsin(R_i/r_{ij})$ centered on the target region's direction as seen from $j$. Each blocker $k$ subtends an angular radius $\beta_k = \arcsin(R_k/r_{jk})$ at an angular offset $\gamma_k$ from the target direction. We then calculate the union area via Monte Carlo. We randomly sample points uniformly within the target's angular disk. For each sample point, we check whether it falls inside any blocker's angular disk. The fraction of covered sample points estimates the occluded fraction $f_{\rm occ}$. The occlusion-corrected solid angle is then
\begin{equation}
\Delta\Omega_{ij}({\rm \texttt{occ}}) = \Delta\Omega_{ij}({\rm \texttt{simple}}) \times (1 - f_{\rm occ}).
\end{equation}

\section{HII region line ratio variations with radiative environment}\label{app:HII_line_ratios_versus_eta}
\setcounter{figure}{0}

\begin{figure*}
	\centering
\includegraphics[width=1\textwidth]{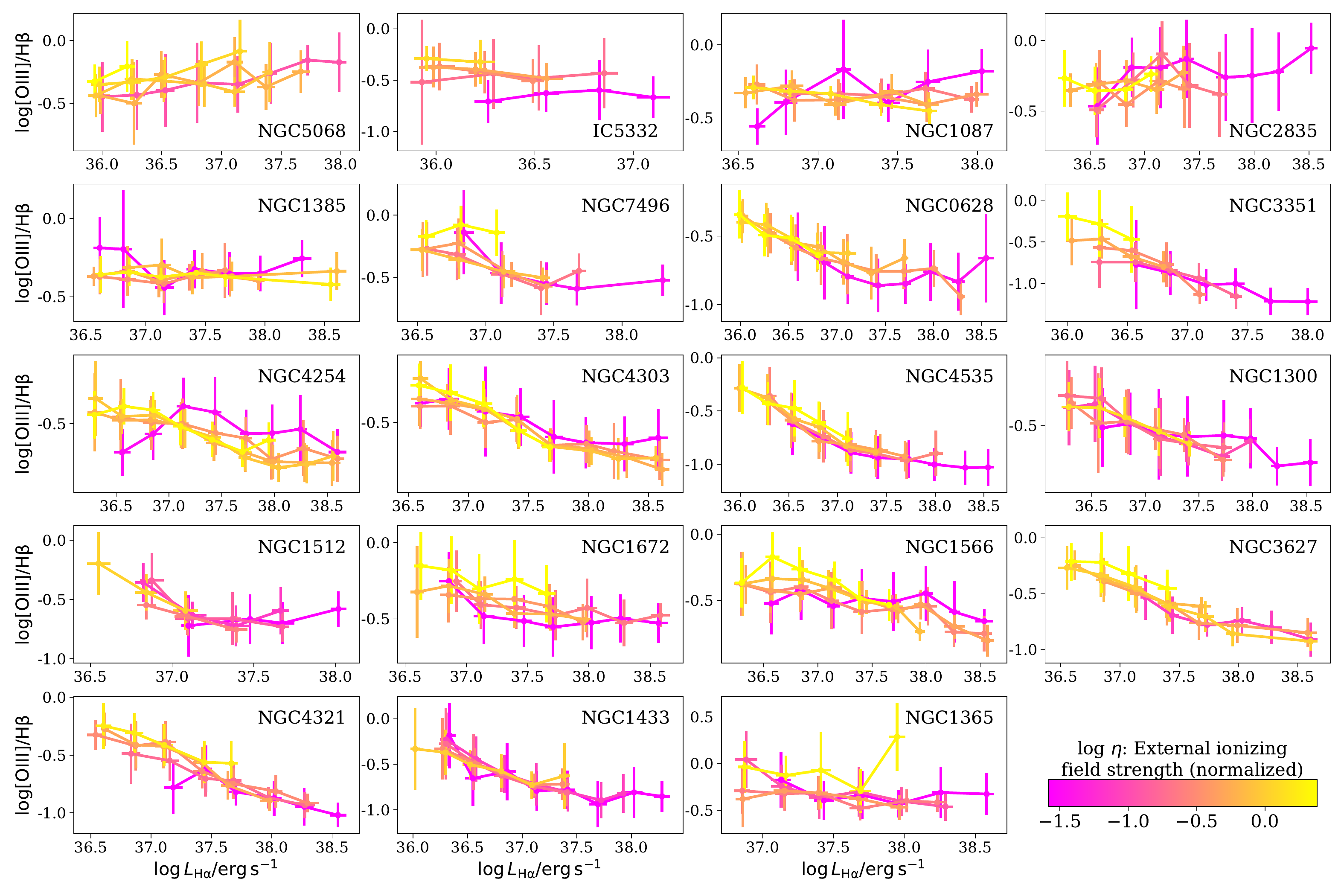}
\caption{\textbf{Dependence of \oiii/\halpha\ on $L_{\mathrm{H}\alpha}$ and radiative environment across PHANGS-MUSE HII regions.}
Each panel shows stacked \oiii/\halpha\ in bins of $L_{\mathrm{H}\alpha}$ and $\eta$, where $\eta$ quantifies the strength of the external ionizing radiation field from nearby HII regions, for individual PHANGS-MUSE galaxies. The galaxies are indicated in each panel, and are ordered according to stellar mass.}
\label{f:OIII_Hbeta_vs_Lhalpha_eta_all_gal}
\end{figure*}

\begin{figure*}
	\centering
\includegraphics[width=1\textwidth]{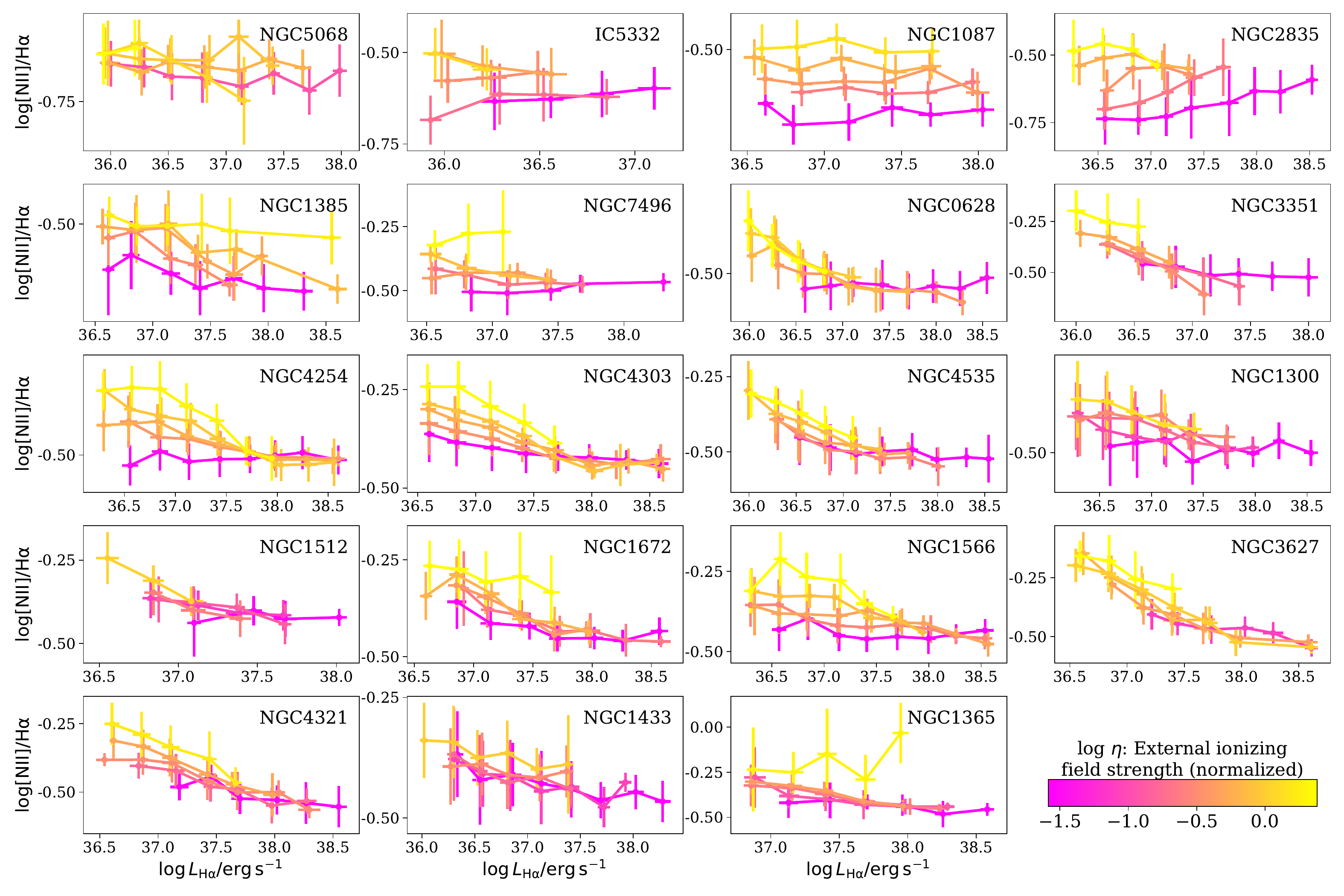}
\caption{\textbf{Dependence of \nii/\halpha\ on $L_{\mathrm{H}\alpha}$ and radiative environment across PHANGS-MUSE HII regions.}
Each panel shows stacked \nii/\halpha\ in bins of $L_{\mathrm{H}\alpha}$ and $\eta$, where $\eta$ quantifies the strength of the external ionizing radiation field from nearby HII regions, for individual PHANGS-MUSE galaxies. The galaxies are indicated in each panel, and are ordered according to stellar mass.}
\label{f:NII_Halpha_vs_Lhalpha_eta_all_gal}
\end{figure*}

\begin{figure*}
	\centering
\includegraphics[width=1\textwidth]{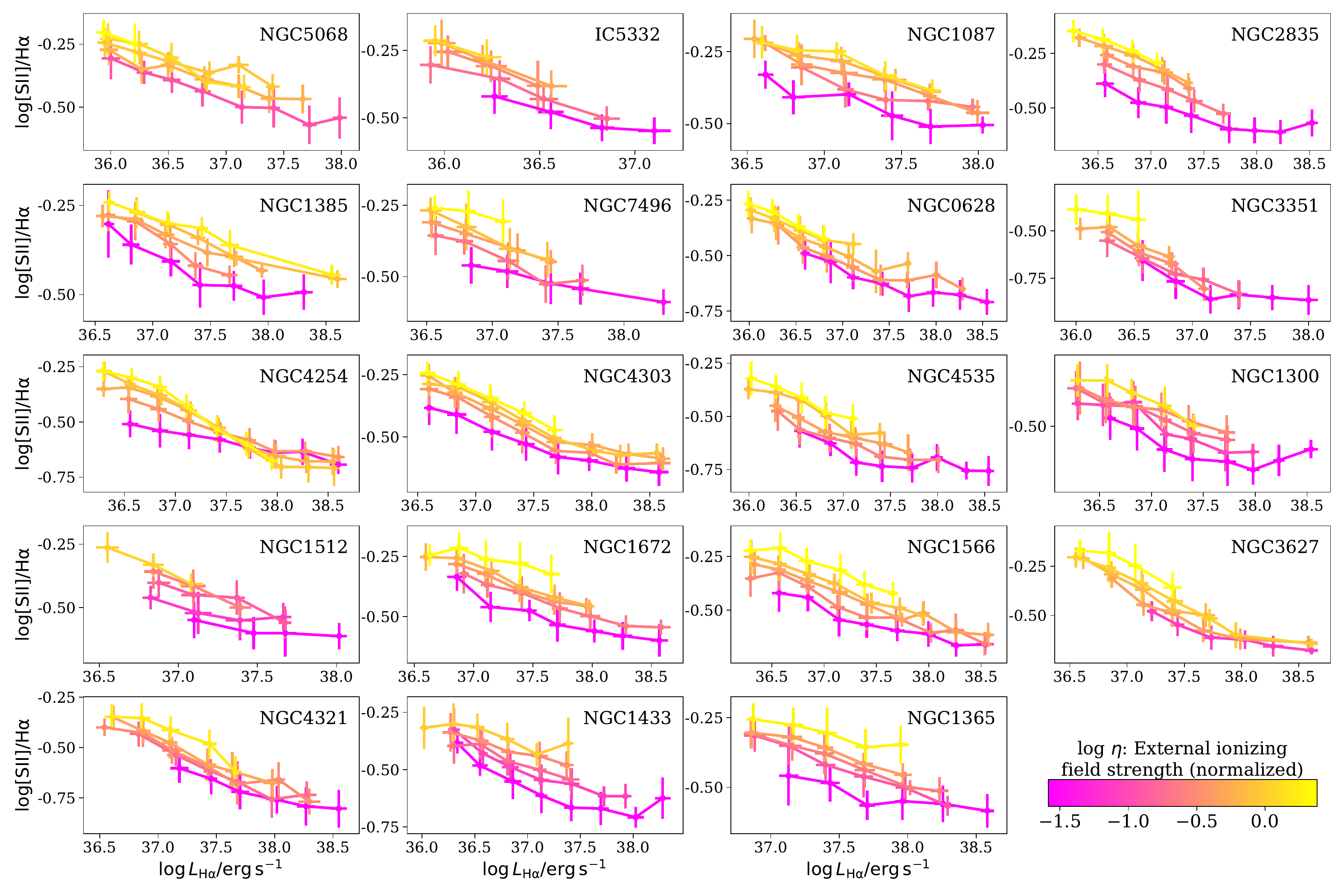}
\caption{\textbf{Dependence of \sii/\halpha\ on $L_{\mathrm{H}\alpha}$ and radiative environment across PHANGS-MUSE HII regions.}
Each panel shows stacked \sii/\halpha\ in bins of $L_{\mathrm{H}\alpha}$ and $\eta$, where $\eta$ quantifies the strength of the external ionizing radiation field from nearby HII regions, for individual PHANGS-MUSE galaxies. The galaxies are indicated in each panel, and are ordered according to stellar mass.}
\label{f:SII_Halpha_vs_Lhalpha_eta_all_gal}
\end{figure*}

\begin{figure*}
	\centering
\includegraphics[width=1\textwidth]{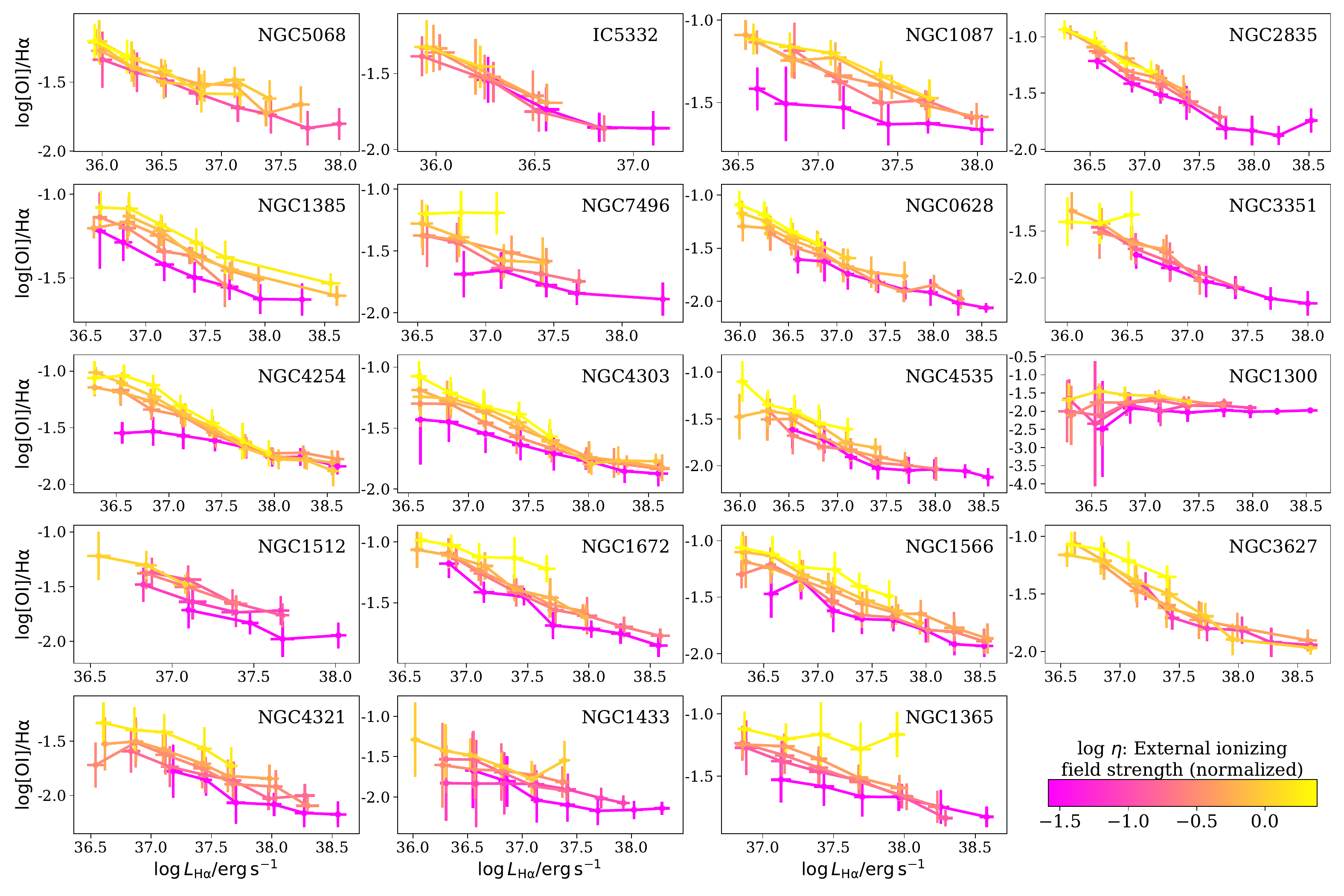}
\caption{\textbf{Dependence of \oi/\halpha\ on $L_{\mathrm{H}\alpha}$ and radiative environment across PHANGS-MUSE HII regions.}
Each panel shows stacked \oi/\halpha\ in bins of $L_{\mathrm{H}\alpha}$ and $\eta$, where $\eta$ quantifies the strength of the external ionizing radiation field from nearby HII regions, for individual PHANGS-MUSE galaxies. The galaxies are indicated in each panel, and are ordered according to stellar mass.}
\label{f:OI_Halpha_vs_Lhalpha_eta_all_gal}
\end{figure*}

Figures~\ref{f:OIII_Hbeta_vs_Lhalpha_eta_all_gal}–\ref{f:OI_Halpha_vs_Lhalpha_eta_all_gal} show the dependence of line ratios (\oiii/\hbeta, \nii/\halpha, \sii/\halpha, and \oi/\halpha) on $L_{\mathrm{H}\alpha}$ and radiative environment for PHANGS-MUSE HII regions. The radiative environment is defined in Equation~\ref{eq:eta}, where the external radiation field is normalized by the luminosity of each HII region; low (high) values of $\eta$ correspond to weak (strong) external radiation fields.

\section{Line-ratio vs $\Sigma$(\halpha) trends in the DIG-subtracted nebular catalog}\label{app:DIGsub_trends}
\setcounter{figure}{0}

\begin{figure*}
	\centering
\includegraphics[width=1\textwidth]{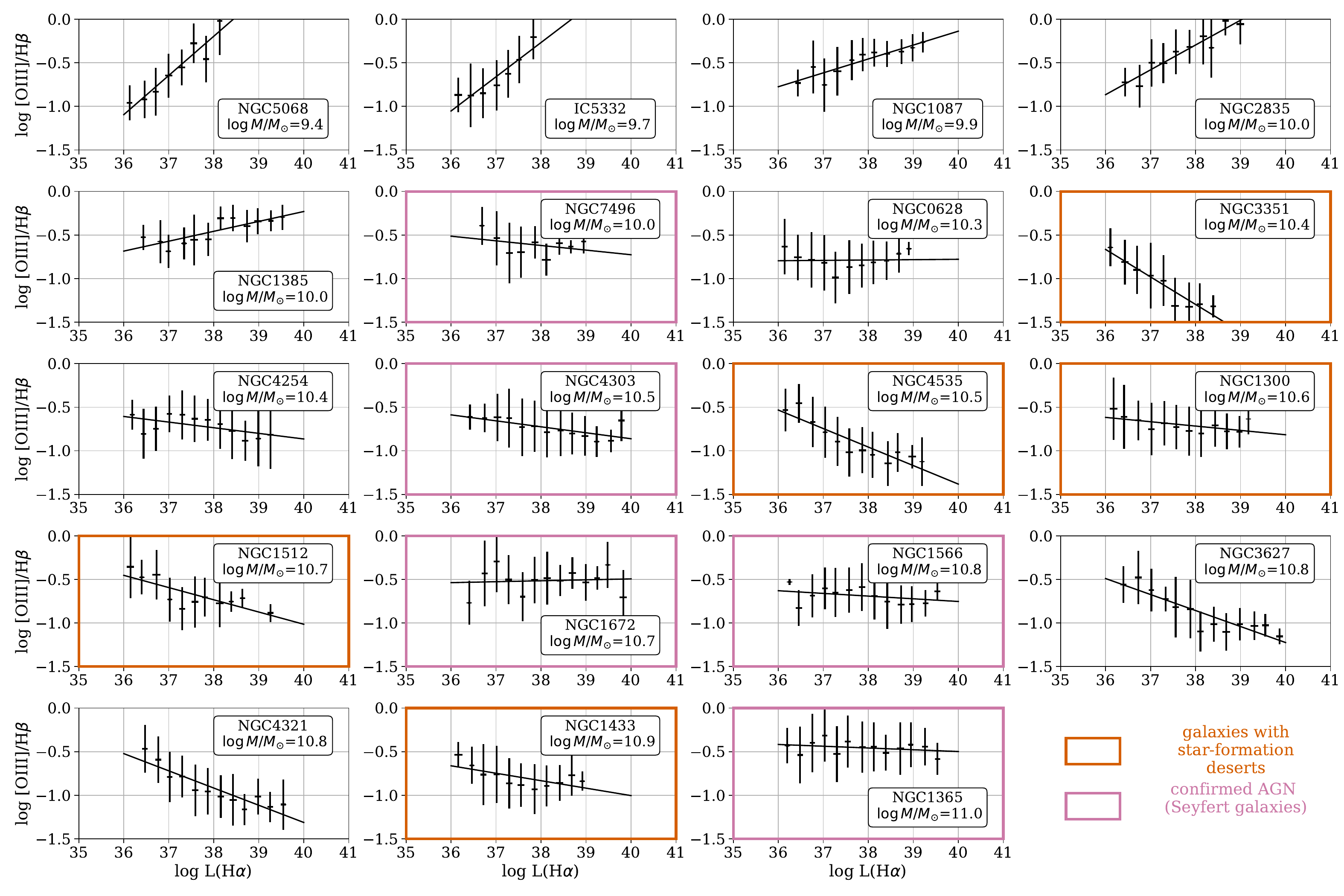}
\caption{\textbf{Variation of \oiii/\hbeta\ with \halpha\ luminosity for DIG-corrected HII regions across the PHANGS-MUSE galaxies.} Each panel shows a galaxy, ordered by increasing stellar mass. Pink and orange borders indicate galaxies hosting an AGN or extended star-formation desert, respectively. Markers show stacked \oiii/\hbeta\ in bins of \halpha\ luminosity for HII regions after subtraction of the DIG contribution using the \citet{congiu23} catalog. Black lines indicate the best-fitting linear relation.}
\label{f:OIIIHbeta_vs_LHa_DIGcorr}
\end{figure*}

\begin{figure*}
	\centering
\includegraphics[width=1\textwidth]{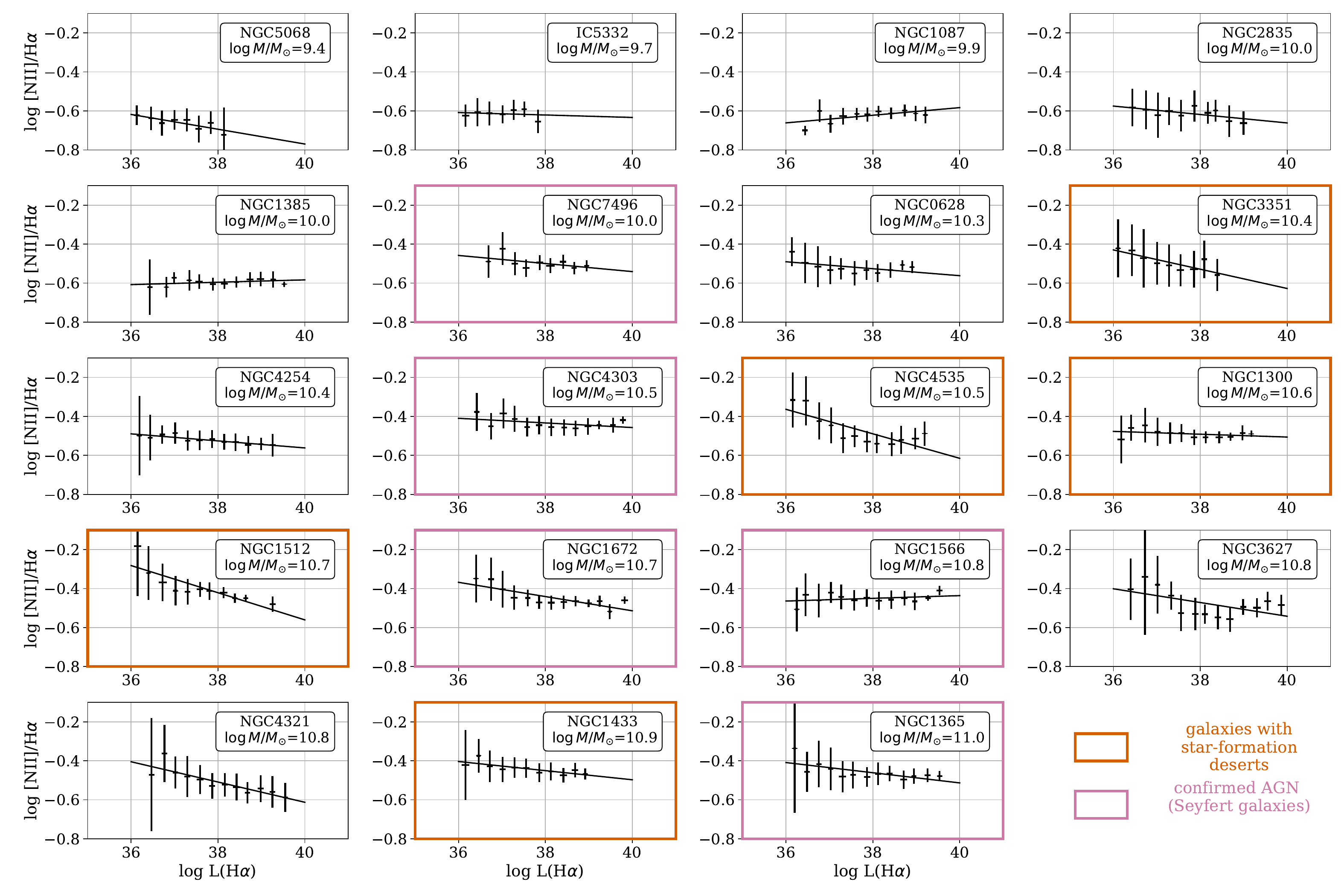}
\caption{\textbf{Variation of \nii/\halpha\ with \halpha\ luminosity for DIG-corrected HII regions across the PHANGS-MUSE galaxies.} Each panel shows a galaxy, ordered by increasing stellar mass. Pink and orange borders indicate galaxies hosting an AGN or extended star-formation desert, respectively. Markers show stacked \nii/\halpha\ in bins of \halpha\ luminosity for HII regions after subtraction of the DIG contribution using the \citet{congiu23} catalog. Black lines indicate the best-fitting linear relation.}
\label{f:NIHalpha_vs_LHa_DIGcorr}
\end{figure*}

\begin{figure*}
	\centering
\includegraphics[width=1\textwidth]{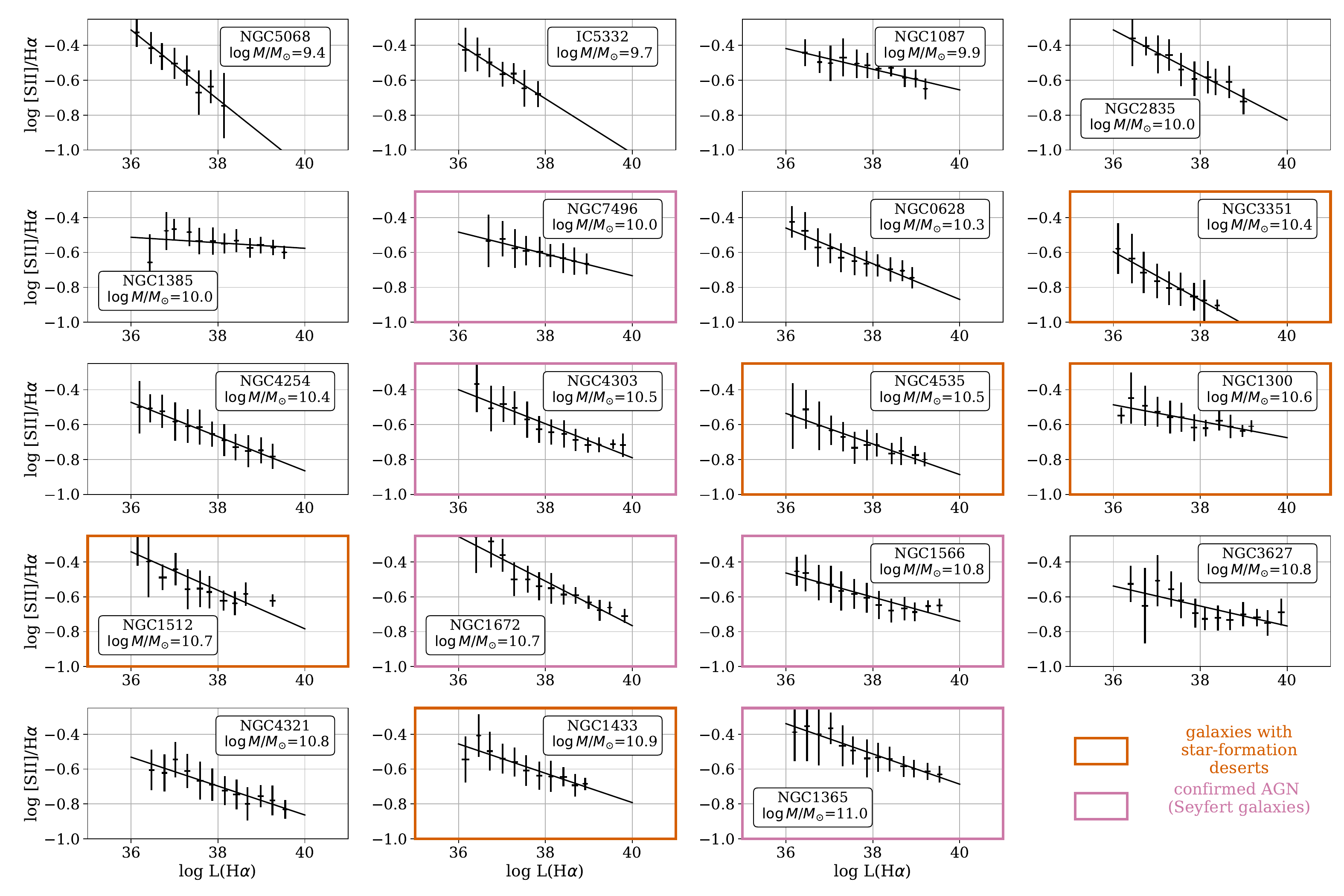}
\caption{\textbf{Variation of \sii/\halpha\ with \halpha\ luminosity for DIG-corrected HII regions across the PHANGS-MUSE galaxies.} Each panel shows a galaxy, ordered by increasing stellar mass. Pink and orange borders indicate galaxies hosting an AGN or extended star-formation desert, respectively. Markers show stacked \sii/\halpha\ in bins of \halpha\ luminosity for HII regions after subtraction of the DIG contribution using the \citet{congiu23} catalog. Black lines indicate the best-fitting linear relation.}
\label{f:SIHalpha_vs_LHa_DIGcorr}
\end{figure*}

\begin{figure*}
	\centering
\includegraphics[width=1\textwidth]{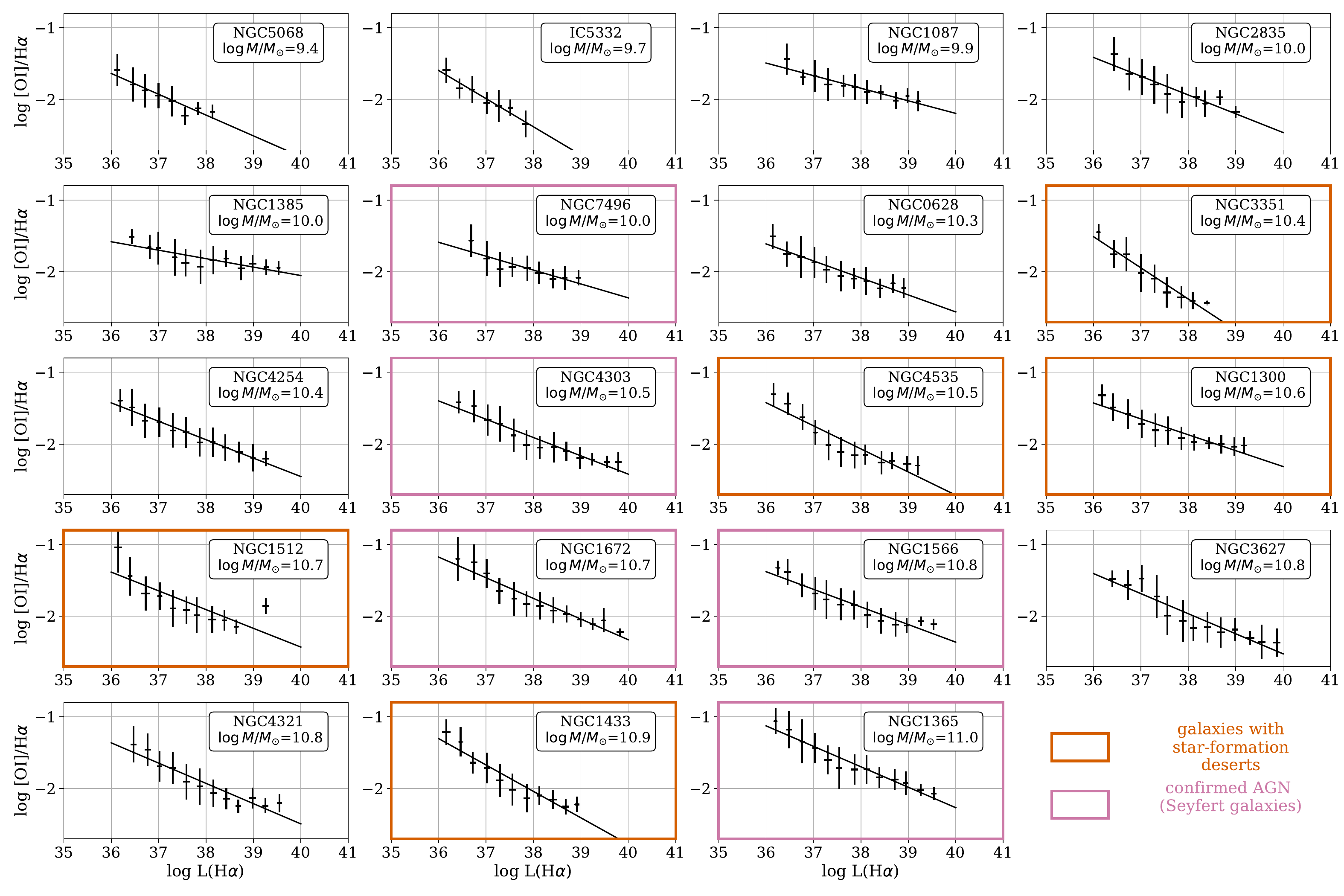}
\caption{\textbf{Variation of \oi/\halpha\ with \halpha\ luminosity for DIG-corrected HII regions across the PHANGS-MUSE galaxies.} Each panel shows a galaxy, ordered by increasing stellar mass. Pink and orange borders indicate galaxies hosting an AGN or extended star-formation desert, respectively. Markers show stacked \oi/\halpha\ in bins of \halpha\ luminosity for HII regions after subtraction of the DIG contribution using the \citet{congiu23} catalog. Black lines indicate the best-fitting linear relation.}
\label{f:OIHalpha_vs_LHa_DIGcorr}
\end{figure*}

Figures~\ref{f:OIIIHbeta_vs_LHa_DIGcorr}--\ref{f:OIHalpha_vs_LHa_DIGcorr} show stacked optical line ratios as a function of \halpha\ luminosity for DIG-corrected HII regions in the PHANGS-MUSE sample, using the \citet{congiu23} catalog.

\section{Radiative coupling between HII regions: numerical methods and parameter exploration}\label{app:HII_rad_coupling_app}
\setcounter{figure}{0}

We explore two classes of solutions to Equation~\ref{eq:HII_prop_eq}. The first assumes that $f_{\mathrm{esc}}$ is known and constant across all HII regions, reducing the problem to a linear system of equations. The second assumes that $f_{\mathrm{esc},i}$ is itself a prescribed function of the unknown ionizing photon production rate, $Q_i$, resulting in a non-linear coupled system.


Section~\ref{app:impact_of_geometry_and_mean_free_path} compares the model results for a range of geometry and mean free path assumptions. Section~\ref{app:impact_of_fesc} examines the impact of varying a constant $f_{\mathrm{esc}}$, while Section~\ref{app:impact_of_fesc_w_Q} explores the impact of different functional forms for $f_{\mathrm{esc}}(Q)$.

\begin{figure*}
	\centering
\includegraphics[width=1\textwidth]{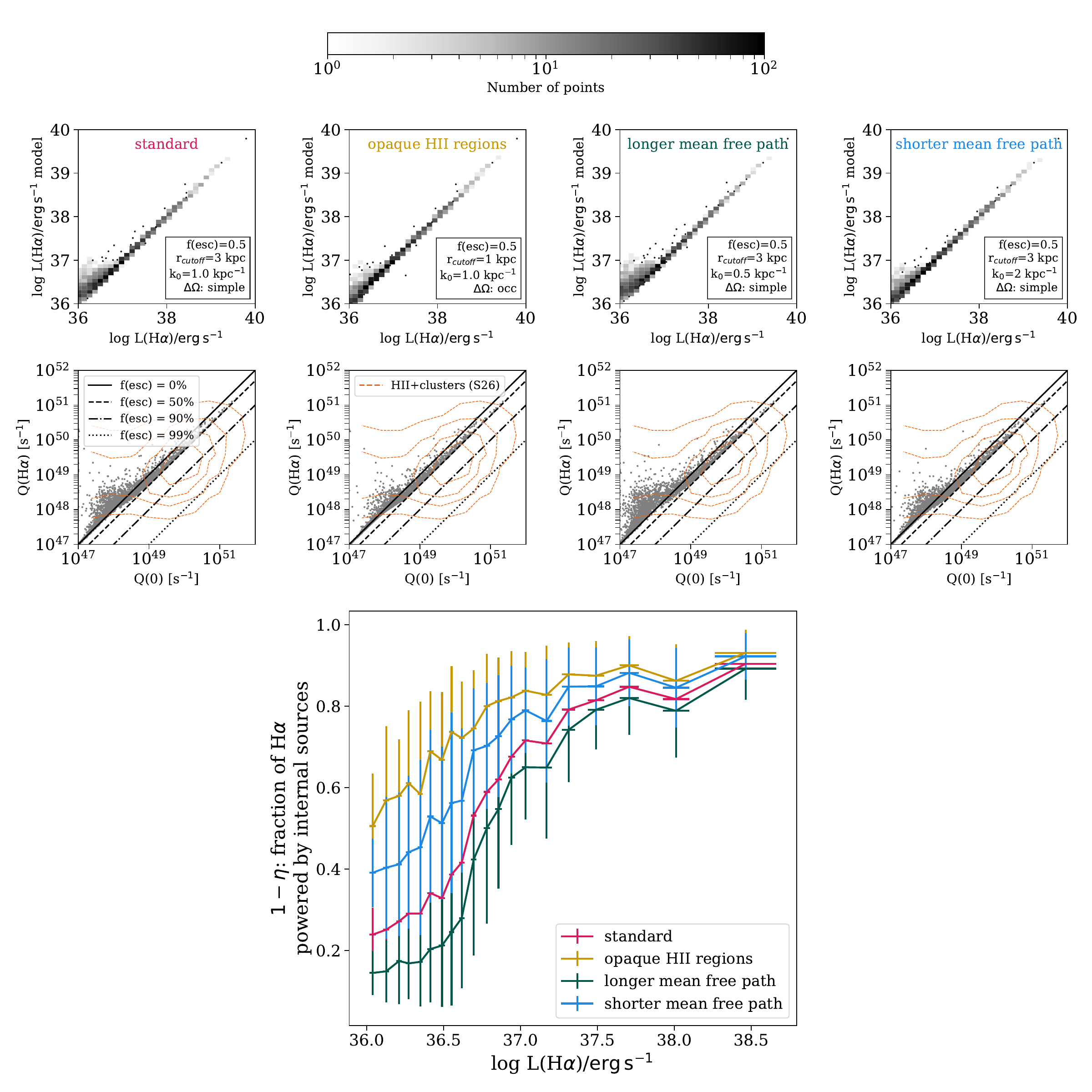}
\caption{\textbf{Models of radiative coupling between neighboring HII regions in NGC~628: I. impact of geometry and mean free path assumptions.} The four columns show different realizations of the radiation propagation model, varying the interaction geometry ($\Delta \Omega_{ij}$) and mean free path ($\lambda_0=1/k_0$; see legends). 
\textbf{Top row:} observed versus model-predicted \halpha\ luminosities for HII regions, shown as 2D histograms with logarithmic color scaling; black points indicate bins containing a single source. 
\textbf{Middle row:} intrinsic ionizing photon production rates derived from the model, $\mathrm{Q(0)}$, compared to $\mathrm{Q(H\alpha)}$ inferred directly from the observed \halpha\ luminosity assuming purely internal powering (as done in \citealt{scheuermann26}; S26). Orange contours show the measurements of star clusters+HII regions in PHANGS-MUSE from S26, and the black lines indicate constant apparent escape fractions, under the same assumption. \textbf{Bottom row:} median and median absolute deviation of the internally powered fraction of the observed \halpha\ luminosity, $1-\eta$, as a function of HII region luminosity. $1-\eta=1$ corresponds to HII regions whose observed \halpha\ emission is entirely powered by internal sources, while $1-\eta=0.5$ corresponds to regions where only half of the observed \halpha\ emission is powered internally, with the remainder supplied by leaking radiation from neighboring HII regions.} 
\label{f:model_comparison_1_DIGcorr}
\end{figure*}

\begin{figure*}
	\centering
\includegraphics[width=1\textwidth]{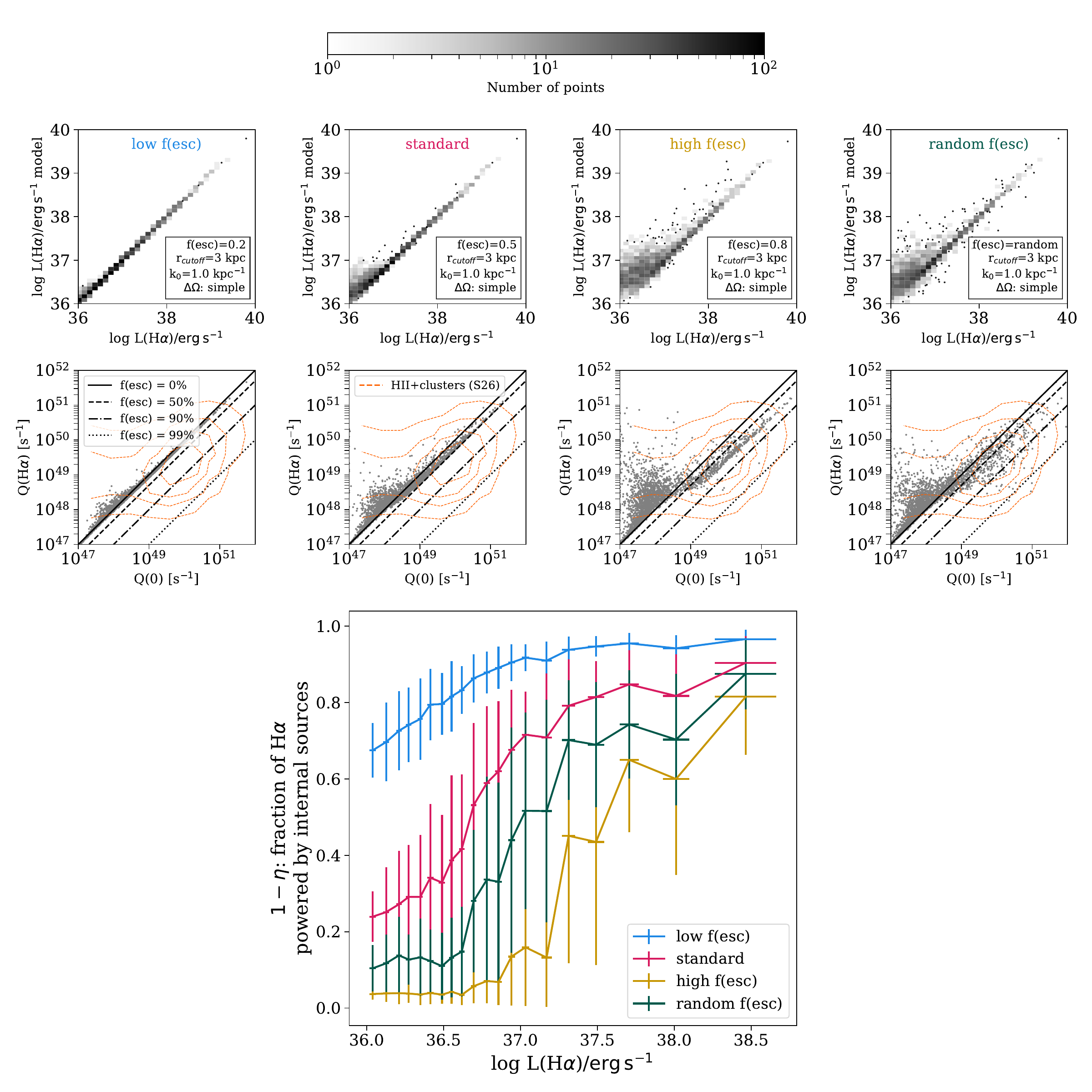}
\caption{\textbf{Models of radiative coupling between neighboring HII regions in NGC~628: II. impact of varying $f_{\mathrm{esc}}$.} Same as Figure~\ref{f:model_comparison_1_DIGcorr}, but exploring models with different assumed escape fractions, as indicated in the legends. The \texttt{random} model assigns each HII region an escape fraction drawn independently from a uniform distribution between 0 and 1.} 
\label{f:model_comparison_2_DIGcorr}
\end{figure*}

\subsection{Impact of geometry and mean free path assumptions}\label{app:impact_of_geometry_and_mean_free_path}

Figure~\ref{f:model_comparison_1_DIGcorr} shows the model results under different realizations of the radiation propagation framework. Our fiducial (``standard'') model assumes $f_{\mathrm{esc}}=0.5$, $k_0 = 1\,\mathrm{kpc^{-1}}$, the \texttt{simple} definition of $\Delta \Omega_{ij}$, and includes neighboring HII regions within $r_{ij} \leq r_{\mathrm{cutoff}} = 3\,\mathrm{kpc}$\footnote{The results reported in this section do not depend significantly on $r_{\mathrm{cutoff}}$ in the range $1-5\,\mathrm{kpc}$.}. We additionally explore models with optically thick HII regions (using the \texttt{occ} definition of $\Delta \Omega_{ij}$), and longer/shorter mean free paths ($k_0 = 0.5\,\mathrm{kpc^{-1}}$ and $k_0 = 1.0\,\mathrm{kpc^{-1}}$). 

The top row compares the observed \halpha\ luminosities of HII regions to those predicted by the model. The goodness of fit is quantified by the scatter about the one-to-one relation, which is excellent for $L_{\mathrm{H}\alpha} \gtrsim 10^{37}\,\mathrm{erg\,s^{-1}}$. At lower luminosities, the scatter increases and the model overpredicts \halpha\ luminosities for a subset of faint HII regions. This deviation arises naturally from radiative coupling between nearby HII regions ($r_{ij} \lesssim 300\,\mathrm{pc}$) spanning a wide luminosity range ($\sim$2.5 dex in NGC~628): faint regions located near bright neighbors can receive sufficient externally supplied ionizing radiation that their predicted recombination luminosity exceeds the observed value.

Reducing this excess requires substantially lower escape fractions ($f_{\mathrm{esc}} \sim 0.2$; see next section). However, such values would be in tension with independent observational constraints, including the DIG \halpha\ fraction of $\sim0.4$ measured by \citet{belfiore22}, which implies that a significant fraction of ionizing photons must escape HII regions to power the diffuse emission.

The middle row compares the model-derived intrinsic ionizing photon production rates, $Q_i$, to the ionizing photon rates inferred directly from the observed \halpha\ luminosities assuming purely internal powering, as in \citet{scheuermann26}. The orange contours show the NGC~628 distribution derived by \citet{scheuermann26} from UV--IR SED fitting of young stellar clusters associated with HII regions. None of the models reproduces the highest-$Q$ systems, consistent with the fact that \citet{scheuermann26} infer escape fractions of $\sim0.8$, whereas we here adopt $f_{\mathrm{esc}}=0.5$. However, the models display a qualitatively comparable low-$Q$ tail where the observed \halpha\ luminosity exceeds that expected from the associated stellar cluster alone. In our framework, these systems correspond to HII regions whose observed \halpha\ emission is significantly supplemented by leaking radiation from neighboring regions.

The bottom row of Figure~\ref{f:model_comparison_1_DIGcorr} shows the median values of $1-\eta$ as a function of HII region \halpha\ luminosity. All models display a characteristic S-shaped trend: bright HII regions are dominated by internal powering, with only $\sim10\%$ external contribution, while progressively fainter HII regions experience increasingly large contamination from externally supplied ionizing radiation. In particular, HII regions with $L_{\mathrm{H}\alpha}\sim10^{37}\,\mathrm{erg\,s^{-1}}$ are predicted to have roughly 20--40\% of their \halpha\ emission powered by external radiation.

The degree of contamination by leaking radiation in faint HII regions depends on the adopted modeling assumptions. The optically thick (\texttt{occ}) case yields the weakest contamination, as a larger fraction of the leaking radiation is blocked by intervening HII regions before reaching neighboring systems. As discussed in Section~\ref{sec:leaking_HII}, however, the \texttt{occ} model likely overestimates the degree of blocking expected in real galaxies, where HII regions should exhibit some vertical scatter. We therefore regard the \texttt{simple} and \texttt{occ} prescriptions as bracketing the plausible range of geometries in real galaxies.

The inferred contamination also depends on the assumed mean free path. Shorter mean free paths lead to systematically weaker contamination, particularly for faint HII regions, while longer mean free paths produce stronger contamination. This behavior is expected, since increasing the mean free path allows ionizing photons to propagate farther through the ISM, increasing the amount of leaking radiation available to power neighboring HII regions.

\begin{figure*}
	\centering
\includegraphics[width=1\textwidth]{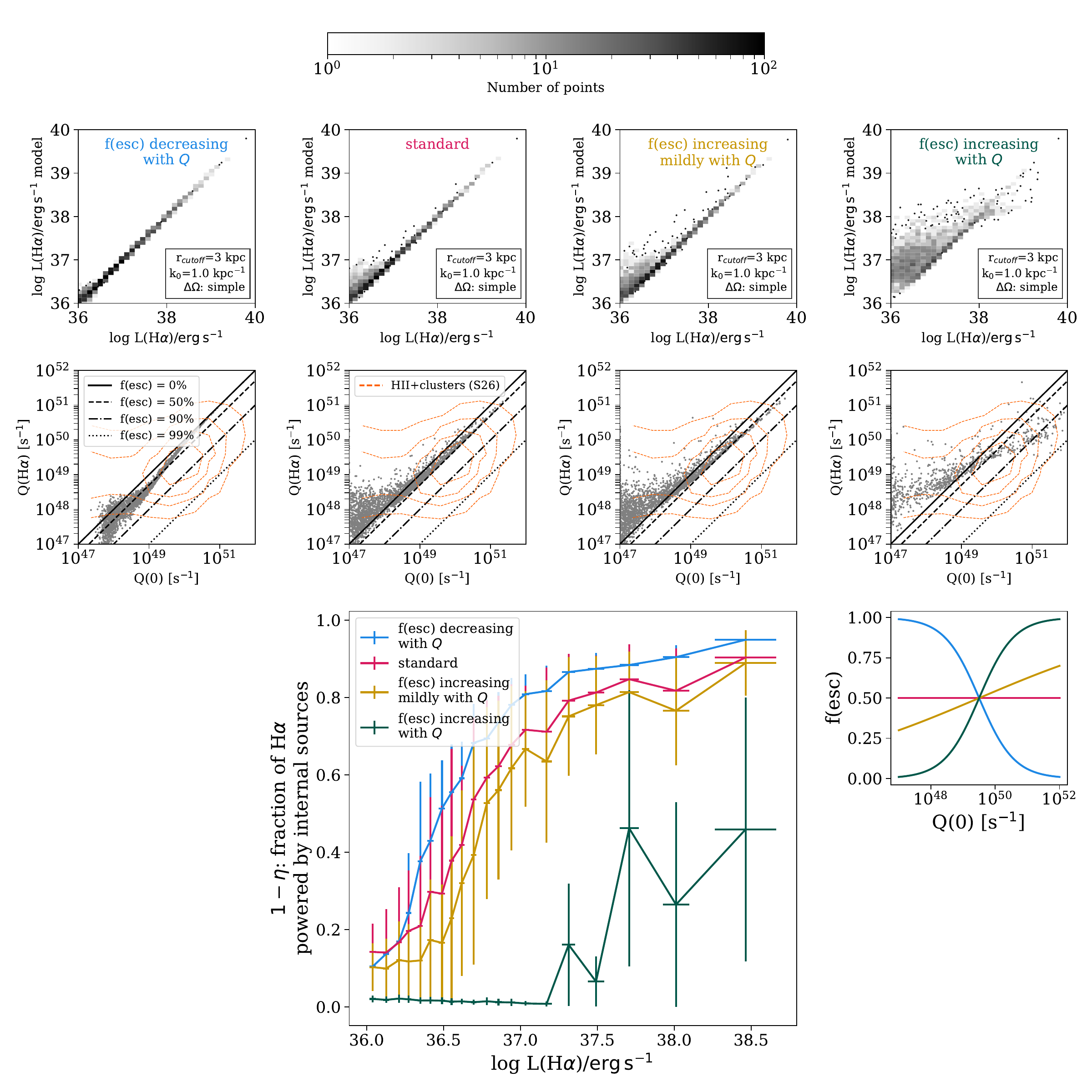}
\caption{\textbf{Models of radiative coupling between neighboring HII regions in NGC~628: III. impact of functional forms for $f_{\mathrm{esc}}(Q_i)$.} Same as Figure~\ref{f:model_comparison_1_DIGcorr}, but exploring models in which $f_{\mathrm{esc}}$ is a function of the unknown $Q_i$. We consider cases where $f_{\mathrm{esc}}$ decreases, remains constant, increases mildly, or increases strongly with $Q_i$.} 
\label{f:model_comparison_3_DIGcorr}
\end{figure*}

\subsection{Impact of $f_\mathrm{esc}$}\label{app:impact_of_fesc}

Figure~\ref{f:model_comparison_2_DIGcorr} explores the impact of varying escape fractions, for the fiducial model parameters ($r_{\mathrm{cutoff}}=3\,\mathrm{kpc}$, $k_0 = 1\,\mathrm{kpc^{-1}}$, and \texttt{simple} $\Delta \Omega_{ij}$). We consider $f_{\mathrm{esc}}=0.2$, 0.5, and 0.8 (the latter corresponding to the average value inferred by \citealt{scheuermann26}), as well as a ``random'' model in which each HII region is assigned an escape fraction drawn from a uniform distribution between 0 and 1.

Increasing $f_{\mathrm{esc}}$ improves agreement with the $Q(\mathrm{H}\alpha)$--$Q(0)$ relation of \citet{scheuermann26}, but at the same time increases the mismatch between the observed and model-predicted \halpha\ luminosities, resulting in a progressively poorer overall goodness of fit. It also strengthens radiative coupling between HII regions, leading to a higher fraction of externally powered emission across the population. In particular, for $f_{\mathrm{esc}}=0.8$, HII regions with $L_{\mathrm{H}\alpha}\sim10^{37}\,\mathrm{erg\,s^{-1}}$ become almost entirely externally powered, and even bright regions with $L_{\mathrm{H}\alpha}\sim10^{38}\,\mathrm{erg\,s^{-1}}$ can have $\sim40\%$ of their \halpha\ emission supplied by external radiation.

In contrast, the low-escape-fraction case ($f_{\mathrm{esc}}=0.2$) yields excellent agreement between observed and predicted \halpha\ luminosities across the full luminosity range and predicts minimal external contamination for most HII regions. However, this model underpredicts both the DIG \halpha\ fraction of $\sim0.4$ measured across PHANGS-MUSE \citep{belfiore22} and the trends reported by \citet{scheuermann26} by large factors.

\subsection{Impact of varying $f_{\mathrm{esc}}(Q_i)$ functional form assumptions}\label{app:impact_of_fesc_w_Q}

Figure~\ref{f:model_comparison_3_DIGcorr} presents the model results for the case where $f_{\mathrm{esc}}$ is a function of the unknown $Q_{i}$. We explore models in which $f_{\mathrm{esc}}$ decreases with $Q$, remains constant with $Q$ (our ``standard'' model), increases mildly with $Q$, and increases significantly with $Q$. The assumed functional forms use the logistic function, $f_{\mathrm{esc},i} = L(a + b \times Q_{i})$, where $L(x)= 1 / (1 + e^{-x})$, and $a$ and $b$ are chosen to produce the different dependencies. The resulting functional forms of $f_{\mathrm{esc}}(Q)$ are shown in the figure.

In terms of the goodness-of-fit between the observed \halpha\ luminosities and those predicted by the model, we find increasing disagreement for models that assign higher escape fractions to brighter HII regions. This occurs because the radiative coupling calculation is most sensitive to the brightest HII regions, whose escape fractions strongly influence the energetics of fainter neighboring systems. The most extreme example is the model in which $f_{\mathrm{esc}}$ increases substantially with $Q$. In this case, the model fails to converge, as the predicted \halpha\ luminosities systematically exceed the observed luminosities for most HII regions. This happens because the brightest HII regions, which dominate the coupling, allow most of their ionizing radiation to escape. We therefore regard this model as unphysical.

Interestingly, the other three models, decreasing, constant, and mildly increasing $f_{\mathrm{esc}}(Q)$, do not differ significantly in the inferred contamination at fixed \halpha\ luminosity. This suggests that the spatial distribution of HII regions and their broad luminosity range are more important than the detailed leakage prescription for individual regions. All three models reproduce the characteristic S-shape in $1-\eta$ versus \halpha\ luminosity. Comparing the $Q(0)$ versus $Q(\mathrm{H}\alpha)$ distributions, the constant and mildly increasing models appear to reproduce the low-$Q$ tail seen in \citet{scheuermann26} better than the model in which $f_{\mathrm{esc}}$ decreases with $Q$.

\end{document}